# SPARSEY®: EVENT RECOGNITION VIA DEEP HIERARCHICAL SPARSE DISTRIBUTED CODES

Gerard Rinkus, Neurithmic Systems LLC

**Abstract**

The visual cortex's hierarchical, multi-level organization is captured in many biologically inspired computational vision models, the general idea being that progressively larger scale (spatially/temporally) and more complex visual features are represented in progressively higher areas. However, most earlier models use localist representations (codes) in each representational field (which we equate with the cortical macrocolumn, "mac"), at each level. In localism, each represented feature/concept/event (hereinafter "item") is coded by a single unit. The model we describe, Sparsey, is hierarchical as well but crucially, it uses sparse distributed coding (SDC) in every mac in all levels. In SDC, each represented item is coded by a small subset of the mac's units. The SDCs of different items can overlap and the size of overlap between items can be used to represent their similarity. The difference between localism and SDC is crucial because SDC allows the two essential operations of associative memory, storing a new item and retrieving the best-matching stored item, to be done in fixed time for the life of the model. Since the model's core algorithm, which does both storage and retrieval (inference), makes a single pass over all macs on each time step, the overall model's storage/retrieval operation is also fixed-time, a criterion we consider essential for scalability to the huge ("Big Data") problems. A 2010 paper described a nonhierarchical version of this model in the context of purely spatial pattern processing. Here, we elaborate a fully hierarchical model (arbitrary numbers of levels and macs per level), describing novel model principles like progressive critical periods, dynamic modulation of principal cells' activation functions based on a mac-level familiarity measure, representation of multiple simultaneously active hypotheses, a novel method of time warp invariant recognition, and we report results showing learning/recognition of spatiotemporal patterns.

## INTRODUCTION

In this paper, we provide the hierarchical elaboration of the macro/mini-column model of cortical computation described in (Rinkus 1996, Rinkus 2010) which is now named Sparsey. We report results of initial experiments involving multi-level models with multiple macrocolumns ("macs") per level, processing spatiotemporal patterns, i.e., "events". In particular, we show: a) single-trial unsupervised learning of sequences where this learning results in the formation of hierarchical spatiotemporal memory traces; and b) recognition of training sequences, i.e., exact or nearly exact reactivation of complete hierarchical traces over all frames of a sequence. The canonical macrocolumnar algorithm—which probabilistically chooses a sparse distributed code (SDC) as a function of a mac's entire input, i.e., its bottom-up (U), horizontal (H), and top-down (D) input vectors, at a given moment—operates similarly, modulo parameters, in both learning and recognition, in all macs at all levels. Computationally, Sparsey's most important property is that a mac both stores (learns) new input items—which in general are temporal-context-dependent inputs, i.e., particular spatiotemporal *moments*—and retrieves the spatiotemporally closest-matching stored item in time that remains fixed as the number of items stored in the mac increases. This property depends critically on the use of SDCs, is essential for scalability to "Big Data" problems, and *has not been shown for any other computational model, biologically inspired or not!*

---

The model has a number of other interesting neurally plausible properties, including the following. 1) A "critical period" concept wherein learning is frozen in a mac's afferent synaptic projections when those projections reach a threshold saturation. In a hierarchical setting, freezing will occur beginning with the lowest level macs (analogous to primary sensory cortex) and progress upward over the course of experience. 2) A "progressive persistence" property wherein the activation duration (persistence) of the "neurons" (and thus of the SDCs which are sets of co-active neurons) increases with level; there is some evidence for increasing persistence along the ventral visual path (Rolls and Tovee 1994, Uusitalo, Jousmäki et al. 1997, Gauthier, Eger et al. 2012). This allows an SDC in a mac at level J to associate with sequences of SDCs in Level J-1 macs with which it is connected, i.e., a chunking (compression) mechanism. In particular, this provides a means to learn in unsupervised fashion perceptual invariances produced by continuous transforms occurring in the environment (e.g., rotation, translation, etc.). Rolls' VisNet model, introduced in Rolls (1992) and reviewed in Rolls (2012), uses a similar concept to explain learning of naturally-experienced transforms, although his trace-learning-rule-based implementation differs markedly from ours. 3) During learning, an SDC is chosen on the basis of signals arriving from all active afferent neurons in the mac's total (U, H, and D) receptive field (RF). However, during retrieval, if the highest-order match, i.e., involving all three (U, H, and D) input sources, falls below a threshold, the mac considers a progression of lower-order matches, e.g., involving only its U and D inputs, but ignoring its H inputs, and if that also falls below a threshold, a match involving only its U inputs. This "back-off" protocol, in conjunction with progressive persistence, allows a protocol by which the model can rapidly—*crucially, the protocol does not increase the time complexity of closest-match retrieval*—compare a test sequence (e.g., video snippet) not only to the set of all sequences *actually* experienced and stored, but to a much larger space of nonlinearly time-warped variants of the actually-experienced sequences. 4) During retrieval, multiple competing hypotheses can momentarily (i.e., for one or several frames) be co-active in any given mac and resolve to a single hypothesis as subsequent disambiguating information enters.

While the results reported herein are specifically for the unsupervised learning case, Sparsey also implements supervised learning in the form of cross-modal unsupervised learning, where one of the input modalities is treated as a label modality. That is, if the same label is co-presented with multiple (arbitrarily different) inputs in another (raw sensory) modality, then a single internal representation of that label can be associated with the multiple (arbitrarily different) internal representations of the sensory inputs. That internal representation of the label then *de facto* constitutes a representation of the class that includes all those sensory inputs regardless of how different they are, providing the model a means to learn essentially arbitrarily nonlinear categories (invariances), i.e., instances of what Bengio terms "AI Set" problems (Bengio 2007). Although we describe this principle in this paper, its full elaboration and demonstration in the context of supervised learning will be treated in a future paper.

Regarding the model's possible neural realization, our primary concern is that all of the model's formal structural and dynamic properties/mechanisms be *plausibly realizable* by known neural principles. For example, we do not give a detailed neural model of the winner-take-all (WTA) competition that we hypothesize to take place in the model's minicolumns, but rather rely on the plausibility of any of the many detailed models of WTA competition in the literature, e.g., (Grossberg 1973, Yu, Giese et al. 2002, Knoblich, Bouvrie et al. 2007, Oster, Douglas et al. 2009, Jitsev 2010). Nor do we give a detailed neural model for the mac's computation of the overall spatiotemporal familiarity of its input (the *"G"* measure), or for the G-contingent modulation of neurons' activation functions. Furthermore, the model relies only upon binary neurons and a simple synaptic learning model. This paper is really most centrally an explanation of why and how the use of SDC in conjunction with hierarchy provides a computationally efficient, scalable, and neurally plausible solution to event (i.e., single- or multimodal spatiotemporal pattern) learning and recognition.

# I.   OVERALL MODEL CONCEPT

The remarkable structural homogeneity across the neocortical sheet suggests a canonical circuit/algorithm, i.e., a core computational module, operating similarly in all regions (Douglas, Martin et al. 1989, Douglas and Martin 2004). In addition, DiCarlo, Zoccolan et al. (2012) present compelling first-principles arguments based on computational efficiency and evolution for a macrocolumn-sized canonical functional module whose goal they describe as "cortically local subspace untangling". We also identify the canonical functional module with the cortical "macrocolumn" (a.k.a. "hypercolumn" in V1, or "barrel"-related volumes in rat/mouse primary somatosensory cortex), i.e., a volume of cortex, ~200-500 um in diameter, and will refer to it as a "mac". In our view, the mac's essential function, or "meta job description", in the terms of DiCarlo, Zoccolan et al. (2012), is to operate as a semi-autonomous content-addressable memory. That is, the mac:

   a) assigns (stores, learns) neural codes, specifically *sparse distributed codes* (SDCs), representing its global (i.e., combined U, H, and D) input patterns; and

   b) retrieves (reactivates) stored codes, i.e., *memories*, on subsequent occasions when the global input pattern matches a stored code sufficiently closely.

If the mac's learning process ensures that *similar inputs map to similar code*s (SISC), as Sparsey's does, then operating as a content addressable memory is functionally equivalent to local subspace untangling.

Although the majority of neurophysiological studies through the decades have formalized the responses of cortical neurons in terms of purely spatial receptive fields (RFs), evidence revealing the truly spatiotemporal nature of neuronal RFs is accumulating (DeAngelis, Ohzawa et al. 1993, DeAngelis, Ghose et al. 1999, Rust, Schwartz et al. 2005, Gavornik and Bear 2014, Ramirez, Pnevmatikakis et al. 2014). In our mac model, time is discrete: U signals arrive from neurons active on the current time step while H and D signals arrive from neurons active on the previous time step. We can view the combined U, H, and D inputs as a 'context-dependent U input' (where the H and D signals are considered the 'context') or more holistically, as an overall particular spatiotemporal *moment* (as suggested earlier).

As will be described in detail, the first step of the mac's canonical algorithm, during both learning and retrieval, is to combine its U, H, and D inputs to yield a (scalar) judgment, $G$, as to the spatiotemporal *familiarity* of the current *moment*. Provided the number of codes stored in the mac is small enough, $G$ measures the *spatiotemporal similarity* of the best matching stored moment, $x$, to the current moment, $I$.

$$G = \arg\max_{x}(sim(I, x))$$

Figure I-1 shows the envisioned correspondence of Sparsey to the cortical macrocolumn. In particular, we view the mac's sub-population of L2/3 pyramidals as the actual repository of SDCs. And even more specifically, we postulate that the approximately 20 L2/3 pyramidals in each of the mac's approximately 70 minicolumns function in winner-take-all (WTA) fashion. Thus, a single SDC code will consist of 70 L2/3 pyramidals, one per minicolumn. Note: we also refer to minicolumns as *competitive modules* (CMs). Two-photon calcium imaging movies, e.g., Ohki, Chung et al. (2005), Sadovsky and MacLean (2014), provide some support for the existence of such macrocolumnar SDCs as they show numerous instances of ensembles, consisting of from several to hundreds of neurons, often spanning several hundred um, turning on and off as tightly synchronized wholes. We anticipate that the recently developed super-fast voltage sensor ASAP1 (St-Pierre, Marshall et al. 2014) may allow much higher fidelity testing of SDCs and Sparsey in general.

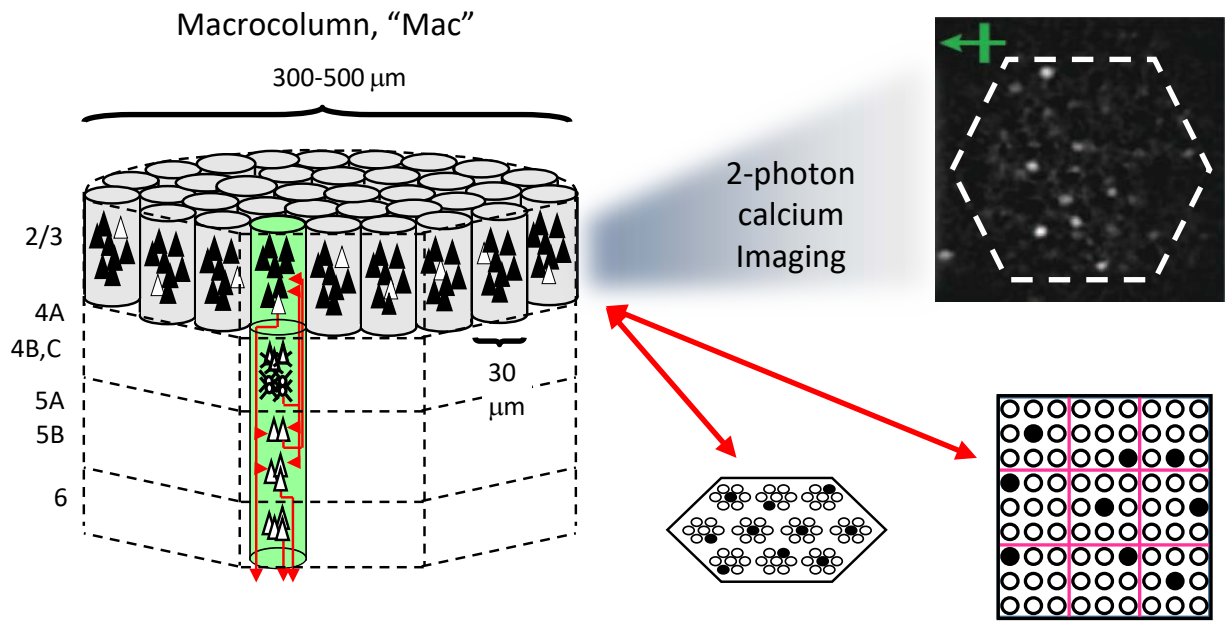

**Figure I-1:** Proposed correspondence between the cortical macrocolumn and Sparsey's mac. Left: schematic of a cortical macrocolumn composed of ~70 minicolumns (green cylinder). SDCs representing context-dependent inputs reside in mac's L2/3 population. An SDC is a set composed of one active L2/3 pyramidal cell per minicolumn. Upper Right: 2-photon calcium image of activity in a mac-sized area of cat V1 given a left-moving vertical bar in the mac's RF; we have added dashed hexagonal boundary to suggest the boundary of macrocolumn/hypercolumn module [adapted from Ohki, Chung et al. (2005)]. Lower Right: two formats that we use to depict macs; they show only the L2/3 cells. The hexagonal format mac has 10 minicolumns each with seven cells. The rectangular format mac has nine minicolumns each with nine cells. Note that in these formats, active cells are black (or red as in many subsequent figures); inactive cells are white.

Figure I-2 (left) illustrates the three afferent projections to a particular mac at level L1 (analog of cortical V1), $M_i^1$ (i.e., the $i^{th}$ mac at level L1). The red hexagon at L0 indicates $M_i^1$'s aperture onto the thalamic representation of the visual space, i.e., its classical receptive field (RF), which we can refer to more specifically as $M_i^1$'s U-RF. This aperture consists of about 40 binary pixels connected all-to-all with $M_i^1$'s cells; black arrows show representative U-weights (U-wts) from two active pixels. Note that we assume that visual inputs to the model are filtered to single-pixel-wide edges and binarized. The blue semi-transparent prism represents the full bundle of U-wts comprising $M_i^1$'s U-RF.

The all-to-all *U-connectivity* within the blue prism is essential because the concept of the RF of a mac *as a whole*, *not of an individual cell*, is central to our theory. This is because the "atomic coding unit", or equivalently, the "atomic unit of meaning" in this theory is the SDC, i.e., a *set* of cells. The activation of a mac, during both learning and recognition, consists in the activation of an entire SDC, i.e., simultaneous activation of one cell in every minicolumn. Similarly, deactivation of a mac consists in the simultaneous deactivation of *all* cells comprising the SDC (though in general, some of the cells contained in a mac's currently active SDC might also be contained in the next SDC to become active in that mac). Thus, in order to be able to view an SDC as *collectively* (or *atomically*) representing the input to a mac as a whole, all cells in a mac must have the same RF (the same set of afferent cells). This scenario is assumed throughout this report.

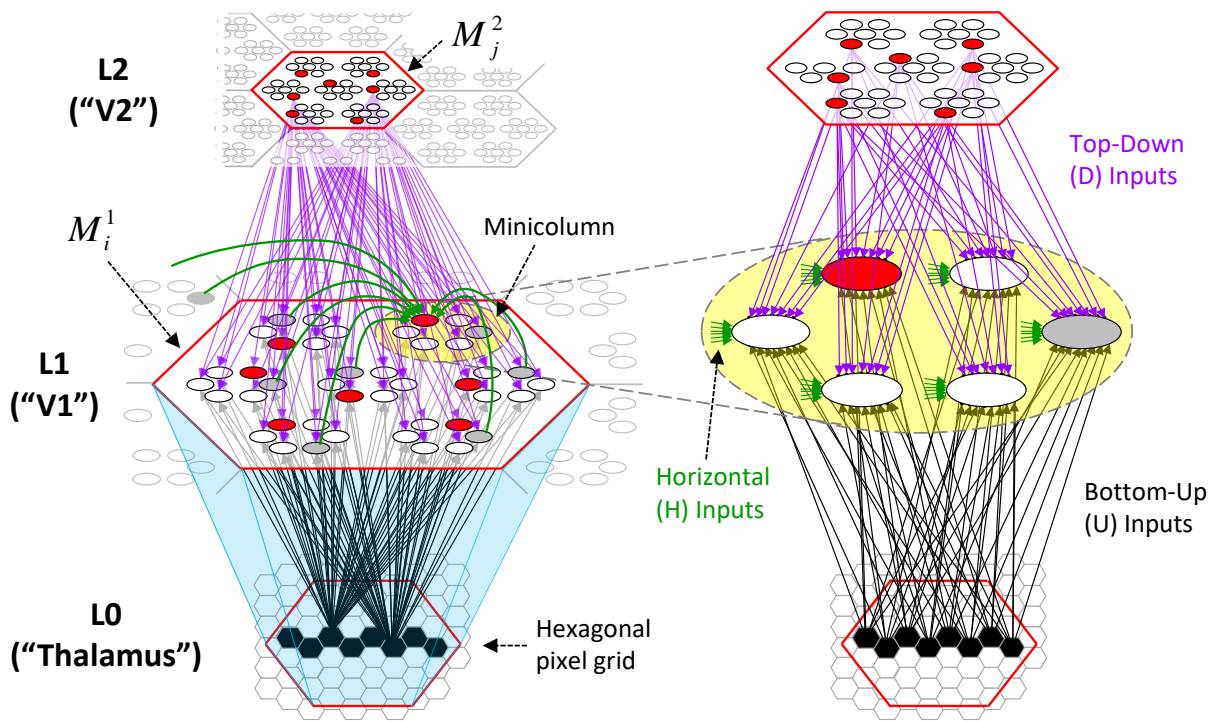

**Figure I-2:** Detail of afferent projections to a mac. See text for description.

In Figure I-2, magenta lines represent the D-wts comprising $M_i^1$'s afferent D projection, or D-RF. In this case, $M_i^1$'s D-RF consists of only one L2 (analog of V2) mac, $M_j^2$, which is all-to-all connected to $M_i^1$ (representative D-wts from just two of $M_j^2$'s cells are shown). Any given mac also receives complete H-projections from all nearby macs in its own level (including itself) whose centers fall within a parameter-specifiable radius of its own center. Signals propagating via H-wts are defined to take one time step (one sequence item) to propagate. Green arrows show a small representative sample of H-wts mediating signals arriving form cells active on the prior time step (gray). Red indicates cells active on current time step. At right of Figure I-2, we zoom in on one of $M_i^1$'s minicolumns (CMs) to emphasize that every cell in a CM has the same H-, U-, and D-RFs. Figure I-3 further illustrates (using the rectangular format for depicting macs) the concept that all cells in a given mac have the same U-, H-, and D-RFs and that those RFs respect the borders of the source macs. Each cell in the L1 mac, $M_{(2,2)}^1$ (here we use an alternate (x,y) coordinate indexing convention for the macs), receives a D-wt from all cells in all five L2 macs indicated, an H-wt from all cells in $M_{(2,2)}^1$ and its N, S, E, and W neighboring macs (green shading), and a U-wt from all 36 cells in the indicated aperture.

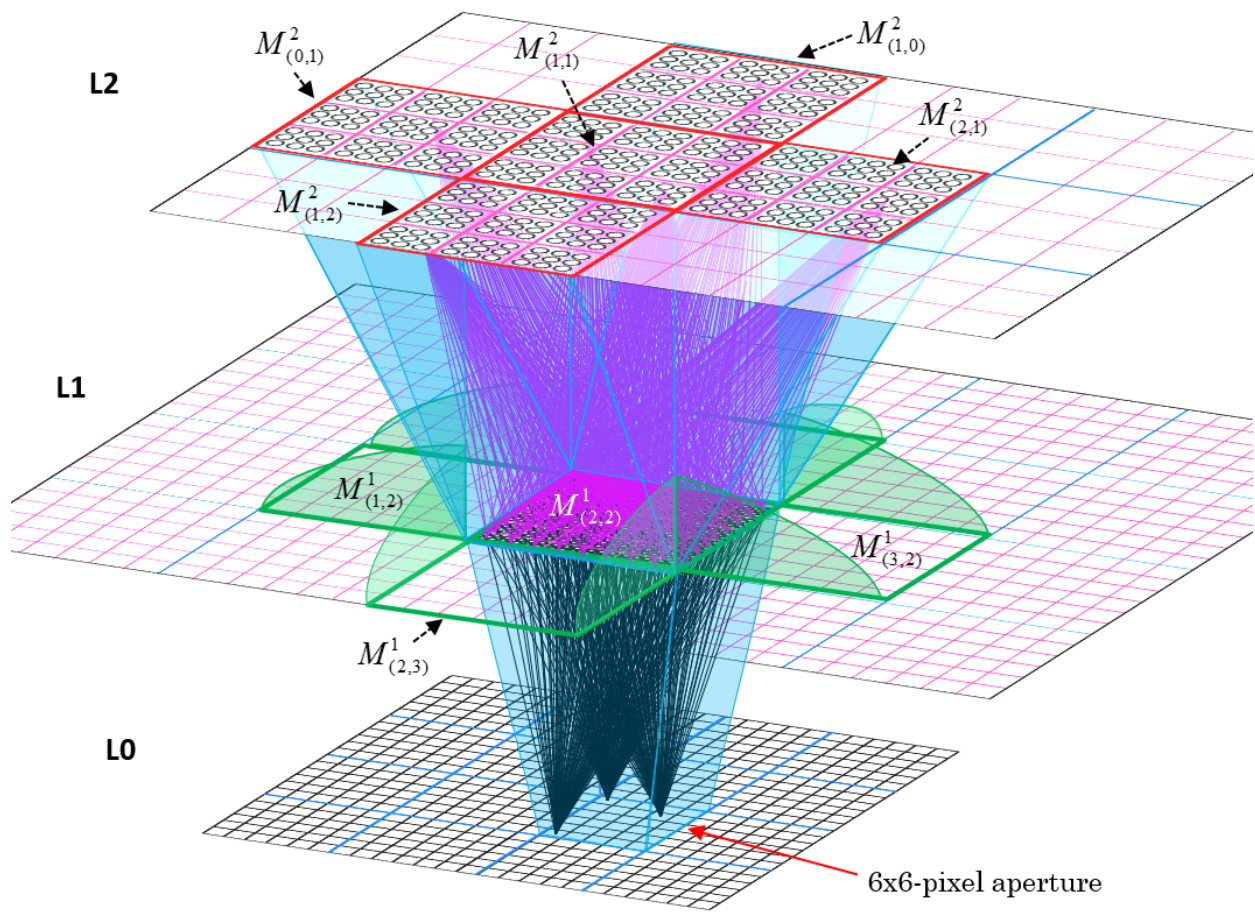

**Figure I-3:** Connectivity scheme. Within each of the three afferent projections, H, U, and D, to a mac, $M^1_{(2,2)}$ (where the mac index is now in terms of (x,y) coordinates in the level, and we have switched to the rectangular mac topology), the connectivity is full and respects mac borders. L1 is a 5x4 sheet of macs (blue borders), each consisting of 36 minicolumns (pink borders), but the scale is too small to see the individual cells within minicolumns. L2 is a 4x3 sheet of macs, each consisting of nine CMs, each consisting of nine cells.

The hierarchical organization of visual cortex is captured in many biologically inspired computational vision models with the general idea being that progressively larger scale (both spatially and temporally) and more complex visual features are represented in progressively higher areas (Riesenhuber and Poggio 1999, Serre, kouh et al. 2005). Our cortical model, Sparsey, is hierarchical as well, but as noted above, a crucial, in fact, the most crucial difference between Sparsey and most other biologically inspired vision models is that Sparsey encodes information at all levels of the hierarchy, and in every mac at every level, with SDCs. This stands in contrast to models that use *localist representations*, e.g., all published versions of the HMAX family of models, e.g., (Murray and Kreutz-Delgado 2007, Serre, Kreiman et al. 2007) and other cortically-inspired hierarchical models (Kouh and Poggio 2008, Litvak and Ullman 2009, Jitsev 2010) and the majority of graphical probability-based models (e.g., hidden Markov models, Bayesian nets, dynamic Bayesian nets). There are several other models for which SDC is central, e.g., SDM (Kanerva 1988, Kanerva 1994, Jockel 2009, Kanerva 2009), Convergence-Zone Memory (Moll and Miikkulainen 1997), Associative-Projective Neural Networks (Rachkovskij 2001, Rachkovskij and Kussul 2001), Cogent Confabulation (Hecht-Nielsen 2005), Valiant's "positive shared" representations (Valiant 2006, Feldman and Valiant 2009), and Numenta's Grok (described in Numenta white papers). However, none of these models has been substantially elaborated or demonstrated in an explicitly hierarchical architecture and most have not been substantially elaborated for the spatiotemporal case.

Figure I-4 illustrates the difference between a localist, e.g., an HMAX-like, model and the SDC-based Sparsey model. The input level (analogous to thalamus) is the same in both cases: each small gray/red hexagon in the input level represents the aperture (U-RF) of a single V1 mac (gray/red hexagon). In Figure I-4a, the representation used in each mac (at all levels) is *localist*, i.e., each feature is represented by a single cell and at any one time, only one cell (feature) is active (red) in any given mac (here the cell is depicted with an icon representing the feature it represents). In contrast, in Figure I-4b, any particular feature is represented by a *set* of co-active cells (red), one in each of a mac's minicolumns: compare the two macs at lower left of Figure I-4a with the corresponding macs in Figure I-4b (blue and brown arrows). Any given cell will generally participate in the codes of many different features. A yellow call-out shows codes for other features stored in the mac, besides the feature that is currently active. If you look closely, you can see that for some macs, some cells are active in more than one of the codes.

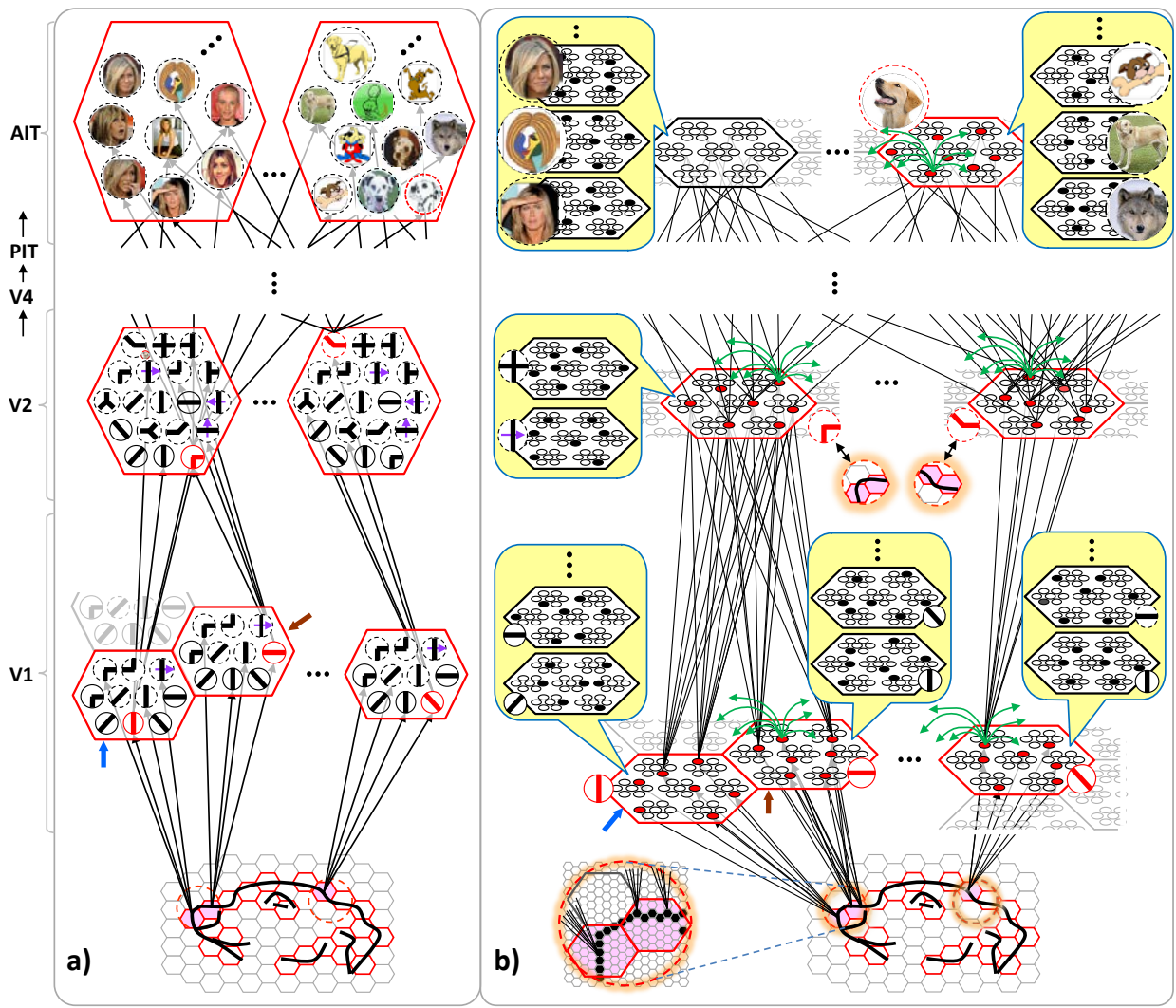

**Figure I-4:** Comparison of a localist (a) and an SDC-based (b) hierarchical vision model. See text.

7                                                                              http://dx.doi.org/10.3389/fncom.2014.00160

Looking at Figure **I-**4a, adapted from Serre, kouh et al. (2005), one can see the basic principle of hierarchical compositionality in action. The two neighboring apertures (pink) over the dog's nose lead to activation of cells representing a vertical and a horizontal feature in neighboring V1 macs. Due to the convergence/divergence of U-projections to V2, both of these cells project to the cells in the left-hand V2 mac. Each of these cells projects to multiple cells in that V2 mac, however, only the red (active) cell representing an "upper left corner" feature, is maximally activated by the conjunction of these two V1 features. Similarly, the U-signals from the cell representing the "diagonal" feature active in the right-hand V1 mac will combine with signals representing features in nearby apertures to activate the appropriate higher-level feature in the V2 mac whose U-RF includes these apertures (small dashed circles in the input level). Note that some notion of competition (e.g., the "max" operation in HMAX models) operates amongst the cells of a mac such that at any one time, only one cell (one feature) can be active.

We underscore that in Figure **I-**4, we depict simple (solid border) and complex (dashed border) features within individual macs, implying that complex and simple features can compete with each other. We believe that the distinction between simple and complex features may be largely due to coarseness of older experimental methods (e.g., using synthetic low-dimensional stimuli): newer studies are revealing far more precise tuning functions (Nandy, Sharpee et al. 2013), including temporal context specificity, even as early as V1 (DeAngelis, Ohzawa et al. 1993, DeAngelis, Ghose et al. 1999), and in other modalities, somatosensory (Ramirez, Pnevmatikakis et al. 2014) and auditory (Theunissen and Elie 2014).

The same hierarchical compositional scheme as between V1 and V2 continues up the hierarchy (some levels not shown), causing activation of progressively higher-level features. At higher levels, we typically call them concepts, e.g., the visual concept of "Jennifer Aniston", the visual concept of the class of dogs, the visual concept of a particular dog, etc. We show most of the features at higher levels with dashed outlines to indicate that they are *complex* features, i.e., features with particular, perhaps many, dimensions of invariance, most of which are learned through experience. In Sparsey, the particular invariances are learned from scratch and will generally vary from one feature/concept to another, including within the same mac. The particular features shown in the different macs in this example are purely notional: it is the overall hierarchical compositionality principle that is important, not the particular features shown, nor the particular cortical regions in which they are shown.

The hierarchical compositional process described above in the context of the localist model of Figure **I-**4a applies to the SDC-based model in Figure **I-**4b as well. However, features/concepts are now represented by sets of cells rather than single cells. Thus, the vertical and horizontal features forming part of the dog's nose are represented with SDCs in their respective V1 macs (blue and brown arrows, respectively), rather than with single cells. The U-signals propagating from these two V1 macs converge on the cells of the left-hand V2 mac and combine, via Sparsey's *code selection algorithm* (CSA) (described in Section II), to activate the SDC representing the "corner" feature, and similarly on up the hierarchy. Each of the orange outlined insets at V2 shows the input level aperture of the corresponding mac, emphasizing the idea that the precise input pattern is mapped into the closest-matching stored feature, in this example, a "upper left 90° corner" at left and a "NNE-pointing 135° angle" at right. The inset at bottom of Figure **I-**4b zooms in to show that the U-signals to V1 arise from individual pixels of the apertures (which would correspond to individual LGN projection cells).

In the past, IT cells have generally been depicted as being *narrowly selective* to particular objects (Desimone, Albright et al. 1984, Kreiman, Hung et al. 2006, Kiani, Esteky et al. 2007, Rust and DiCarlo 2010). However, as DiCarlo, Zoccolan et al. (2012) point out, the data overwhelmingly support the view of individual IT cells as having a "diversity of selectivity"; that is, individual IT cells generally respond to many different objects and in that sense are much more broadly tuned. This diversity is notionally suggested in Figures **I-**4b and **I-**5 in that individual cells are seen to participate in multiple SDCs representing different images/concepts. However, the particular input (stimulus) dimensions for which any given cell ultimately demonstrates some degree of invariance is not prescribed a priori. Rather they emerge essentially idiosyncratically over the history of a cell's inclusions in SDCs of particular experienced moments. Thus, the dimensions of invariance in the tuning functions of even immediately neighboring cells may generally end up quite different.

Figure I-5 embellishes the scheme shown in Figure I-4b and (turning it sideways) casts it onto the physical brain. We add paths from V1 and V2 to an MT representation as well. We add a notional PFC representation in which a higher-level concept involving the dog, i.e., the fact that it is being walked, is active. We show a more complete tiling of macs at V1 than in Figure I-4b to emphasize that only V1 macs that have a sufficient fraction of active pixels, e.g., an edge contour, in their aperture become active (pink). In general, we expect the fraction of active macs to decrease with level. As this and prior figures suggest, we currently model the macs as having no overlap with each other (i.e., they tile the local region), though their RFs [as well as their *projective fields* (PFs)] can overlap. However, we expect that in the real brain, macs can physically overlap. That is, any given minicolumn could be contained in multiple overlapping macs, where only one of those macs can be active at any given moment. The degree of overlap could vary by region, possibly generally increasing anteriorly. If so, then this would partially explain (in conjunction with the extremely limited view of population activity that single/few-unit electrophysiology has provided through most of the history of neuroscience) why there has been little evidence thus far for macs in more frontal regions.

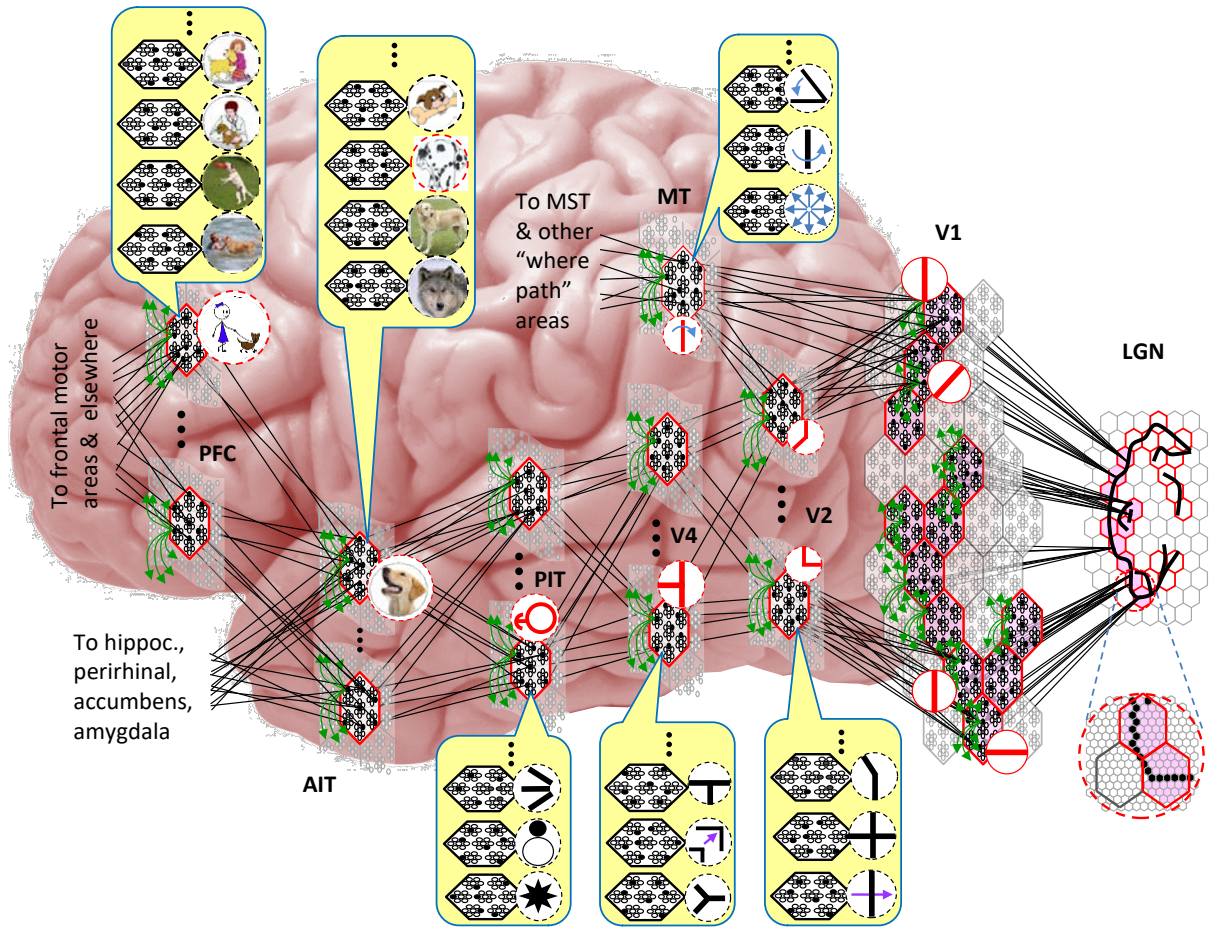

**Figure I-5:** Notional mapping of Sparsey to brain.

## I.A.   SPARSE DISTRIBUTED CODES VS. LOCALIST CODES

One important difference between SDC and localist representation is that the space of representations (codes) for a mac using SDC is exponentially larger than for a mac using a localist representation. Specifically, if $Q$ is the number of CMs in a mac and $K$ is the number of cells per CM, then there are $K^Q$ unique SDC codes for that mac. A localist mac of the same size only has $Q \times K$ unique codes. Note that it is not the case that an SDC-based mac can use that entire code space, i.e., store $K^Q$ features. Rather, the limiting factor on the number of codes storable in an SDC-based mac is the fraction of the mac's afferent synaptic weights that are set high (our model uses effectively binary weights), i.e., degree of *saturation*. In fact, the number of codes storable such that all stored codes can be retrieved with some prescribed average retrieval accuracy (error), is probably a vanishingly small fraction of the entire code space. However, real macrocolumns have $Q \approx 70$ minicolumns, each with $K \approx 20$ L2/3 principal cells: a "vanishingly small fraction" of $20^{70}$ can of course still be a large absolute number of codes.

While the difference in code space size between localist and SDC models is important, it is the distributed nature of the SDC codes per se that is most important. Many have pointed out a key property of SDC which is that since codes overlap, the number of cells in common between two codes can be used to represent their similarity. For example, if a given mac has $Q=100$ CMs, then there are 101 possible degrees of intersection between codes, and thus 101 degrees of similarity, which can be represented between concepts stored in that mac. The details of the process/algorithm that assigns codes to inputs determines the specific definition of similarity implemented. We will discuss the similarity metric(s) implemented and implementable in Sparsey throughout the sequel.

However, as stated earlier, the most important distinction between *localism* and SDC is that SDC allows the two essential operations of associative (content-addressable) memory, storing new inputs and retrieving the *best-matching* stored input, to be done *in fixed time for the life of the model*. That is, given a model of a fixed size (dominated by the number of weights), and which therefore has a particular limit on the amount, C, of information that it can store and retrieve subject to a prescribed average retrieval accuracy (error), the time it takes to either store (learn) a new input or retrieve the best-matching stored input (memory) remains constant regardless of how much information has been stored, so long as that amount remains less than C. *There is no other extant model, including all HMAX models, all convolutional network (CN) models, all Deep Learning (DL) models, all other models in the class of graphical probability models (GPMs), and the locality-sensitive hashing models, for which this capability—constant storage and best-match retrieval time over the life of the system—has been demonstrated.* All these other classes of models realize the benefits of hierarchy per se, i.e., the principle of hierarchical compositionality which is critical for rapidly learning highly nonlinear category boundaries, as described in Bengio, Courville et al. (2012), but only Sparsey also realizes the speed benefit, and therefore ultimately, the scalability benefit, of SDC. We state the algorithm in Section II. The reader can see by inspection of the CSA (Table II-1) that it has a fixed number of steps; in particular, it does *not* iterate over stored items.

Another way of understanding the computational power of SDC compared to localism is as follows. We stated above that in a localist representation such as in Figure I-5a, only one cell, representing one hypothesis can be active at a time. The other cells in the mac might, at some point prior to the choice of a final winner, have a distribution of sub-threshold voltages that reflects the likelihood distribution over all represented hypotheses. But ultimately, only one cell will win, i.e., go supra-threshold and spike. Consequently, only that one cell, and thus that one hypothesis, will materially influence the next time step's decision process in the same mac (via the recurrent H matrix) and in any other downstream macs.

In contrast, because SDCs physically overlap, if one particular SDC (and thus, the hypothesis that it represents) is *fully* active in a mac, i.e., if all $Q$ of that code's cells are active, *then all other codes (and thus, their associated hypotheses) stored in that mac are also simultaneously physically partially active in proportion to the size of their intersections with the single fully active code*. Furthermore, if the process/algorithm that assigns the codes to inputs has enforced the *similar-inputs-to-similar-codes* (SISC) property, then all stored inputs (hypotheses) are active with strength in descending order of similarity to the fully active hypothesis. We assume that more similar inputs generally reflect more similar world states and that world state similarity correlates with likelihood. In this case, the single fully active code also physically functions as the *full likelihood distribution over all SDCs (hypotheses) stored in a mac*. Figure I-6 illustrates this concept. We show five hypothetical SDCs, denoted with $\phi()$, for five input items, A-E (the actual input items are not shown here), which have been stored in the mac shown. At right, we show the decreasing intersections of the codes with $\phi(A)$. Thus, when code $\phi(A)$ is (fully) active, $\phi(B)$ is 4/7 active, $\phi(C)$ is 3/7 active, etc. Since cells representing *all* of these hypotheses, not just the most likely hypothesis, A, actually spike, it follows that *all of these hypotheses physically influence the next time step's decision processes*, i.e., the resulting likelihood distributions, active on the next time step in the same and all downstream macs.

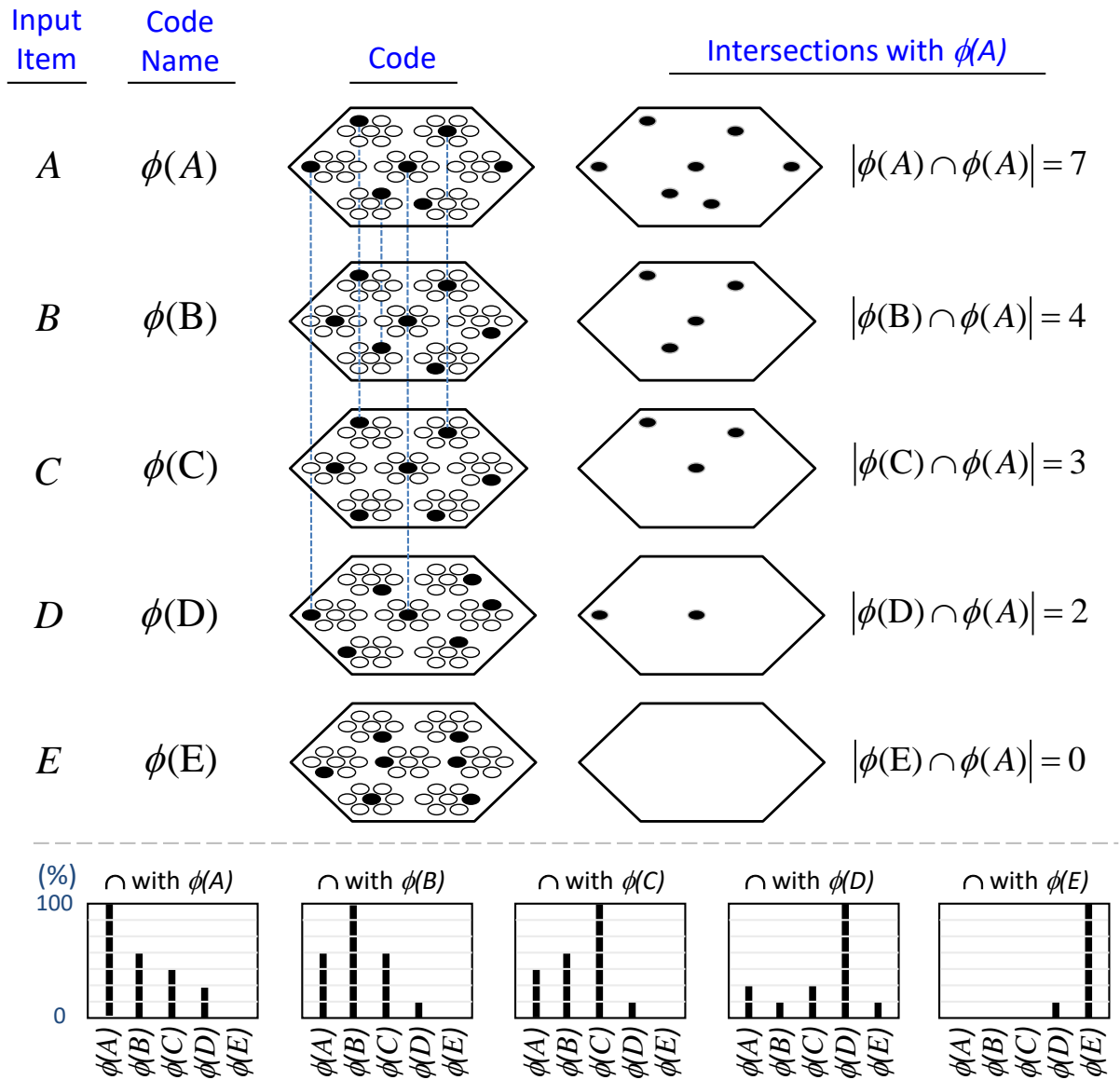

**Figure I-6:** If the process that assigns SDCs to inputs enforces the *similar-input-to-similar-codes* (SISC) property, then the currently active code in a mac simultaneously physically functions as the entire likelihood distribution over all hypotheses stored in the mac. At bottom, we show the activation strength distribution

over all five codes (stored hypotheses), when each of the five codes is fully active. If SISC was enforced when these codes were assigned (learned), then these distributions are interpretable as likelihood distributions. See text for further discussion.

We believe this difference to be fundamentally important. In particular, it means that performing a single execution of the *fixed-time* CSA transmits the influence of *every* represented hypothesis, regardless of how strongly active a hypothesis is, to every hypothesis represented in downstream macs. We emphasize that the representation of a hypothesis's probability (or likelihood) in our model—i.e., as the fraction of a given hypothesis's full code (of *Q* cells) that is active—differs fundamentally from existing representations in which single neurons encode such probabilities in their strengths of activation (e.g., firing rates) as described in the recent review of (Pouget, Beck et al. 2013).

## II. SPARSEY'S CORE ALGORITHM

During learning, Sparsey's core algorithm, the code selection algorithm (CSA), operates on every time step (frame) in every mac of every level, resulting in activation of a set of cells (an SDC) in the mac. The CSA can also be used, with one major variation, during retrieval (recognition). However, there is a much simpler retrieval algorithm, essentially just the first few steps of the CSA, which is preferable if the system "knows" that it is in retrieval mode. Note that this is not the natural condition for autonomous systems: in general, the system must be able to decide for itself, on a frame-by-frame basis, whether it needs to be in learning mode (if, and to what extent, the input is novel) or retrieval mode (if the input is completely familiar). We first describe the CSA's learning mode, then its variation for retrieval, then its much simpler retrieval mode.

### II.A. CSA: LEARNING MODE

The overall goal of the CSA when in learning mode (Table II-1) is to assign codes to a mac's inputs in adherence with the SISC property, i.e., more similar overall inputs to a mac are mapped to more highly intersecting SDCs. With respect to each of a mac's individual afferent RFs, U, H, and D, the similarity metric is extremely primitive: the similarity of two patterns in an afferent RF is simply an increasing function of *the number of features* in common between the two patterns, thus embodying only what Bengio, Courville et al. (2012) refer to as the weakest of priors, the *smoothness* prior. However, the CSA multiplicatively combines these component similarity measures and, because the H and D signals carry temporal information reflecting the history of the sequence being processed, the CSA implements a spatiotemporal similarity metric. Nevertheless, the ability to learn arbitrarily complex *nonlinear* similarity metrics (i.e., category boundaries, or invariances), requires a hierarchical network of macs and the ability for an individual SDC, e.g., active in one mac, to associate with multiple (perhaps arbitrarily different) SDCs in one or more other macs. We elaborate more on Sparsey's implementation of this capability in Section II.A.14.

The CSA has 12 steps which can be broken into two phases. Phase 1 (Steps 1-7) culminates in computation of the *familiarity*, *G* (normalized to [0,1]), of the overall (H, U, and D) input to the mac as a whole, i.e., *G* is a function of the *global* state of the mac. To first approximation, *G* is the similarity of the current overall input to the closest-matching previously stored (learned) overall input. As we will see, computing *G* involves a round of deterministic (hard max) competition resulting in one winning cell in each of the *Q* CMs. In Phase 2 (Steps 8-12), the activation function of the cells is modified based on *G* and a second round of competition occurs, resulting in the final set of *Q* winners, i.e., the activated code in the mac on the current time step. The second round of competition is *probabilistic* (*soft* max)*,* i.e., the winner in each CM is chosen as a draw from a probability distribution over the CM's *K* cells.

In neural terms, each of the CSA's two competitive rounds entail the principal cells in each CM integrating their inputs, engaging the local inhibitory circuitry, resulting in a single spiking winner. The difference is that the cell activation functions (F/I-curves) used during the second round of integration will generally be very different from those used during the first round. Broadly, the goal is as follows: as $G$ approaches 1, make cells with larger inputs compared to others in the CM increasingly likely to win in the second round, whereas as G approaches 0, make all cells in a CM equally likely to win in the second round. We discuss this further in Section II.A.15.

We now describe the steps of the CSA in learning mode. We will refer to the generic "circuit model" in Figure II-1 in describing some of the steps. The figure has two internal levels with one small mac at each level, but the focus, in describing the algorithm, will be on the L1 mac, $M_j^1$, highlighted in yellow. $M_j^1$ consists of $Q$=4 CMs, each with $K$=3 cells. Gray arrows represent the U-wts from the input level, L0, consisting of 12 binary pixels. Magenta arrows represent the D-wts from the L2 mac. Green lines depict a subset of the H-wts. The representation of where the different afferents arrive on the cells is not intended to be veridical. The depicted "Max" operations are the hard max operations of CSA Step 7. The blue arrows portray the mac-global $G$-based modulation of the cellular $V$-to-$\psi$ map (essentially, the F/I curve). The probabilistic draw operation is not explicitly depicted in this circuit model.

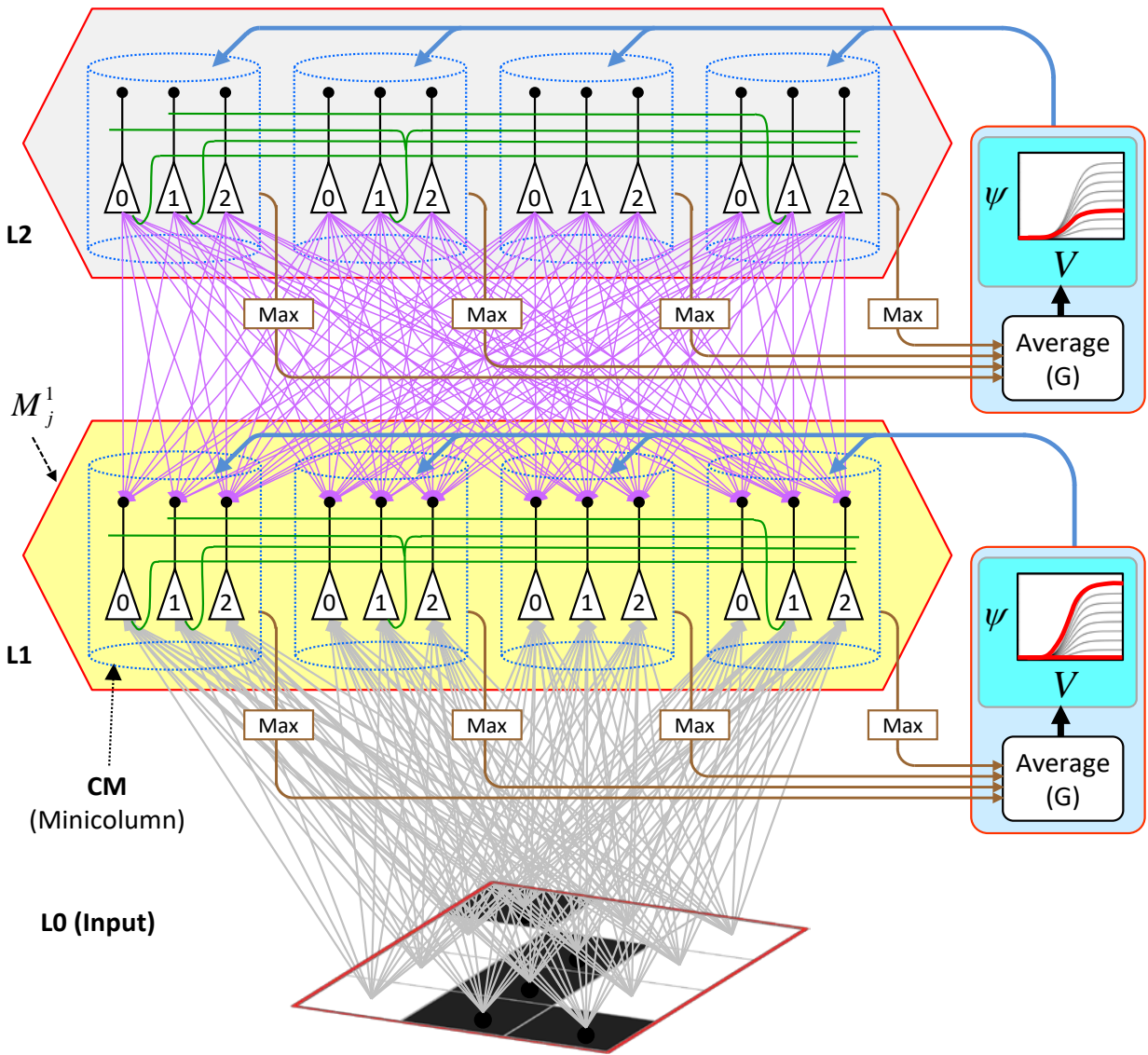

**Figure II-1:** Generic "circuit model" for reference in describing the some steps of the CSA.

### II.A.1. Step 1: Determine if the mac will become active

As shown in Eq. 1, during learning, a mac, $m$, becomes active if either of two conditions hold: a) if the number of active features in its U-RF, $\pi_U(m)$, is between $\pi_U^-$ and $\pi_U^+$; or b) if it is already active but the number of frames that it has been on for, i.e., its *code age*, $\Upsilon(m)$, is less than its persistence, $\delta(m)$. That is, during learning, we want to ensure that codes remain on for their entire prescribed persistence durations. We currently have no conditions on the number of active features in the H and D RFs.

$$Active(m) = \begin{cases} true & \Upsilon(m) < \delta(m) \\ true & \pi_U^- \leq \pi_U(m) \leq \pi_U^+ \\ false & otherwise \end{cases} \qquad \text{(Eq. 1)}$$

### II.A.2. Step 2: Compute raw U, H, and D-summations for each cell, $i$, in the mac

Every cell, $i$, in the mac computes its three weighted input summations, $u(i)$, as in Eq. 2a. $RF_U$ is a synonym for U-RF. $a(j,t)$ is pre-synaptic cell $j$'s activation, which is binary, on the current frame. Note that the synapses are *effectively binary*. Although the weight range is [0,127], pre-post correlation causes a weight to increase immediately to $w_{max} = 127$ and the asymptotic weight distribution will have a tight cluster around 0 (for weights that are effectively "0") and around 127 (for weights that are effectively "1"). The learning policy and mechanics are described in Section II.A.13. $F(\zeta(j,t))$ is a term needed to adjust the weights of afferent signals from cells in macs in which multiple competing hypotheses (MCHs) are active. If the number of MCHs ($\zeta$) is small then we want to boost the weights of those signals, but if it gets too high, in which case we refer to the source mac as being *muddled*, those signals will generally only serve to decrease SNR in target macs and so we disregard them. Computing and dealing with MCHs is described in Steps 5 and 6. $h(i)$ and $d(i)$ are computed in analogous fashion (Eqs. 2b,c), with the slight change that H and D signals are modeled as originating from codes active on the previous time step ($t$-1).

$$u(i) = \sum_{j \in RF_U} a(j,t) \times F(\zeta(j,t)) \times w(j,i) \qquad \text{(Eq. 2a)}$$

$$h(i) = \sum_{j \in RF_H} a(j,t-1) \times F(\zeta(j,t-1)) \times w(j,i) \qquad \text{(Eq. 2b)}$$

$$d(i) = \sum_{j \in RF_D} a(j,t-1) \times F(\zeta(j,t-1)) \times w(j,i) \qquad \text{(Eq. 2c)}$$

### II.A.3. Step 3: Normalize and filter the raw summations

The summations, $u(i)$, $h(i)$, and $d(i)$, are normalized to [0,1] interval, yielding $U(i)$, $H(i)$, and $D(i)$. We explained above that a mac $m$ only becomes active if the number of active features in its U-RF, $\pi_U(m)$, is between $\pi_U^-$ and $\pi_U^+$, referred to as the lower and upper *mac activation bounds*. Given our assumption that visual inputs to the model are filtered to single-pixel-wide edges and binarized, we expect relatively straight or low-curvature edges roughly spanning the diameter of an L0 aperture to occur rather frequently in natural imagery. Figure II-2 shows two examples of such inputs, as frames of sequences, involving either only a single L0 aperture (panel a) or a region consisting of three L0 apertures, i.e., as might comprise the U-RFs of an L2 mac (e.g., as in Figure I-5b). The general problem, treated in this figure, is that the number of features present in a mac's U-RF, $\pi_U(m)$, may vary from one frame to the next. Note that for macs at L2 and higher, the number of features present in an RF is the *number of active macs* in that RF, not the total number of active cells in that RF. The policy implemented in Sparsey is that inputs with different numbers of active features compete with each other on an equal footing. Thus, normalizers (denominators) in Eqs. 3a,b,c use the lower mac activation bound, $\pi_U^-$, $\pi_H^-$, and $\pi_D^-$. This necessitates hard limiting the maximum possible normalized value to 1, so that inputs with between $\pi_U^-$ and $\pi_U^+$ active features yield normalized values confined to [0,1]. There is one additional nuance. As noted above, if a mac in $m$'s U-RF is muddled, then we disregard all signals from it, i.e., they are not included in the $u$-summations of $m$'s cells. However, since that mac is active, it will be included in the number of active features, $\pi_U(m)$. Thus, we should normalize by the number of active, *non-muddled* macs in $m$'s U-RF (not simply the number of active macs): we denote this value as $\pi_U^*$. Finally, note that when the afferent feature is represented by a mac, that feature is actually being represented by the simultaneous activation of, and thus, inputs from, $Q$ cells; thus the denominator must be adjusted accordingly, i.e., multiplied by $Q$ and by the maximum weight of a synapse, $w_{max}$.

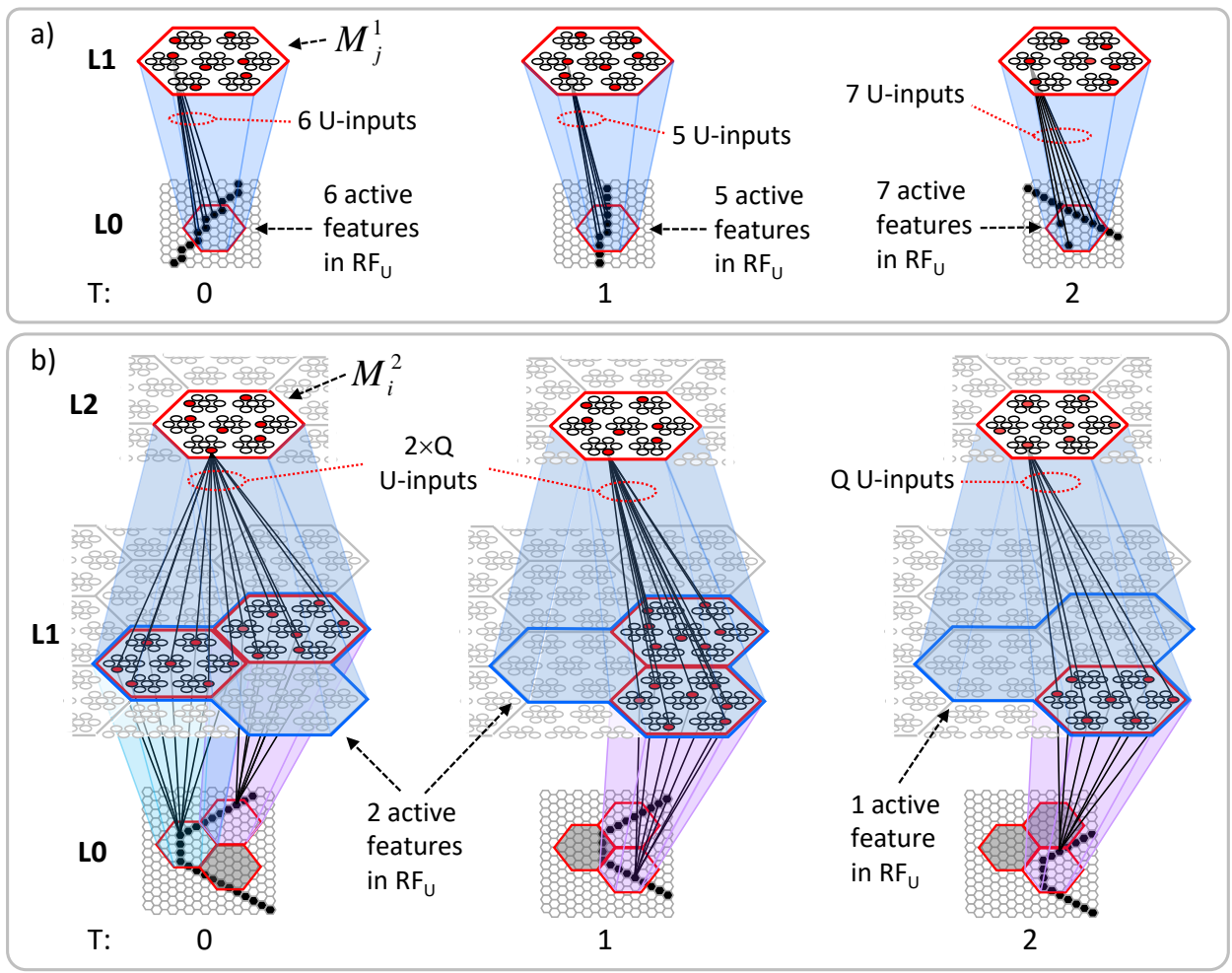

**Figure II-2:** The mac's normalization policy must be able to deal with inputs of different sizes, i.e., inputs having different numbers of active features. (a) An edge rotates through the aperture over three time steps, but the number of active features (in this case, pixels) varies from one time step (moment) to the next. In order for the mac to be able to recognize the 5-pixel input (T=1) just as strongly as the 6 or 7-pixel inputs, the $u$-summations must be divided by 5. (b) The U-RFs of macs at L2 and higher consist of an integer number of subjacent level macs, e.g., here, $M_i^2$'s U-RF consists of three L1 macs (blue border). Each active mac in $M_i^2$'s U-RF represents one feature. As for panel a, the number of active features varies across moments, but in this case, the variation is in increments/decrements of $Q$ synaptic inputs. Grayed-out apertures have too few active pixels for their associated L1 macs to become active.

$$U(i) = \begin{cases} \max(1, u(i)/\pi_U^- \times w_{max}) & L = 1 \\ \max(1, u(i)/\min(\pi_U^-, \pi_U^*) \times Q \times w_{max}) & L > 1 \end{cases} \quad \text{(Eq. 3a)}$$

$$H(i) = \max(1, h(i)/\min(\pi_H^-, \pi_H^*) \times Q \times w_{max}) \quad \text{(Eq. 3b)}$$

$$D(i) = \max(1, d(i)/\min(\pi_D^-, \pi_D^*) \times Q \times w_{max}) \quad \text{(Eq. 3c)}$$

### II.A.4. Step 4: Compute overall *local* support for each cell in the mac

The overall *local* (to the individual cell) measure, $V(i)$, of evidence/support that cell $i$ should be activated is computed by multiplying filtered versions of the normalized inputs as in Eq. 4. $V(i)$ can also be viewed as the normalized degree of match of cell $i$'s total afferent (including U, H, and D) synaptic weight vector to its total input pattern. We emphasize that the $V$ measure is *not* a measure of support for a *single* hypothesis, since an individual cell does not represent a single hypothesis. Rather, in terms of hypotheses, $V(i)$ can be viewed as the local support for the *set* of hypotheses whose representations (codes) include cell $i$. The individual normalized summations are raised to powers ($\lambda$), which allows control of the relative sensitivities of $V$ to the different input sources (U, H, and D). Currently, the U-sensitivity parameter, $\lambda_U$, varies with time (index of frame with respect to beginning of sequence). We will add time-dependence to the H and D sensitivity parameters as well and explore the space of policies regarding these schedules in the future. In general terms, these parameters (along with many others) influence the shapes of the boundaries of the categories learned by a mac.

$$V(i) = \begin{cases} H(i)^{\lambda_H} \times U(i)^{\lambda_U(t)} \times D(i)^{\lambda_D} & t \geq 1 \\ U(i)^{\lambda_U(0)} & t = 0 \end{cases} \quad \text{(Eq. 4)}$$

As described in Section II.B, during retrieval, this step is significantly generalized to provide an extremely powerful, general, and efficient mechanism for dealing with arbitrary, nonlinear invariances, most notably, nonlinear time-warping of sequences.

### II.A.5. Step 5: Compute the number of competing hypotheses that will be active in the mac once the final code for this frame is activated.

To motivate the need for keeping track of the number of competing hypotheses active in a mac, we consider the case of *complex* sequences, in which the same input item occurs multiple times and in multiple contexts. Figure II-3 portrays a minimal example in which item B occurs as the middle state of sequences [ABC] and [DBE]. Here, the model's single internal level, L1, consists of just one mac, with $Q$=4 CMs, each with $K$=4 cell. Figure II-3a shows notional codes (SDCs) chosen on the three time steps of [ABC]. The code name convention here is that $\phi$ denotes a code, the superscript "1" indicates the model level at which code resides. The subscript indicates the specific moment of the sequence that the code represents; thus, it is necessary for the subscript to specify the full temporal context, from start of sequence, leading up to the current input item. Successively active codes are chained together, resulting in spatiotemporal memory traces that represent sequences. Green lines indicate the H-wts that are increased from one code to the next. Black lines indicate the U-wts that are increased from currently active pixels to currently active

L1 cells (red). Thus, as described earlier, e.g., in Figure I-2, individual cells learn spatiotemporal inputs in correlated fashion, as whole SDCs Learning is described more thoroughly in Section II.A.13.

As portrayed in Figure II-3b, if [ABC] has been previously learned, then when item B of another sequence, [DBC], is encountered, the CSA will generally cause a different SDC, here, $\phi^1_{DB}$, to be chosen. $\phi^1_{DB}$ will be H-associated with whatever code is activated for the next item, in this case $\phi^1_{DBE}$ for item E. This choosing of codes in a context-dependent way (where the dependency has no fixed Markov order and in practice can be extremely long), enables subsequent recognition of complex sequences without confusion.

However, what if in some future recognition test instance, we prompt the network with item B, i.e., as the first item of the sequence, as shown in Figure II-3c? In this case, there are no active H-wts and so the computation of local support (Eq. 4) depends only on the U-wts. But, the pixels comprising item B have been fully associated with the two codes, $\phi^1_{AB}$ and $\phi^1_{DB}$, which have been assigned to the two moments when item B was presented, [A**B**] and [D**B**]. We show the two maximally implicated (more specifically, maximally U-implicated) cells in each CM as orange to indicate that a choice between them in each CM has not yet been made. However, by the time the CSA completes for the frame when item B is presented, one winner must be chosen in each CM (as will become clear as we continue to explain the CSA throughout the remainder of Section II). And, because it is the case in each CM, that both orange cells are equally implicated, we choose winners randomly between them, resulting in a code that is an equal mix of the winners from $\phi^1_{AB}$ and $\phi^1_{DB}$. In this case, we refer to the mac as having multiple competing hypotheses active (MCHs), where we specifically mean that all the active hypotheses (in this case, just two) are approximately equally strongly active.

The problem can now be seen at the right of Figure II-3c when C is presented. Clearly, once C is presented, the model has enough information to know which of the two learned sequences, or more specifically, which particular moment is intended, [AB**C**] rather than [DB**E**]. However, the cells comprising the code representing that learned moment, $\phi^1_{ABC}$, will, at the current test moment (lower inset in Figure II-3c), have only half the active H-inputs that they had during the original learning instance (i.e., upper inset in Figure II-3c). This leads, once processed through steps 2b, 3b, and 4, to *V* values that will be far below *V*=1, for simplicity, let's say *V*=0.5, for the cells comprising $\phi^1_{ABC}$. As will be explained in the remaining CSA steps, this ultimately leads to the model *not* recognizing the current test trial moment [B**C**] as equivalent to the learning trial moment [AB**C**], and consequently, to activation of a new code that could in general be arbitrarily different from $\phi^1_{ABC}$.

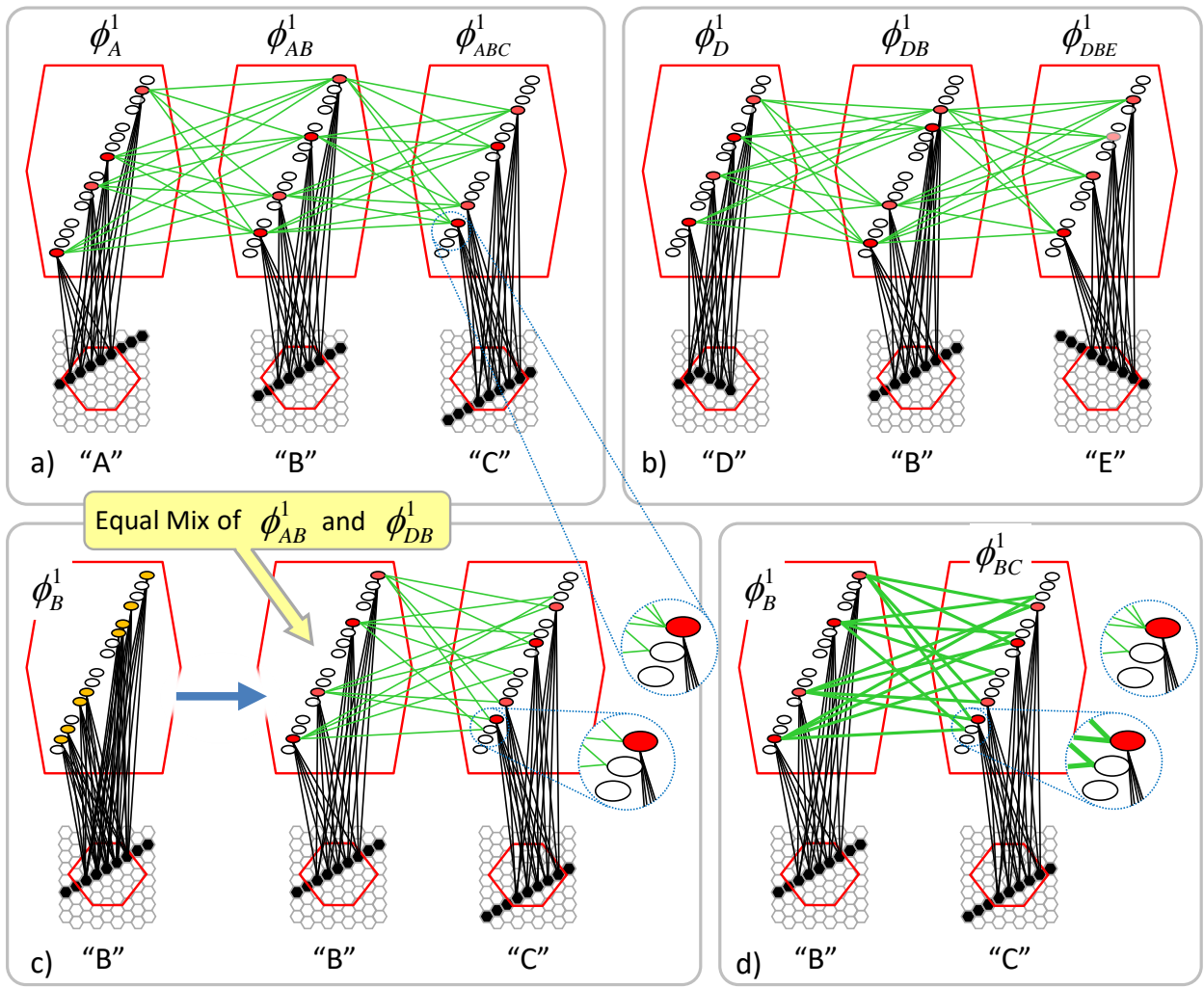

**Figure II-3:** Portrayal of reason why macs need to know how many multiple competing hypotheses (MCHs) are/were active in their afferent macs.

However, there is a fairly general solution to this problem where multiple competing hypotheses are present in an active mac code, e.g., in the code for B indicated by the yellow call-out. The mac can easily detect when an MCH condition exists. Specifically, it can tally the number cells with *V*=1—or, allowing some slight tolerance for considering a cell to be *maximally* implicated, cells with $V(i) > V_\zeta$, where $V_\zeta$ is close to 1, e.g., $V_\zeta = 0.95$—in each of its *Q* CMs, as in Eq. 5a. It can then sum $\zeta_q$ over all *Q* CMs and divide by *Q* (and round to the nearest integer, "**rni**"), resulting in the number of MCHs active in the mac, $\zeta$, as in Eq. 5b. In this example, $\zeta = 2$, and the principle by which the H-input conditions, specifically the h-summations, for the cells in $\phi^1_{ABC}$ on this test trial moment [B**C**] can be made the same as they were during the learning trial moment [AB**C**], is simply to multiply all outgoing H-signals from $\phi^1_B$ by $\zeta = 2$. We indicate the *inflated* H-signals by the thicker green lines in the lower inset at right of Figure II-3d. This ultimately leads to *V*=1 for all four cells comprising $\phi^1_{ABC}$ and, via the remaining steps of the CSA, reinstatement of $^1\phi_{ABC}$ with very high probability (or with certainty, in the simple retrieval mode described in Section II.C), i.e., with recognition of test trial moment [B**C**] as equivalent to learning trial moment [AB**C**]. The model has successfully gotten through an ambiguous moment based on presentation of further, disambiguating inputs.

We note here that uniformly boosting the efferent H-signals from $\phi_B^1$ also causes the h-summations for the four cells comprising the code $\phi_{DBE}^1$ to be the same as they were in the learning trial moment [DB**E**]. However, by Eq. 4, the *V* values depend on the U-inputs as well. In this case, the four cells of $\phi_{DBE}^1$ have u-summations of zero, which leads to *V*=0, and ultimately to essentially zero probability of any of these cells winning the competitions in their respective CMs. Though we don't show the example here, if on the test trial, we present E instead of C after B, the situation is reversed; the u-summations of cells comprising the code $\phi_{DBE}^1$ are the same as they were in the learning trial moment [DB**E**] whereas those of the cells comprising the code $\phi_{ABC}^1$ are zero, resulting with high probability (or certainty) in reinstatement of $\phi_{DBE}^1$.

$$\zeta_q = \sum_{i=0}^{K}\left[V(\text{i}) > \text{V}_\zeta\right] \quad \text{(Eq. 5a)}$$

$$\zeta = \mathbf{rni}\left(\sum_{j=0}^{Q-1}\zeta_q\Big/Q\right) \quad \text{(Eq. 5b)}$$

### II.A.6. Step 6: Compute correction factor for multiple competing hypotheses to be applied to efferent signals from this mac

The example in Figure II-3 was rather clean in that it involved only two sequences having been learned, containing a total of six moments, [**A**], [A**B**], [AB**C**], [**D**], [D**B**], and [DB**E**], and very little pixel-wise overlap between the items. Thus, cross-talk between the stored codes was minimized. However, in general, macs will store far more codes. If for example, the mac of Figure II-3 was asked to store 10 moments where B was presented, then, if we prompted the network with B as the first sequence item, we would expect almost all cells in all CMs to have *V*=1. As discussed in Step 2, when the number of MCHs ($\zeta$) in a mac gets too high, i.e., when the mac is *muddled*, its efferent signals will generally only serve to decrease SNR in target macs (including itself on the next time step via the recurrent H-wts) and so we disregard them. Specifically, when $\zeta$ is small, e.g., two or three, we want to boost the value of the signals coming from all active cells in that mac by multiplying by $\zeta$ (as in Figure II-3d). However, as $\zeta$ grows beyond that range, the expected overlap between the competing codes increases and to approximately account for that, we begin to diminish the boost factor as in Eq. 6, where *A* is an exponent less than 1, e.g., 0.7. Further, once $\zeta$ reaches a threshold, *B*, typically set to 3 or 4, we multiply the outgoing weights by 0, thus effectively disregarding the mac completely in downstream computations. We denote the correction factor for MCHs as $F(\zeta)$, defined as in Eq. 6. We also use the notation, $F(\zeta(j,t))$, as in Eq. 2, where $\zeta(j,t)$ is the number of hypotheses tied for maximal activation strength in the owning mac of a pre-synaptic cell, *j*, at time (frame) *t*.

$$F(\zeta) = \begin{cases} \zeta^A & 1 \leq \zeta \leq B \\ 0 & \zeta > B \end{cases} \quad \text{(Eq. 6)}$$

### II.A.7. Step 7: Determine the maximum local support in each of the mac's CMs

Operationally, this step is quite simple: simply find the cell with the highest $V$ value, $\hat{V}_j$, in each CM, $C_j$, as in Eq. 7. Multiple cells in a CM may be tied for $\hat{V}_j$.

$$\hat{V}_j = \max_{i \in C_j} \{V(i)\} \tag{Eq. 7}$$

Conceptually, the cell with $\hat{V}_j$ in a CM is the cell most implicated by the mac's total input (multiple cells can be tied for $\hat{V}_j$), or in other words, the *most likely* winner in the CM. In fact, in the simple retrieval mode (Section II.C), the cell with $\hat{V}_j$ in each CM is chosen winner.

### II.A.8. Step 8: Compute the familiarity of the mac's overall input

The average, $G$, of the maximum $V$'s across the mac's $Q$ CMs is computed as in Eq. 8: $G$ is a measure of the *familiarity* of the macs overall input. This is done on every time step (frame), so we sometimes denote $G$ as a function of time, $G(t)$. And, $G$ is computed independently for each activated mac, so we may also use more general notation that indicates mac as well.

$$G = \sum_{q=1}^{Q} \hat{V}_k / Q \tag{Eq. 8}$$

The main intuition motivating the definition and use of $G$ is as follows. If the mac's current input moment has been experienced in the past, then all active afferent weights (U, H, and D) to the code activated in that instance would have been increased. Thus, in the current moment, all $Q$ cells comprising that code will have $V=1$. Thus, $G=1$. Thus, a *familiar* moment must always result in $G=1$ (assuming that MCHs are accounted for as described above). On the other hand, suppose that the current *overall* input moment is novel, even if sub-components of the current overall input have been experienced exactly before. In this case, provided that few enough codes have been stored in the mac (so that crosstalk remains sufficiently small), there will be at least some CMs, $C_j$, for which $\hat{V}_j$ is significantly less than 1. Thus, $G<1$. Moreover, as the examples in the Results section will show, $G$ correlates with the familiarity of the overall mac input. Thus, $G$ measures the familiarity, or inverse novelty, of the global input to the mac.

Note that in the brain, this step requires that the $Q$ cells with $V = \hat{V}_j$ become active (i.e., spike) so that their outputs can be summed and averaged. This constitutes the first of two rounds of competition that occurs within the mac's CMs on each execution of the CSA. However, as explained herein, this set of $Q$ cells will, in general, *not* be identical to (and can often be substantially different from, especially when $G\approx 0$) the finally chosen code for this execution of the CSA (i.e., the code chosen in Step 12).

### II.A.9. Step 9: Determine the expansivity/compressivity of the I/O function to be used for the second and final round of competition within the mac's CMs

Determine the range, $\eta$, of the sigmoid activation function, which transforms a cell's $V$ value into its relative (within its own CM) probability of winning, $\psi$. We refer to that transform as the $V$-to-$\psi$ map. We refer to $\chi$ as the *sigmoid expansion factor* and $\gamma$ as the *sigmoid expansion exponent*.

$$\eta = 1 + \left(\left[\frac{G - G^-}{1 - G^-}\right]^+\right)^\gamma \times \chi \times K \tag{Eq. 9}$$

As noted several times earlier, the overall goal of the CSA when in learning mode is to assign codes to a mac's inputs in adherence with the SISC property, i.e., more similar overall inputs to a mac are mapped to more highly intersecting SDCs. Given that $G$ represents, to first approximation, the similarity of the closest-matching stored input to the current input, we can restate the goal as follows.

1. as $G$ goes to 1, meaning the input X is completely familiar, we want the probability of reinstating the code $\phi_X$ that was originally assigned to represent X, to go to 1. It is the cells comprising $\phi_X$, which are causing the high $G$ value. But these are the cells with the maximal $V$'s ($V = \hat{V}_j = 1$) in their respective CMs. Thus, within each CM, $C_j$, we want to increase the probability of picking the cell with $V = \hat{V}_j$ relative to cells with $V < \hat{V}_j$, i.e., we want to transform the $V$'s via an *expansive* nonlinearity

2. as $G$ goes to 0 (completely novel input), we want the set of winners chosen to have the minimum average intersection with all stored codes. We can achieve that by choosing the winner in each CM from the uniform distribution, i.e., by making all cells in a CM equally likely to win, i.e., transform the $V$'s via a maximally *compressive* nonlinearity.

The first goal is met by making the activation function a very expansive nonlinearity. Figure II-4 shows how the expansivity of the $V$-to-$\psi$ map affects cell win probability, and indirectly, whole-code reinstatement probability. All nine panels concern a small example mac with $Q$=6 CMs each comprised of $K$=7 cells. Each panel shows hypothetical $V$ and $\rho$ vectors over the cells of the CMs, across two parametrically varying conditions: model "age" (across columns), which we can take as a correlate of the number of stored codes and thus, of the amount of interference (crosstalk) between codes during retrieval, and expansivity ($\eta$) (across rows) of the $V$-to-$\psi$ map. As described shortly, the $V$ values are first transformed to relative probabilities ($\psi$) (Step 10), which are then normalized to absolute probabilities ($\rho$) (Step 11). In all panels, the example $V$ vector in each CM has one cell with $V$=1 (pink bars). Thus, by Step 8, all panels correspond to a $G$=1 condition. The other six cells (black bars) in each CM are assigned uniformly randomly chosen values in defined intervals that depend on the age of the model. The intervals for "Early", "Middle", and "Late", are [0.0, 0.1], [0.1, 0.5], and [0.2, 0.8], respectively, simulating the increasing crosstalk with age.

For each age condition, we show the effects of using a $V$-to-$\psi$ map with three different $\eta$ values. Note that in actual operation (specifically, Step 9), all panels would be processed with a $V$-to-$\psi$ map with the maximal $\eta$ value (again, because $G$=1 in all panels). But our purpose here is just to show the consequences on the final $\rho$ distribution for a given $V$ distribution (the $V$ distribution is the same for all three rows in any given column) as a function of $\eta$. And, note that the minimum $\psi$ value in all cases is 1. Thus, for the "Early" column, the highly expansive $V$-to-$\psi$ map ($\eta$=300) (top row) results in a 300/306≈98% probability of selecting the cell with $V$=1 (pink) in each CM. This results in a (300/306)$^6$≈89% probability of choosing the pink cell in *all* $Q$=6 CMs, i.e., of reinstating the *entire* correct code. In the second row, $\eta$ is reduced to 30. Each of the six black cells ultimately ends up with a 1/36 probability of winning and the pink cell, with a 30/36=5/6 win probability. In this case the likelihood of reinstating the *entire* correct code, is (5/6)$^6$≈33%. In the bottom row, $\eta$=1, i.e., the $V$-to-$\psi$ map has been collapsed to the constant function, $\psi$=1. As can be seen, all cells, including the cell with $V$=1 become equally likely to be chosen winner in their respective CMs.

Greater crosstalk can clearly be seen in the "Middle" condition. Consequently, even for $\eta$=300, several of the cells with non-maximal $V$ end up with significant final probability $\rho$ of being chosen winner in their respective CMs. The $\rho$-distributions are slightly further compressed (flatter) when $\eta$=30, and completely compressed when $\eta$=1 (bottom row). The "Late" condition is intended to model a later period of the life of the model, after many memories (codes) have been stored in this mac. Thus, when the input pattern associated with any of those stored codes is presented again, many of the cells in each CM will have an appreciable *V* value (again, here they are drawn uniformly from [0.2, 0.8]). In this condition, even if $\eta$=300, the probability of selecting the correct cell (pink) in each CMs is close to chance, as is the chance of reinstating the entire correct code. And the situation only gets worse for lower $\eta$ values.

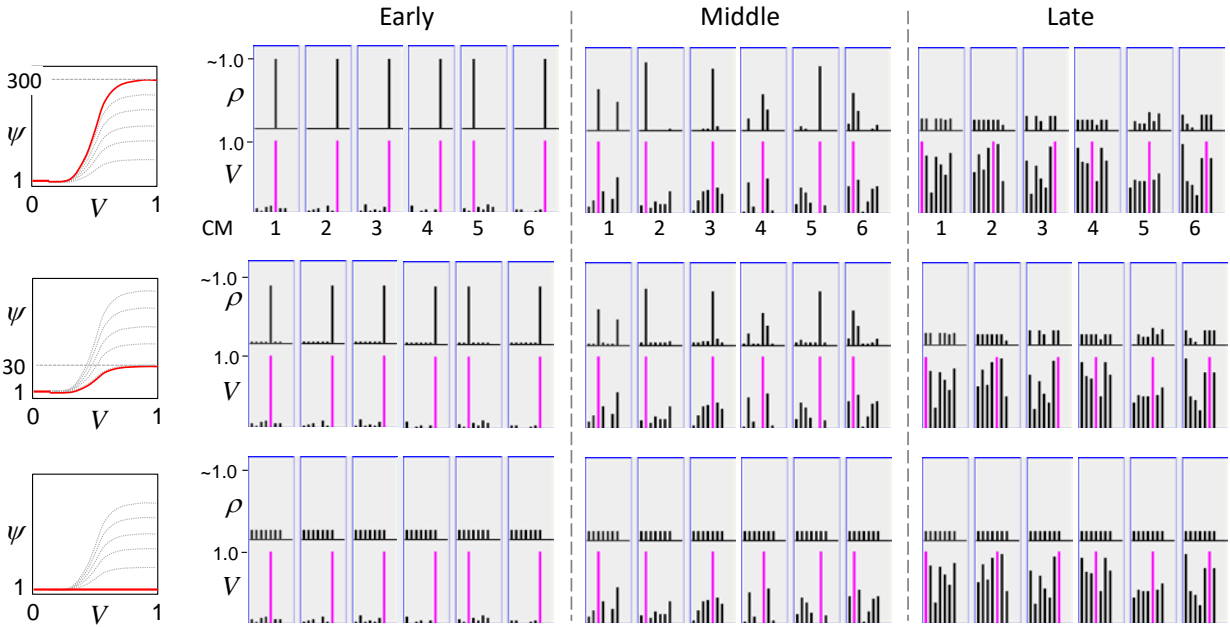

**Figure II-4:** *G*-based sigmoid transform characteristics. All panels show hypothetical *V* and $\rho$ vectors over the *K*=7 cells in each of the *Q*=6 CMs comprising the mac. In all nine panels, the *V* vector in each CM has one cell prescribed to have *V*=1 (pink bars). The *V*'s of the other six cells (black bars) in each CM are drawn randomly from defined intervals that depend on the age (amount of inputs experienced) of the model. For each age condition, we show the effects of using a *V*-to-$\psi$ map with three different $\eta$ values. But our purpose here is just to show the consequences on the final $\rho$ distribution for a given *V* distribution (the *V* distribution is the same for all three rows in any given column) as a function of the expansivity/compressivity ($\eta$) of the *V*-to-$\psi$ map. See text for details.

Note that for any particular *V* distribution in a CM, the relative increase to the final probability of being chosen winner is a smoothly and faster-than-linearly increasing (typically, $\gamma \geq 2$) function of *G*. Thus, in each CM, the probability that the most highly implicated (by the mac's total input) cell (those corresponding to the pink bars in Figure II-4) wins increases smoothly as *G* goes to 1. (Strictly, this is true only for the portion of the sigmoid nonlinearity with slope > 1). The initial (left) and final (right) portions of the sigmoid are compressive ranges.) And since the overall code is just the result of the *Q* independent draws, it follows that the expected intersection of the code consisting of the *Q* most highly implicated cells, i.e., the code of the closest-matching stored input, with the finally chosen code is also an increasing function of *G*, i.e., thus realizing the "SISC" property.

II.A.10. Step 10: Apply the modulated activation function to all the mac's cells, resulting in a relative probability distribution of winning over the cells of each CM

Apply sigmoid activation function to each cell. Note: the sigmoid collapses to a constant function, $\psi(i) = 1$, when $\eta = 1$ (i.e., when $G < G^-$).

$$\psi(i) = \frac{(\eta - 1)}{(1 + \sigma_1 e^{-\sigma_2 (V(i) - \sigma_3)})^{\sigma_4}} + 1 \qquad \text{(Eq. 10)}$$

In a more general development, the CSA could include additional prior steps for setting any of the other sigmoid parameters, $\sigma_1$, $\sigma_2$, $\sigma_3$, and $\sigma_4$, all of which interact to control overall sigmoid expansivity and shape. In particular, in the current implementation, the horizontal position of the sigmoid's inflection point is moved rightward as additional codes are stored in a mac. Figure II-5 shows that doing so greatly increases the probability of choosing the correct cell in each CM and thus, of reinstating the entire correct code, even when many codes have been stored in the mac. In the "Middle" condition, even if $\eta=30$, the probability of choosing the pink cell in each CM is very close to 1. For the "Late" condition, setting $\eta=30$ significantly improves the situation relative to the top right panel of Figure II-4 and setting $\eta=300$ makes the probability of choosing the correct cell close to 1 in four of the six CMs. Thus, we have a mechanism for keeping memories accessible for longer lifetimes.

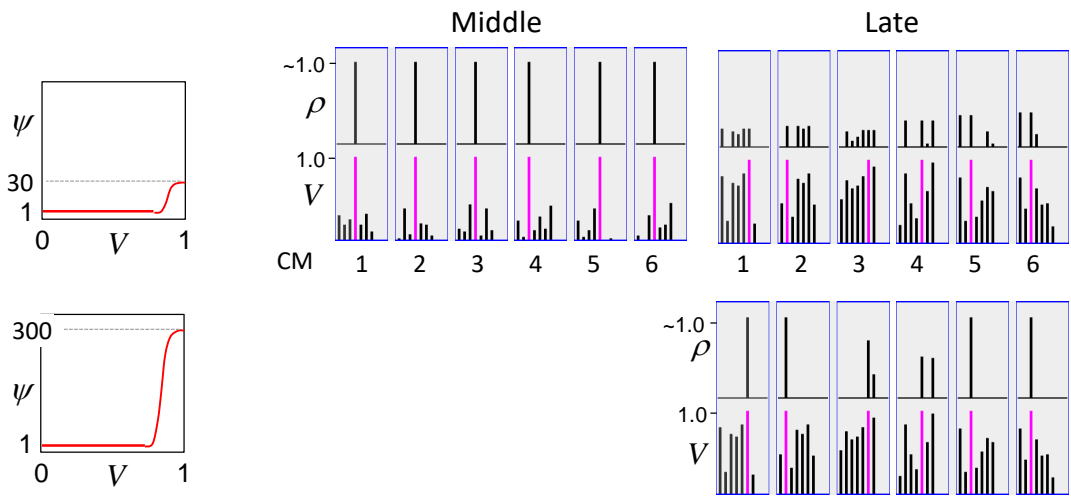

**Figure II-5:** Moving the inflection point of the sigmoidal $V$-to-$\psi$ map to the right greatly increases the probability of selecting the correct cell despite mounting crosstalk due to a growing number of codes stored in superposition.

II.A.11. Step 11: Convert relative win probability distributions to absolute distributions

In each of the mac's CMs, the $\psi$ values of the cells are converted to true probabilities of winning ($\rho$) and the winner is selected by drawing from the $\rho$ distribution, resulting in a final SDC, $\phi$, for the mac, as in Eq. 11.

$$\rho(i) = \frac{\psi(i)}{\sum_{k \in CM} \psi(k)} \qquad \text{(Eq. 11)}$$

II.A.12. Step 12: Pick winners in the mac's CMs, i.e., activate the SDC

The last step of the CSA is just selecting a final winner in each CM according to the $\rho$ distribution in that CM, i.e., soft max. This is the second round of competition. Our hypothesis that the canonical cortical computation involves two rounds of competition is a strong and falsifiable prediction of the model with respect to actual neural dynamics, which we would like to explore further.

**Table II-1: The CSA during Learning**

| | Equation | Short Description |
|---|---|---|
| 1 | $Active(m) = \begin{cases} true & \Upsilon(m) < \delta(m) \\ true & \pi_U^- \leq \pi_U(m) \leq \pi_U^+ \\ false & otherwise \end{cases}$ | Determine if mac *m* will become active. |
| 2 | $u(i) = \sum_{j \in RF_U} x(j,t) \times F(\zeta(j,t)) \times w(j,i)$ <br> $h(i) = \sum_{j \in RF_H} x(j,t-1) \times F(\zeta(j,t-1)) \times w(j,i)$ <br> $d(i) = \sum_{j \in RF_D} x(j,t-1) \times F(\zeta(j,t-1)) \times w(j,i)$ | Compute the raw U, H, and D input summations. |
| 3 | $U(i) = \begin{cases} \max(1, u(i)/\pi_U^- \times w_{max}) & L=1 \\ \max(1, u(i)/\min(\pi_U^-, \pi_U^*) \times Q \times w_{max}) & L>1 \end{cases}$ <br> $H(i) = \max(1, h(i)/\min(\pi_H^-, \pi_H^*) \times Q \times w_{max})$ <br> $D(i) = \max(1, d(i)/\min(\pi_D^-, \pi_D^*) \times Q \times w_{max})$ | Compute normalized, filtered input summations. |
| 4 | $V(i) = \begin{cases} H(i)^{\lambda_H} \times U(i)^{\lambda_U(t)} \times D(i)^{\lambda_D} & t \geq 1 \\ U(i)^{\lambda_U(0)} & t = 0 \end{cases}$ | Compute local evidential support for each cell. |
| 5a | $\zeta_q = \sum_{i=0}^{K} \left[ V(i) > V_\zeta \right]$ | (a) Compute #cells representing a maximally competing hypothesis in each CM. (b) Compute # of maximally active hypotheses, $\zeta$, in the mac. |
| 5b | $\zeta = \sum_{j=0}^{Q-1} \zeta_q / Q$ | |
| 6 | $F(\zeta) = \begin{cases} \zeta^A & 1 \leq \zeta \leq B \\ 0 & \zeta > B \end{cases}$ | Compute the *multiple competing hypotheses* (MCH) correction factor, $F(\zeta)$, for the mac. |
| 7 | $\hat{V}_j = \max_{i \in C_j} \{V(i)\}$ | Find the max *V*, $\hat{V}_j$, in each CM, $C_j$. |
| 8 | $G = \sum_{q=1}^{Q} \hat{V}_k / Q$ | Compute *G* as the average $\hat{V}$ value over the *Q* CMs. |
| 9 | $\eta = 1 + \left( \left[ \frac{G - G^-}{1 - G^-} \right]^+ \right)^\gamma \times \chi \times K$ | Determine the expansivity of the sigmoid activation function. |
| 10 | $\psi(i) = \frac{(\eta - 1)}{(1 + \sigma_1 e^{-\sigma_2 (V(i) - \sigma_3)})^{\sigma_4}} + 1$ | Apply sigmoid activation function (which collapses to the constant function when $G < G^-$) to each cell. |
| 11 | $\rho(i) = \frac{\psi(i)}{\sum_{k \in CM} \psi(k)}$ | In each CM, normalize the relative probabilities of winning ($\psi$) to final probabilities ($\rho$) of winning. |
| 12 | Select a final winner in each CM according to the $\rho$ distribution in that CM, i.e., soft max. | |

### II.A.13. Learning Policy and Mechanics

Broadly, Sparsey's learning policy can be described as Hebbian with passive weight decay. As noted earlier, the model's synapses are *effectively binary*. By this we mean that although the weight range is [0,127], the several learning related properties conspire to cause the asymptotic weight distribution to tend towards having two spikes, one at 0 and the other at $w_{max} = 127$, thus effectively being binary.

In actuality, a synapse's weight, $w(j,i)$, where $j$ and $i$ index the pre- and post-synaptic cells, respectively, is determined by two primary variables, its *age*, $\sigma(j,i)$, which is the number of time steps (e.g., video frames) since it was last increased, and its *permanence*, $\theta(j,i)$, which measures how resistant to decrease the weight is (i.e., the passive decay rate). The learning law is implemented as follows. Whenever a synapse's pre- and postsynaptic cells are coactive [i.e., a "pre-post correlation", $a(j) = 1 \wedge a(i) = 1$], its age is set to zero, as in Eq. 13a., which has the effect of setting its weight to $w_{max}$ (as can be seen in the "weight table" of Figure II-6, an age of zero always maps to $w_{max}$). Otherwise, $\sigma(j,i)$ increases by one on each successive time step (across all frames of all sequences presented) on which there is no pre-post correlation (Eq. 13c), stopping when it gets to the maximum age, $\sigma_{max}$ (Eq. 13d). Also note that once a synapse has reached maximum permanence, $\theta_{max}$, its age stays at zero, i.e., its weight stays at $w_{max}$ (Eq. 13b). At any point, the synapse's weight, $w(j,i)$, is gotten by dereferencing $\sigma(j,i)$ and $\theta(j,i)$ from the weight table shown in Figure II-6.

The intent of the decay schedule (for any permanence value) is to keep the weight at or near $w_{max}$ for some initial window of time (number of time steps), $T_\sigma(\theta)$, and then allow it to decay increasingly rapidly toward zero. Thus, the model "assumes" that a pre-post correlation reflects an important / meaningful event in the input space and therefore strongly embeds it in memory (consistent with the notion of episodic memory). If the synapse experiences a second pre-post correlations within the window $T_\sigma(\theta)$, its permanence is incremented as in Eq. 14 and $\sigma(j,i)$ is set back to 0 (i.e., its weight is set back to $w_{max}$); otherwise the age, $\sigma(j,i)$, increases by one with each time step and the weight decreases according to the decay schedule in effect. Thus, pre-post correlations due to noise or spurious events, which will have a much longer expected time to recurrence, will tend to fade from memory. Sparsey's permanence property is closely related to the notion of synaptic tagging (Frey and Morris 1997, Morris and Frey 1999, Sajikumar and Frey 2004, Moncada and Viola 2007, Barrett, Billings et al. 2009)).

$$\sigma(j,i) = \begin{cases} 0 & , \ a(j) = 1 \wedge a(i) = 1 & \text{(Eq. 13a)} \\ 0 & , \ \theta(j,i) = \theta_{max} & \text{(Eq. 13b)} \\ \sigma(j,i)+1 & , \ a(j) = 0 \vee a(i) = 0 & \text{(Eq. 13c)} \\ \sigma(j,i) & , \ \sigma(j,i) = \sigma_{max} & \text{(Eq. 13d)} \end{cases}$$

$$\theta(j,i) = \begin{cases} \theta(j,i)+1 & , \ a(j) = 1 \wedge a(i) = 1 \wedge \ \sigma(j,i) \leq T_\sigma(\theta(j,i)) \\ \theta(j,i) & , \ \text{otherwise} \end{cases} \quad \text{(Eq. 14)}$$

The exact parametric details are less important, but as can be seen in the weight table, the decay rate decreases with $\theta(j,i)$ and the window, $T_\sigma(\theta)$, within which a second pre-post correlation will cause an increase in permanence, increases with $\theta(j,i)$ (three example value shown). Permanence can only increase and in our investigations thus far, we typically make a synaptic weight completely permanent on the second or third within-window pre-post correlation [$\theta_{max}=1$ or $\theta_{max}=2$, respectively]. The justification of this policy derives from two facts: a) a mac's input is a sizable set of co-active cells; and b) due to the SISC property, the probability that a weight will be increased correlates with the strength of the statistical regularity of the input (i.e., the structural permanence of the input feature) causing that increase. These two facts conspire to make the expected time of recurrence of a pre-post correlation exponentially longer for spurious / noisy events than for meaningful (i.e., due to structural regularities of the environment) events.

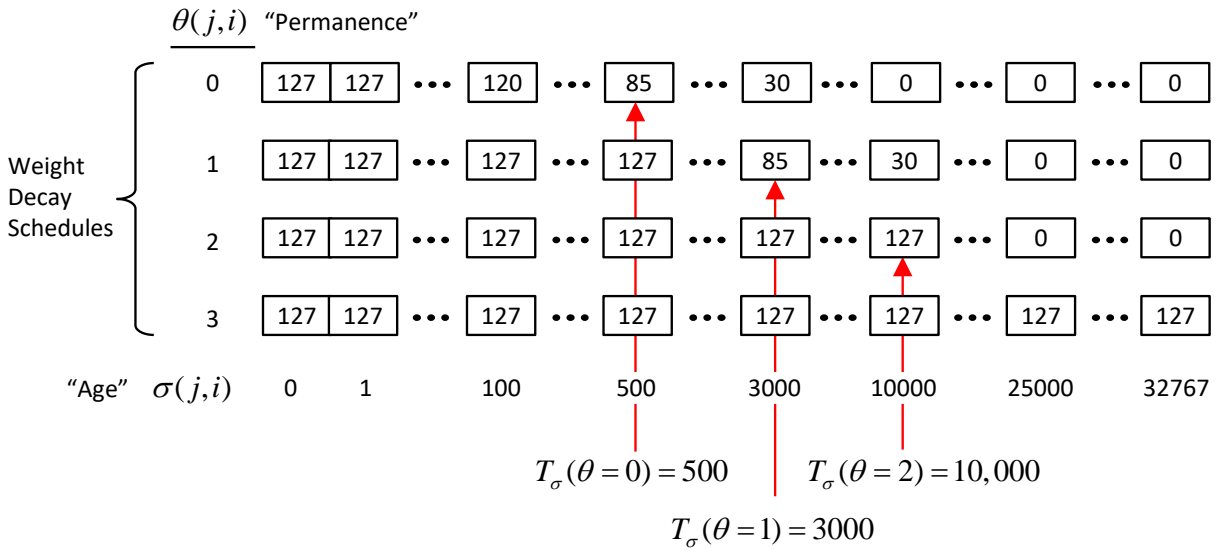

**Figure II-6:** The "weight table": Indexed by age (columns) and permanence (rows). A synapse's weight is gotten by dereferencing its age, $\sigma(j,i)$, and its permanence, $\theta(j,i)$. See text for details.

If we run the model indefinitely, then eventually every synapse will experience two successive pre-post correlations occurring within any predefined window, $T_\sigma$. Thus, without some additional mechanism in place, eventually *all* afferent synapses into a mac will be permanently increased to $w_{max}=127$ at which point (total saturation of the afferent weight matrices) all information will be lost from the afferent matrices. Therefore, Sparsey implements a "critical period" concept, in which *all* weights leading to a mac are "frozen" (no further learning) once the fraction of weights that have been increased in any *one* of its afferent matrices crosses a threshold. This may seem a rather drastic solution to the classic trade-off that Grossberg termed the "stability-plasticity dilemma" (Grossberg 1980). However, note that: a) 'critical periods' have been demonstrated in the real brain in vision and other modalities (Wiesel and Hubel 1963, Barkat, Polley et al. 2011, Pandipati and Schoppa 2012); b) model parameter settings can readily be found such that in general, all synaptic matrices afferent to a mac approach their respective saturation thresholds roughly at the same time (so that the above rule for freezing a mac does not result in significantly underutilized synaptic matrices); and c) in Sparsey, freezing of learning is applied on a mac-by-mac basis. We anticipate that in actual operation, the statistics of natural visual input domains (filtered as described earlier, i.e., to binary 1-pixel wide edges) in conjunction with model principles/parameters will result in the tendency for the lowest level macs to freeze earliest, and progressively higher macs to freeze progressively later, i.e., a "progressive critical periods" concept. Though clearly, if the model as a whole is to be able to learn new inputs throughout its entire "life", parameters must be set so that some macs, logically those at the highest levels, never freeze. We are still in the earliest stages of exploring the vast space of model parameters that influence the pattern of freezing across levels.

The ultimate test of whether the use critical periods as described above is too drastic or not is how well a model can continue to perform recognition/retrieval (or perform the specific recognition/retrieval-contingent tasks with which it is charged) over its operational lifetime (which will in general entail large numbers of novel inputs), in particular, after many of its lower levels have been frozen.

### II.A.14. Learning arbitrarily complex nonlinear similarity metrics

The essential property needed to allow learning of arbitrarily complex *nonlinear* similarity metrics (i.e., category boundaries, or invariances) is the ability for an individual SDC in one mac to associate with multiple, perhaps arbitrarily different, SDCs in one or more other macs. This ability is present *a priori* in Sparsey in the form of the *progressive persistence* property wherein code duration, or *persistence* ($\delta$), (measured in frames) increases with hierarchical level (in most experiments so far, $\delta$ doubles with level). For example, the V2 code $\phi_X^{2,j}$ in Figure II-7a becomes associated with the V1 code $\phi_Y^{1,i}$ at time $t$, and because it persists for two time steps, it also becomes associated with $\phi_Z^{1,i}$ at $t+1$. By construction of this example, $\phi_X^{2,j}$ represents (a particular instance of) the spatiotemporal concept, "rightward-moving vertical edge". However, if for the moment, we ignore the fact that these two associations occurred on successive time steps, then we can view $\phi_X^{2,j}$ as representing $XOR(\phi_Y^{1,i}, \phi_Z^{1,i})$, i.e., just two different (in fact, pixel-wise disjoint) instances of a vertical edge falling within the U-RF of $M_j^2$. That is, the U-signals from either of these two input patterns alone (but not together[1]) can cause reinstatement of $\phi_X^{2,j}$. This provides an unsupervised means by which *arbitrarily different*, but temporally contiguous, input images, which may in principle portray any transformation that can be carried out over a two-time-step period and over the spatial extent of the RF in question, can be associated with the *same* object or class (the identity of which is carried by the persisting code, $\phi_X^{2,j}$).

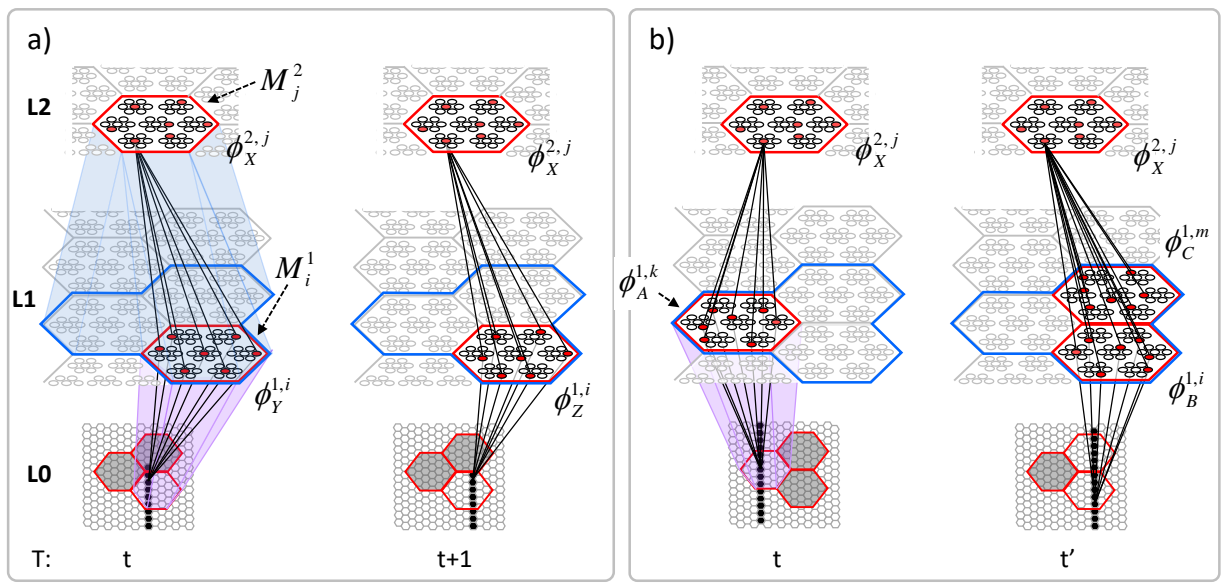

**Figure II-7:** (a) The basic model property, progressive persistence, allows SDCs in higher level macs to associate with sequences of temporally contiguous SDCs active in macs in their U-RFs. (b) More generally, any mechanism which allows a particular code, e.g., $\phi_X^{2,j}$, to be activated under the control of a supervisory

---

[1] Indeed, the two codes, $\phi_Y^{1,i}$ and $\phi_Z^{1,i}$, cannot occur together since they occur in the same L1 mac, $M_i^1$.

signal can cause $\phi_X^{2,j}$ to associate with two or more arbitrarily different codes presented at arbitrarily different times, thus allowing $\phi_X^{2,j}$ to represent arbitrary invariances (classes, similarity metrics).

Figure II-7b shows two more instances in which $\phi_X^{2,j}$ is active, denoted *t* and *t'* to suggest that they may occur at arbitrary times. If there is a supervisory signal by which $\phi_X^{2,j}$ can be activated whenever desired, then $\phi_X^{2,j}$ will associate with whatever codes are active in its RF (in this example, specifically, its U-RF) at such times. In this case, the two inputs associated with $\phi_X^{2,j}$ are just two different instances of a vertical edge falling within $\phi_X^{2,j}$'s RF. Furthermore, note that the number of active codes (features) in the RF can vary across association events. Thus, $\phi_X^{2,j}$ can serve as a code representing *any invariances* present in the set of codes with which it has been associated.

This is in fact how supervised learning is implemented in Sparsey. That is, the supervised learning signal (label) is essentially just another input modality and supervised learning is therefore treated as a special case of cross-modal unsupervised learning. We have conducted preliminary supervised learning studies involving the MNIST digit recognition database (LeCun, Bottou et al. 1998) using a model architecture like that in Figure II-8. However, to adequately describe the supervised learning architecture, protocol, and theory, would add too much length to this paper and so we save that work for a separate paper. Nevertheless, we are confident that the general framework described here will allow arbitrarily complex *nonlinear* similarity metrics, e.g., functions described as comprising the "AI Set", by Bengio, Courville et al. (2012), to be efficiently learned as *unions*, where each element of the union is a *hierarchical spatiotemporal composition* of the locally primitive (i.e., smoothness prior only) similarity metrics embedded in individual macs.

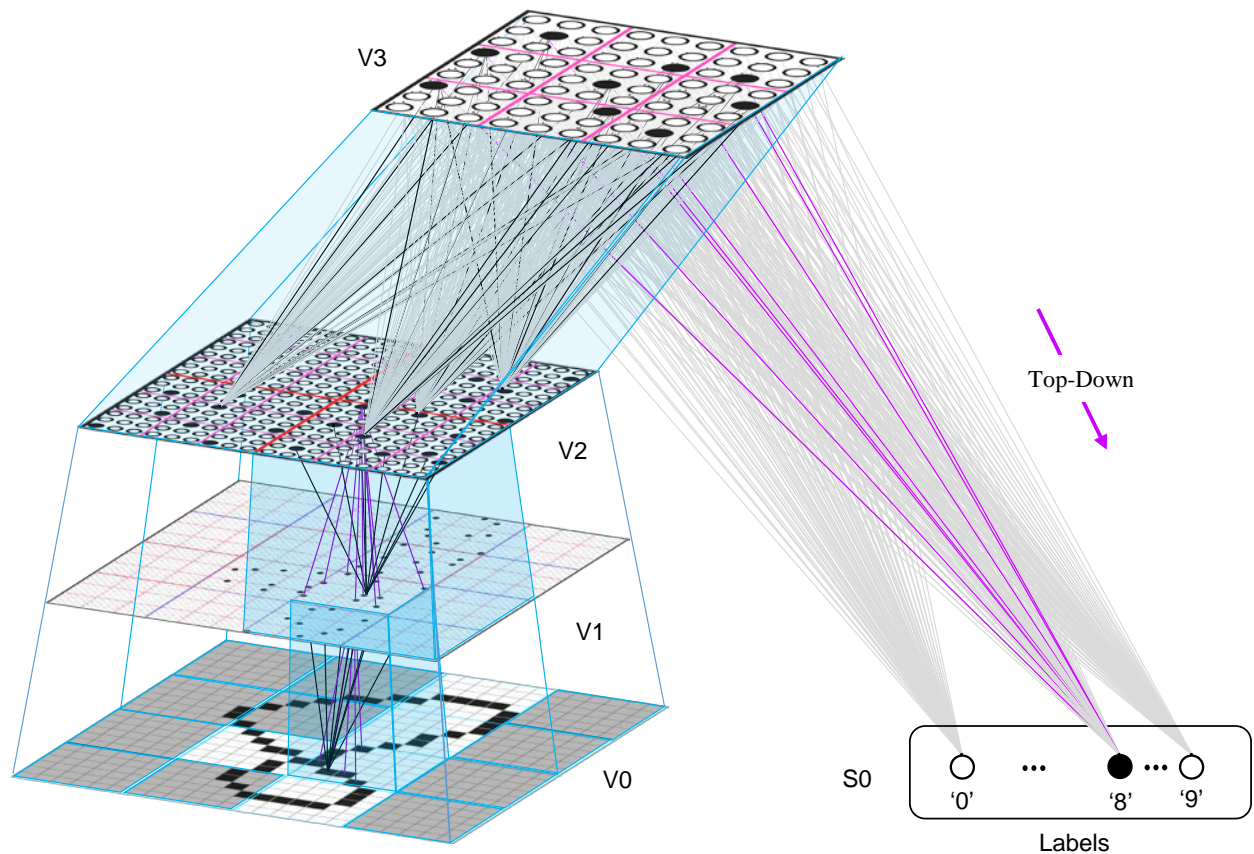

**Figure II-8:** The 4-level model used in preliminary supervised learning studies involving the MNIST digit recognition task. This shows the recognition test trial in which the '8' was presented, giving rise to a flow of U-signals activating codes in macs throughout the hierarchy, and finally a top-down flow from the activated V3 code to the Label field, where the unit with maximal D-summation, the '8' unit, wins.

### II.A.15. Neural Implementation of CSA

Though we identify the broad correspondence of model structures and principles to biological counterparts throughout the paper, we have thus far been less concerned with determining precise neural realizations. Our goal has been to elucidate computationally efficient and biologically plausible mechanisms for generic functions, e.g., the ability to form large numbers of permanent memory traces of arbitrary spatiotemporal events on-the-fly and based on single trials, the ability to subsequently directly (i.e., without any serial search) retrieve the best-matching or most relevant memories, invariance to nonlinear time warping, coherent handling of simultaneous activation of multiple hypotheses, etc. We believe that Sparsey meets these criterion so far. For one thing, it does not require computing any gradients or sampling of distributions, as do the Deep Learning models (Hinton, Osindero et al. 2006, Salakhutdinov and Hinton 2012). Nevertheless, we make the following points about Sparsey's relation to the brain.

First, we believe it is quite important, both for distinguishing Sparsey from other canonical cortical microcircuit models and for falsifiability, that the CSA really does entail two rounds, in quick succession, in which the mac's principal cells integrate their inputs, resulting in at least one of the cells in each CM reaching threshold and sending action potentials to the local inhibitory circuitry, which then fires, thus keeping all other cells in the CM from spiking (according to any number of detailed biophysical mechanisms, e.g., Jitsev (2010)). The first round winners' outputs (in addition to engaging the local inhibitory circuitry to suppress the other cells in their respective CMs) are averaged to yield *G*. And *G* then drives a modulation of the cell activation functions (as described in Sections II.A.9 and II.A.10) in preparation for the second round of competition. Due to the modulated activation functions, the second (and final) round winners will generally differ from the first round winners. Specifically, the intersection of the set of second round winners with the first round winners increases with $G$. In (Rinkus 2010), we speculated that some combination of neuromodulators could underlie this behavior, but we have not yet refined that hypothesis.

Second, we note that Sparsey is a highly simplified/reduced model of cortical processing. It lacks analogs of layers 4, 5, or 6, and does not explicitly model inhibitory cells. In addition, it uses binary (non-spiking) neurons, effectively binary weights with variable permanence, and a simple Hebbian learning scheme with passive decay. The general consensus is that L4 is the main recipient of feedforward signals (from thalamus or from earlier cortical stages), whereas L2/3 receives horizontal (intrinsic) and top-down inputs. And, L5 and L6 project to earlier cortical stages and to subcortical structures and are involved in local feedback loops with L2/3. While numerous studies provide more detailed specifications fitting the above supra/infra-granular canonical circuit motif (Douglas and Martin 2004), numerous details are yet to be understood and various new studies force significant modification/clarification of the canonical view, e.g., that L5/6 cells are also activated directly by U (specifically, thalamic) input (Constantinople and Bruno (2013) and that thalamic input to L1 is much more substantial than previously thought (Rubio-Garrido, Pérez-de-Manzo et al. 2009).

In any case, while realizing the generic functionalities noted above has thus far required only a single population (layer) of principal cells, which best matches the L2/3 pyramidals, we anticipate incorporating modeling of other layers as needed. In particular, in its current "1-layer" form, Sparsey can be viewed as carrying out spatiotemporal processing underlying perception / recognition of spatiotemporal patterns and thinking, but without the accompanying motor output. Incorporating a "motor side" to the model naturally suggests a move to a "2-layer" concept, i.e., supragranular (L2/3 and L4) and infragranular (L5 and L6).

## II.B.    CSA: Retrieval Mode

In this section, we will first motivate the need for introducing some complexity to the computation of $G$ when in retrieval mode and then describe the modification. We begin by thinking about how the model should respond to test trials involving previously learned sequences corrupted in particular ways. For example, if the model has learned the sequence S1=[BOUNDARY] in the past and is now presented with S2=[BOUNDRY], should it decide that S2 is functionally equivalent to S1? That is, should it respond equivalently to S2 and S1? More precisely, should its internal state at the end of processing S2 be the same as it was at the end of processing S1? The reader will probably agree that it should. We all encounter spelling errors like this all the time and read right through them.  Similarly, if one encountered S3=[BBOUNDARY], S4=[BBOOUUNNDDAARRYY], S5=[BOUNNNNNNDARY], or any of numerous other variations, he/she would likely decide it was an instance of S1. We could think of all these variations (corruptions) simply as omissions/repetitions. However, we prefer to think of this class of corruptions as instances of the class of *nonlinearly time-warped* instances of (discrete) sequences. Thus, S2 can be thought of as an instance of S1 that is presented at the same speed as during learning up until item "D" is reached, at which time the process presenting the items momentarily speeds up (e.g., doubles its speed) so that "A" is presented but then replaced by "R" before the model's next sampling period. Then the process slows back down to its original speed and item "Y" is sampled. Thus S2 is a nonlinearly time-warped instance of S1. We can construct similar explanations, involving the underlying process producing the sequences undergoing a schedule of speedups and slowdowns relative to the original learning speed, for S3, S4, etc. In fact, S4 is even simpler; it's just a uniform slowing down, to half speed, of the whole process.

Of course, there are limits to how much we want a system to generalize regarding these warpings. And the final equivalence classes, in particular for processing language, must be experience-dependent and idiosyncratic. For example, should a model think that S6=[COD] is just an instance of S7=[CLOUDS], produced twice as fast as during the learning instance? In general, probably not. Furthermore, we have not even considered in these examples the fact that the individual sequence items are actually pixel patterns which can themselves by noisy, partially occluded, etc., which would of course influence the normative category decisions. Nevertheless, the ubiquity of instances such as described above, not just in the realm of language, but of lower-level raw sensory inputs, suggests that a model have some mechanism for dealing with them, i.e., some mechanism for treating moments produced by nonlinearly time-warping as equivalent.

Our explanation of the modified *G* computation in retrieval mode uses an example involving a 3-level model that has only one mac at each level. Figure II-9 shows representative samples of the U, H, and D learning that occurs as the model is presented with the sequence, [BOTH]. Note that the model is unrolled in time here, i.e., the model is pictured at four successive time steps and in particular, the origin and destination cell populations of the increased H synapses (green) are the same. This figure illustrates several key concepts. First, learning a sequence involves increasing the H-wts from the previously active code to the currently active code. The D-wts (magenta) are also increased from the previously active code (in this case, in the L2 mac) to the currently active destination code in the L1 mac. Note however that the U-wts (blue) are increased from the currently active input (L0 code) to the currently active L1 code. We show the full set of afferent U, H, and D wts that are increased for one cell—the winner in the upper left CM of the L1 mac—at each time step. Thus, this figure emphasizes that, on each moment, individual cells become associated with their entire afferent input (spatiotemporal context) in one fell swoop. Though we only show this occurring for one cell on each frame, all winners in a mac code will receive the same weight increases simultaneously. Thus we can say not only that individual cells become associated with the mac's entire spatiotemporal contexts but that whole mac codes become associated with the mac's entire spatiotemporal contexts.

The second key concept illustrated is progressive persistence, in this case, that L2 codes persist for twice as long as L1 codes. Cell color in this figure is used to make persistence clear. Thus, the first L2 code that becomes active D-associates with two L1 codes. And, because of the modeling decision that D-wts are increased from previously active to currently active codes, the two L1 codes are those at $t$=2 and $t$=3. The second L2 code to become active (orange) D-associates with the L2 code at $t$=3 and would associate with a $t$=4 L1 code if one occurred.

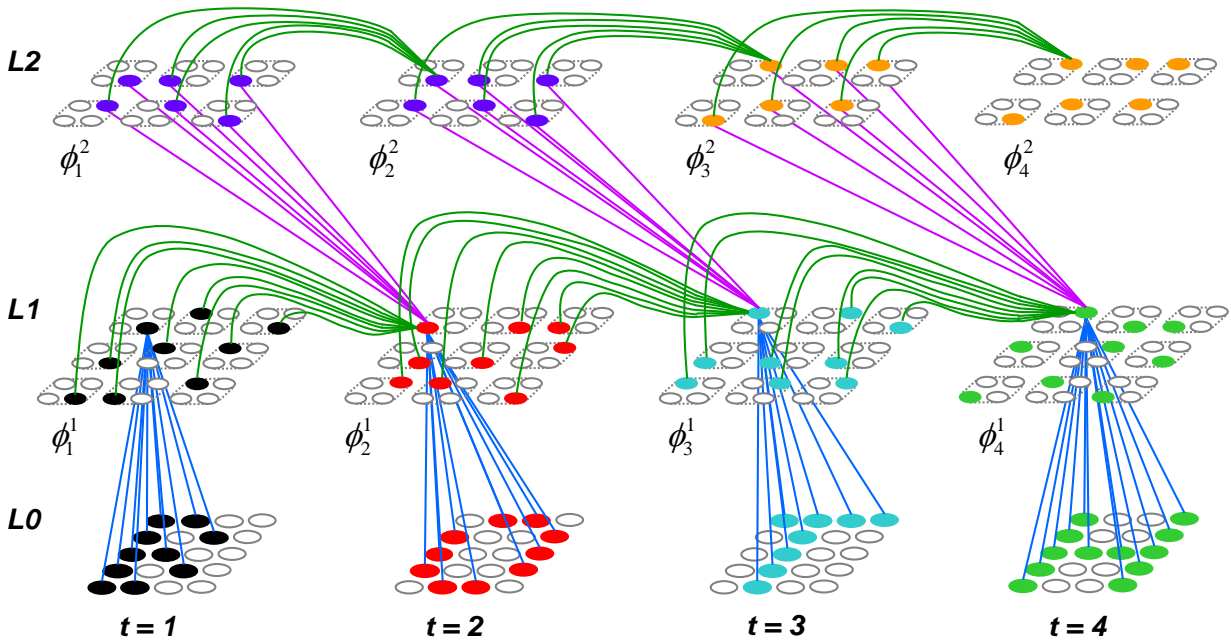

**Figure II-9:** The formation of a hierarchical spatiotemporal memory trace, unrolled in time, of the input sequence, [BOTH]. We only show representative samples of the increased weights on each frame. The model has one L1 mac with $Q_1$=9 CMs, each with $K$=4 cells and one L2 mac with $Q_2$=6 CMs, each with $K$=4 cells. The resulting trace can be said to have been produced using both *chaining* (increasing H-wts between successively active codes at the same level) and *chunking* (increasing U and D wts between single higher-level (L2) codes and multiple lower-level (L1) codes. See text for detailed explanation.

Having illustrated (in Figure II-9) the nature of the hierarchical spatiotemporal memory trace that the model forms for [BOTH], Figure II-10 compares model conditions when processing one particular moment—the second moment—of a test trial that is identical to the learning trial (Figure II-10a) to conditions when processing the second moment of a time-warped instance of the learning trial—specifically, a moment at which the item that originally appeared as the third item of the learning trial, "T", now appears as the second item immediately after "B", i.e., "O" has been omitted (Figure II-10b). We can represent the two test trial moments as [B**O**] and [B**T**], respectively, where bolding indicates the frame currently being processed and the non-bolded letters indicate the context leading up to the current moment. The first thing to say is that the second moment of the time-warped instance is simply a novel moment. Thus, the caveat we mentioned above applies. That is, deciding whether a particular novel input moment should be considered a time-warped instance of a known moment or as a new moment altogether cannot be done absolutely.

Figure II-10a shows the case where the test trial moment [B**O**] is identical to the learning trial moment [B**O**]. The main point to see here is that, given the weight increases that will have occurred on the learning trial, all three input vectors, U, H, and D, will be maximal (equal to 1) for the red cell (which is in $\phi_2^1$) in each L1 CM. At right (yellow), we zoom in on the conditions only for the upper left L1 CM, but the conditions are statistically similar for all L1 CMs. We show that for the red cell, $U$=1, $H$=1, and $D$=1. The blue cell (which is in $\phi_3^1$) also has maximal D-support and the blue, green, and black cells have non-zero U inputs (their U-inputs are not shown in the main figure to minimize clutter), due to the pixel overlap amongst the four input patterns, but they all have $H$=0. Thus, according to Eq. 4 of the CSA (Table II-1), the red cell has $V=U\times H\times D$=1, whereas the others have $V$=0. We refer to red cell as having a "3-way match" in that all three evidence vectors are maximal and agree. Also, we refer to the *G* version computed using all three input vectors as $G_{HUD}$. Thus, in this case, where the test moment is identical to a learned moment, CSA Eq. 4 is sufficient as is.

However as shown in Figure II-10b, when an item ("O") has been omitted with respect to the learning trial, the *H* and *D* vectors to the red cell will no longer agree with its *U* vector. Various policies could be imagined for handling this situation. The model could simply consider such a case as being a novel moment, [B**T**]. This would require no modification to the CSA. Or, as discussed earlier, the model could *check* to see whether the current moment could have resulted from a nonlinear time-warping process, and should therefore be judged identical to some previously learned moment. In this case, the current moment [B**T**] is identical to the learning trial moment [BO**T**] if we assume that the process presenting the sequence to the model sped up by 2x at t=1, causing the "O" to be missed.

So, how does the model *check* this possibility? It is quite simple. All it needs to do is disregard the H signals when computing the *V*'s (CSA Step 4). In other words, it "backs off" from the more stringent 3-way $G_{HUD}$ match criterion to the more permissive 2-way $G_{UD}$ criterion. Note that the model begins by computing the highest-order *G* available at the current moment, in this case, using all three input vectors. Only if that highest-order *G* falls below a threshold, which we typically set rather high, e.g., $G_{HUD}^{+} = 0.9$, does it bother to compute the next lower order version(s) of *G*, i.e., $G_{UD}$, $G_{HU}$, and $G_{HD}$. Similarly, only if whichever 2-way version has been considered falls below another threshold, which is typically set even higher than the first, e.g., $G_{UD}^{+} = 0.95$, does the model *back-off* to the next lower order match criterion.

In this example, $G_{UD} = 1$, meaning that there is a code stored in the L1 mac—specifically, the set of blue cells assigned as the L1 code at *t*=3 of the learning trial (Figure II-9)—which yields a perfect 2-way match. Thus, there is no need to back-off to the "1-way" match criterion, $G_{U}$. However, there are many naturally occurring instances in which backing all the way off to the lowest-order criterion (i.e., basing the *V* values and thus, the *G*, on only the U signals, ignoring the H and D signals) is appropriate. There are myriad policy considerations regarding possible precedence orders of the different *G* versions and whether or not and under what conditions the various versions should be considered. We are actively exploring these issues, but cannot delve into this topic in this paper.

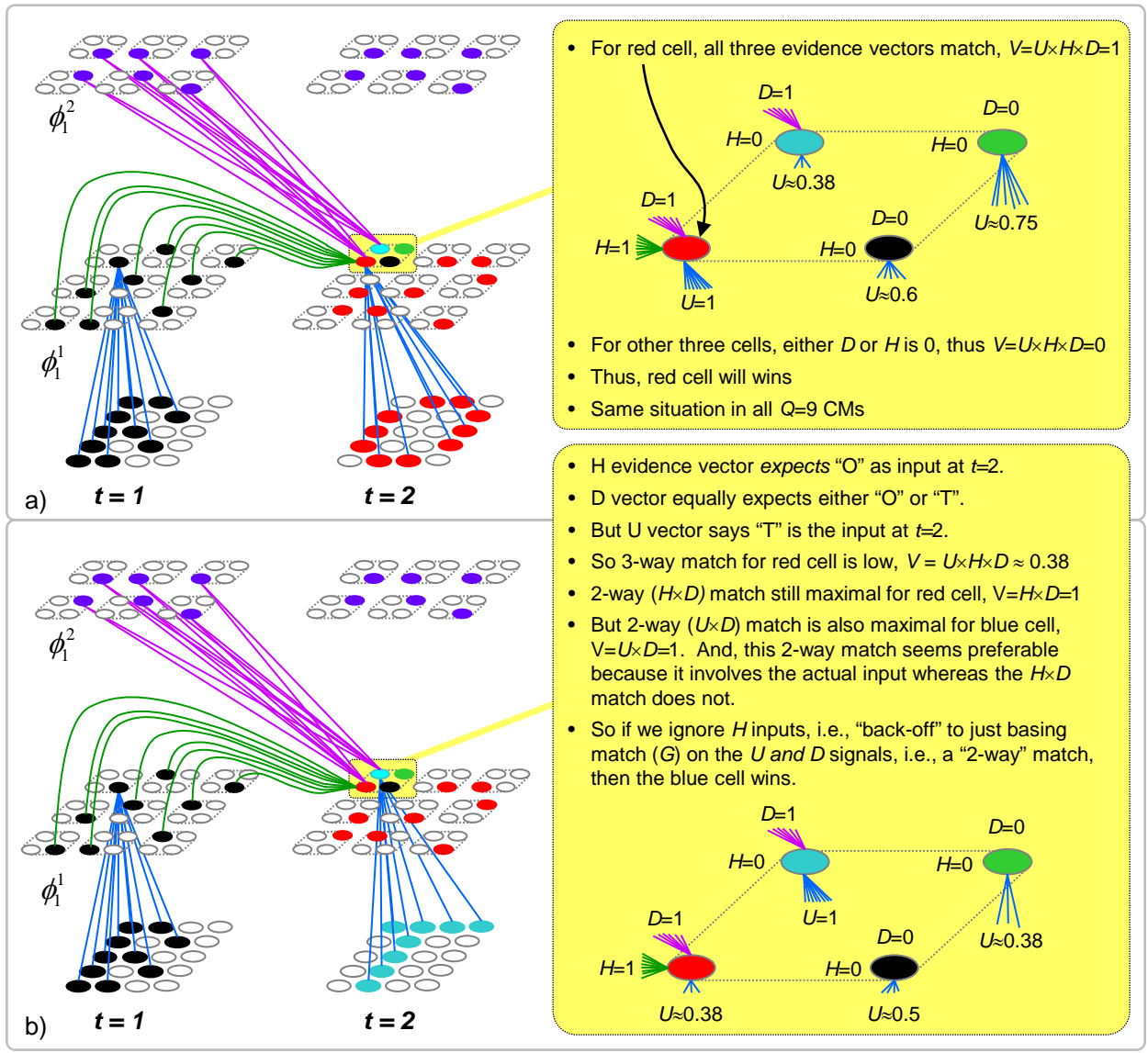

**Figure II-10:** Motivation for the Back-off Strategy for computing $G$ in retrieval mode. (a) Detail of conditions that exist at L1 when processing the second moment [B**O**] of a test trial that is identical to the learning trial (in Figure II-8). (b) Detail of conditions that exist at L1 when processing the second moment [B**T**] of a test trial that is a time-warped version of the learning trial, specifically, a sequence that is sped up by 2x at $t=1$, causing the "O" to be missed and the "T" to occur immediately after the "B". See text for detailed discussion.

Figure II-11 completes this example by showing that the back-off policy allows the model to keep pace with nonlinearly time-warped instances of previously learned sequences. That is, the model's internal state (i.e., the codes active in the macs) can either advance more quickly (as in this example) or slow down (not demonstrated herein) to stay in sync with the sequence being presented. Figure II-11a is given for comparison, showing the full memory trace that becomes active during a retrieval trial for an exact duplicate of the training trial, [BOTH]. In this case, no back-off would be required because all signals at all times would be the same during retrieval as they were during learning. Figure II-11b shows the trace that obtains, using the back-off protocol, throughout presentation of the nonlinearly time-warped instance of the training trial, [BTH].

The back-off from $G_{HUD}$ to $G_{UD}$ occurs in the L1 mac at $t=2$ (as was described in Figure II-10b). Since $G_{UD}=1$, the *V*-to-$\psi$ map is made very expansive, resulting in activation, at $t=2$ of the test trial, of the code, $\phi_3^1$ (blue cells), which was originally activated at $t=3$ in the learning trial. Thus, the back-off has allowed the model's internal state (in L1) to "catch up" to the momentarily sped up process that is producing the input sequence. Once $\phi_3^1$ is activated, it sends U-signals to L2 (blue signals converging on orange cell in rose highlight box). This results in the L2 code, $\phi_3^2$ (orange cells), being activated without requiring any back-off. That's because the L2 code from which H signals arrive at $t=2$, $\phi_1^2$ (purple cells) increased its weights not only onto itself (at $t=2$ of the learning trial) but also onto $\phi_3^2$ at $t=3$ of the learning trial. Thus, the six cells comprising $\phi_3^2$ (orange) yield $G_{HU}=1$ (note that $G_{HU}$ is the highest order *G* version available at L2 since there is no higher level). Consequently, a maximally expansive *V*-to-$\psi$ map is used in the L2 mac, resulting in reinstatement of $\phi_3^2$. At this point—$t=2$ of the test trial—the entire internal state of the model (i.e., at L1 and L2) is identical to its state at $t=3$ of the learning trial (two central dashed boxes connected by double-headed black arrow): the model, as a whole, has "caught up" with the momentary speed up of the sequence. The remainder of the sequence proceeds the same as it did during learning, i.e., state at $t=3$ of retrieval trial equals state at $t=4$ of learning trial.

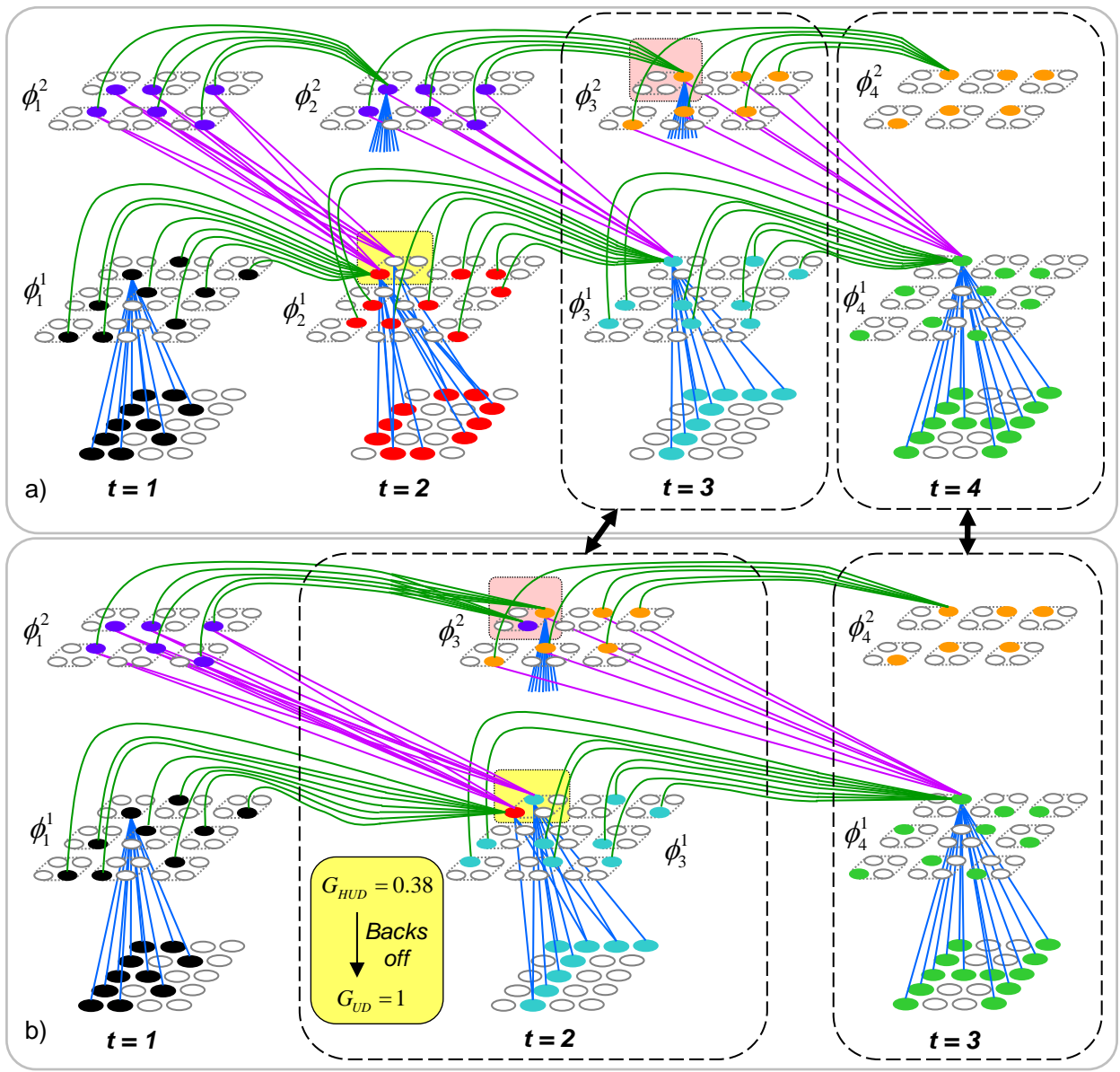

**Figure II-11:** This figure shows the complete test trial traces for: (a) an exact duplicate of the training trial, [BOTH]; and (b) the nonlinearly time-warped instance of the training trial, [BTH]. In (b), back-off from $G_{HUD}$ to $G_{UD}$ occurs in the L1 mac at $t=2$ (as was described in Figure II-10b), which allows entire internal state of the model (i.e., at L1 and L2) to "catch up" with the momentary speed up of the sequence. The remainder of the sequence and the associated internal trace then obtains the same as during learning. See text for detailed description.

The final, and really the most important, point of this section is that Sparsey's back-off policy *does not change* the time complexity of the CSA: it still runs with *fixed time complexity*, which is essential in terms of scalability to real-world problems. True, expanding the logic to compute multiple versions of $G$ increases the absolute number of computer operations required by a single execution of the CSA. However, the number of possible $G$ versions is small and more to the point, fixed. Thus, adding the back-off logic adds only a fixed number of operations to the CSA and so does not change the CSA's time complexity.

During each execution of the CSA, *all stored codes* compete with each other. In general, the set of stored codes will correspond to moments spanning a large range of *Markov orders*. For example, in Figure II-9, the four moments, [**B**], [B**O**], [B**OT**], and [B**OTH**], are stored, which are of progressively greater Markov order. *During each moment of retrieval, they all compete*. More specifically, they all compete first using the highest-order *G*, and then if necessary, using progressively lower-order *G*'s. However, it is crucial to see that with back-off, not only are the explicitly stored (i.e., actually experienced) moments compared, but so are a far larger number of time-warped versions of the actually-experienced moments. For example in Figures II-10b and II-11b, the moment [B**T**], which never actually occurred competes and wins (by virtue of back-off) over the moment [B**O**], which did occur. And crucially, as noted above, all these comparisons take place with fixed time complexity! Space does not permit here, but the above mechanism and reasoning generalizes to arbitrarily deep hierarchies. As the number of levels increases, with persistence doubling at each level, the space of hypothetical nonlinearly time-warped versions of actually experienced moments, which will materially compete with the actual moments (on every frame and in every mac) grows exponentially. And, we emphasize that these exponentially increasing spaces of never-actually-experienced hypotheses are *envelopes* around the actually-experienced moments: thus, the invariances implicitly represented by these envelopes are (a) learned and (b) idiosyncratic to the specific experience of the model.

### II.C. CSA: SIMPLE RETRIEVAL MODE

Both the learning mode CSA and the retrieval mode CSA described above, which is just the learning mode CSA augmented by the back-off protocol, involve the *G*-based modification of the cell activation functions and the second, probabilistic round of competition for choosing the final code (CSA Steps 8-12, Table II-1). If the model is operating as a truly autonomous agent, then it, or rather any of its constituent macs, may be presented with a truly novel input pattern at every moment experienced. Thus, a mac must be prepared to learn, i.e., assign a new SDC, at every moment.[2] As described in earlier sections, the CSA's two competitive stages, with the second, probabilistic stage using the *G*-modulated cell activation functions, satisfies the requirements for autonomous operation. That is, as *G* decreases, the expected intersection of the final code (for the current frame) chosen with the closest matching stored code decreases to chance, which results in the occurrence of novel pre-post correlations, and thus new learning. On the other hand, as *G* increases towards 1, the expected intersection of the finally chosen code with the closest matching stored code increases to complete, which results in no (or at least, statistically, very few) novel pre-post correlations and thus no new learning.

However, if the model "knows" that is operating in pure retrieval mode, i.e., that at each moment each mac should simply activate the code of the learned moment that most closely matches its current input moment, then there is no advantage to having the second *G*-dependent probabilistic stage of competition. In fact, the optimal strategy in this case is simply to choose the cell with the highest *V* value in each CM. The transfer of global information (*G*) back into the local (within each CM) winner selection processes, which occurs in steps 8-12, does not help and in fact, can only hurt (i.e., it can only reduce the probability of the maximally likely cell in a given CM winning). Thus, in this "simple retrieval mode", in which the model knows that it will not be asked to learn anything new, the optimal algorithm is just the first seven steps of the CSA given in Table II-1, but augmented with the back-off protocol described in the previous section. Thus, we do not state the simple retrieval mode of the CSA separately. We will clearly indicate which of the two retrieval modes is used in the studies reported in the next section.

We emphasize that the *deterministic* "simple retrieval mode" algorithm cannot be used during learning because it would result in essentially mapping all of the mac's input patterns to one or a very small number of codes, vastly over-utilizing only a tiny fraction of the mac's cells and vastly decreasing the number of codes (amount of information) that can be stored in the mac.

---

[2] Actually, in a hierarchical model faced with the prospect of possibly having to learn something new on every moment of its operational lifetime, its sufficient only that at least one mac (which would typically be at the highest level) be prepared to learn at every moment (cf. earlier discussion of cirtical periods).

However, based on first principles, it seems plausible that for the vast majority of Sparsey's envisioned operational regime, i.e., the regime in which the number of codes stored in the macs (or more specifically, the faction of synapses that have been increased) remains below a threshold, the simple retrieval mode should always do better (on average) than the probabilistic retrieval mode  Specifically, recall that in probabilistic retrieval mode, the winner in a CM is chosen *as a draw* from the *V* distribution. Depending on the particular shape/statistics of the *V* distribution, the cell with the maximum *V* might therefore be chosen winner only a small fraction of the time. Yet, that max-*V* cell is the most likely cell given the total evidence (from the U, H, and D signals) arriving at the mac. In simple retrieval mode, the max-*V* cell always wins. Again, provided that the fraction of the mac's afferent synapses that have been increased remains low enough, simply choosing the max-*V* cell as winner yields higher expected accuracy.

## II.D.    DEFINITIONS OF SYMBOLS USED HEREIN

**Table II-2: Major Symbols in CSA Equations.**

| Symbol | Definition | Symbol | Definition |
|---|---|---|---|
| *Active(m)* | Whether mac m is active or not | $\lambda_{U(t)}$ | Power to which U is raised prior to being multiplied with H and D signals. It can vary as a function of time from beginning of the sequence (snippet) being processed. |
| $\Upsilon(m)$ | Age, in number of time steps (frames), of the currently active code in mac *m*. | $\lambda_H$, $\lambda_D$ | Analogous to $\lambda_{U(t)}$ except that for now they are not functions of time. |
| $Q, Q_i$ | Number of CMs per mac; same but for a specific level, *i*. | $\delta(m)$ | Persistence, in number of time steps, of mac *m*. Currently, all macs of a given level have the same persistence. |
| $K, K_i$ | Number of cells per CM; ; same but for a specific level, *i* | $M_{2,3}$ $M_4^3$ | The mac at coordinates (2,3) (when the level is unambiguous). Alternate notation: Mac with index "4" at level "3". |
| $u(i)$, $h(i), d(i)$ | Raw sum of weighted signals from cells comprising cell *i*'s U-RF. $h(i), d(i)$ are analogous | $U(i)$, $H(i)$, $D(i)$ | $U(i)$ is the normalized $u(i)$, to [0,1] range. $H(i), D(i)$ are analogous. |
| $U^-$ | Lower threshold below which a cell's U value is considered 0. $H^-, D^-$ analogous | $U^+$ | Upper threshold above which a cell's U value is considered 1. $H^+, D^+$ analogous |
| $\pi_U$ | The # of active features in a mac's U-RF. | $\pi_U^*$ | Number of active features in a mac's U-RF, which are active in macs with $\zeta \leq B$. |
| $\pi_U^-$, $\pi_U^+$ | Lower and upper bounds on the number of active features that must be present in a mac's U-RF for that mac to activate. | $V(i)$ | Overall local evidence that cell *i* should become active. Product of functions of $U(i)$, $H(i)$, and $D(i)$. |
| $\hat{V}_j$ | Maximum *V(i)* in CM, C*j*. | $V_\zeta$ | Threshold for a cell to be considered as part of an active hypothesis |
| $G$ $G(t)$ | Average $\hat{V}$ value over a mac's *Q* CMs. It is a measure of the familiarity of a mac's total input, normalized to [0,1]. | $G^-$ | Threshold below which the mac's *G* value is considered effectively zero. |
| $\chi$ | The sigmoid expansion factor | $\gamma$ | The sigmoid expansion exponent |
| $\eta$ | Range of the $V$-to-$\psi$ map, which transforms a cell's $V$ value into its relative (within its own CM) probability of winning, $\psi$. | $\sigma_1$, $\sigma_2$ $\sigma_3$, $\sigma_4$ | Parameters that interact to control overall sigmoid expansivity and shape, e.g., horizontal position of inflection pt., etc. |
| $a(j,t)$ | Activation (0,1) of cell $j$ at time $t$. | $M_j$ | Number of macs in Level $j$. |
| $\zeta_q$ | # of cells in CM $q$ with $V(i) > V_\zeta$. Typically, $V_\zeta$ is set close to 1, e.g., 0.95. | $\zeta$ | The number of maximally active hypotheses, $\zeta$, in a mac. |
| $F(\zeta)$ | The correction factor for increasing the weights of outgoing signals from cells in macs that have *multiple competing hypotheses* (MCHs), i.e., $\zeta > 1$. | $F(\zeta(j,t))$ | The MCH correction factor $F(\zeta)$ at time $t$ for mac that contains cell $j$. |
| $A$ | Exponent (<1.0) for discounting MCH correction factor when $\zeta > 1$. | $B$ | Threshold on $\zeta$ above which we ignore completely signals from the source mac. |
| $\psi(i)$ | The relative probability of activating cell $i$ in a mac. | $\rho(i)$ | The absolute probability of activating cell $i$ in a mac. |
| $G_U$ | $G$ computed based only on the U signals to a mac. Similarly, $G_{HUD}$ is $G$ computed based on all three input vectors, U, H, and D. Similarly, for $G_{HU}$, $G_{UD}$, $G_{HD}$, $G_H$ | | |
| $G_{HU}^+(t)$ | Threshold below which we back off to the next lower-order (or more generally, the next-considered) version of G. Here, we suggest that this threshold can be a function of time (frame). | | |
| U-RF, H-RF, D-RF | U-RF is a bottom-up receptive field. Can be applied to single cells or to whole macs. For cells/macs at L1 the U-RF is a set (or array) of individual binary cells (e.g., pixels). For cells/macs at higher levels, the U-RF is a set (array) of macs. H-RF and D-RF are analogous, but they always consist of a set (array) of macs. | | |

# III. RESULTS

## III.A. STUDY 1: SPATIOTEMPORAL SISC PROPERTY

Study 1 is an *unsupervised learning* study that demonstrates that Sparsey maps spatiotemporally more similar inputs to more highly intersecting SDCs, i.e., the *similar-inputs-to-similar-codes* (SISC) property. This is an instance of what others have referred to as the "smoothness prior" (Bengio, Courville et al. 2012). The model instance used here has a 12x12-pixel input level (L0) and one internal level (L1) consisting of one mac with $Q$=25 CMs, each with $K$=9 cells. The set of six 2-frame sequences (S0-S5) used in this study are shown in Figure III-1a. All sequences have the same second item, X, while the pixel-wise overlap of the sequence-initial item with S0's first item, A, decreases across sequences, S1=[BX], S2=[CX], etc. Thus, the spatiotemporal similarity of the second frame of each sequence with the second frame of S0 drops across sequences (even though the *purely spatial* similarity of the second frame remains the same at 100%). We will show that the codes assigned to the second frame of the progressively spatiotemporally less similar sequences have progressively smaller intersection with the code assigned to the second frame of S0.

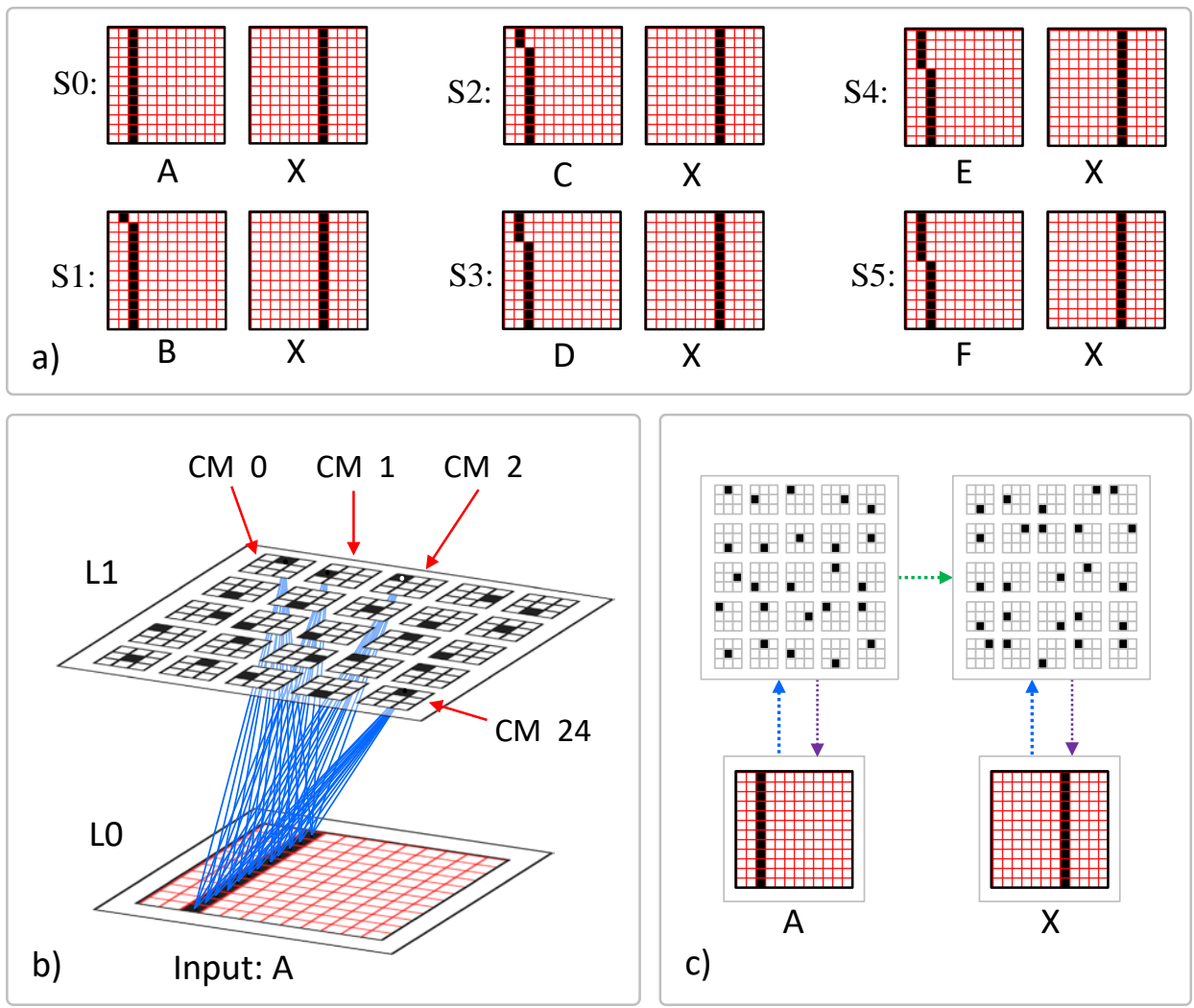

**Figure III-1:** (a) The six 2-frame sequences used in Study 1. (b) The model whose internal level consists of one mac comprised of $Q$=25 CMs, each with $K$=9 cells. A subset of the U-wts (blue) increased at T=0

of sequence S0 = [AX] from the active pixels to the cells comprising the winning SDC. (c) The memory trace assigned to S0 to which we will compare (in Figure III-2) the memory traces assigned to the other five sequences in Figure III-1a. The green arrow represents the learning that occurs in the recurrent H-matrix from the 25 winners at T=0, when A is presented, to the 25 winners at T=1, when X is presented. The blue (magenta) arrows represent the learning in the U (D) matrix on each of the two time steps.

During learning, on each frame of an input sequence, an L1 code is chosen using the learning mode CSA (Table II-1). Then, associative learning occurs from active L0 units (active pixels) to active L1 units: these U-wts are set high, i.e., they are effectively binary. Also, on the second frame (T=1), H wts from L1 units active at T=0 to currently active L1 units are set high. Figure III-1c shows the memory trace assigned to S0. The trace consists of two SDCs. One might also refer to the set of weight increases made during presentation of S0 as the "memory trace", however, it is the sequence of SDCs across time steps which, unless otherwise stated, we refer to as the memory trace of a sequence. Note that because [AX] is the first sequence presented to the model, the particular units chosen on both frames of S0 are chosen at random.

Figure III-2 shows, in panels b-f, the memory traces assigned to five sequences, [BX], [CX], [DX], [EX], and [FX], which are progressively less spatiotemporally similar to [AX]. In addition, Figure III-2a shows the memory trace reactivated in response to a second presentation of [AX]. For each of the experiments represented by the six panels of Figure III-2, the sequence shown is presented as the second sequence experienced by the model. For example, when S4=[EX] is presented, it is presented to the model after the model has only learned [AX], *not* the rest of the intervening sequences, S1-S3.

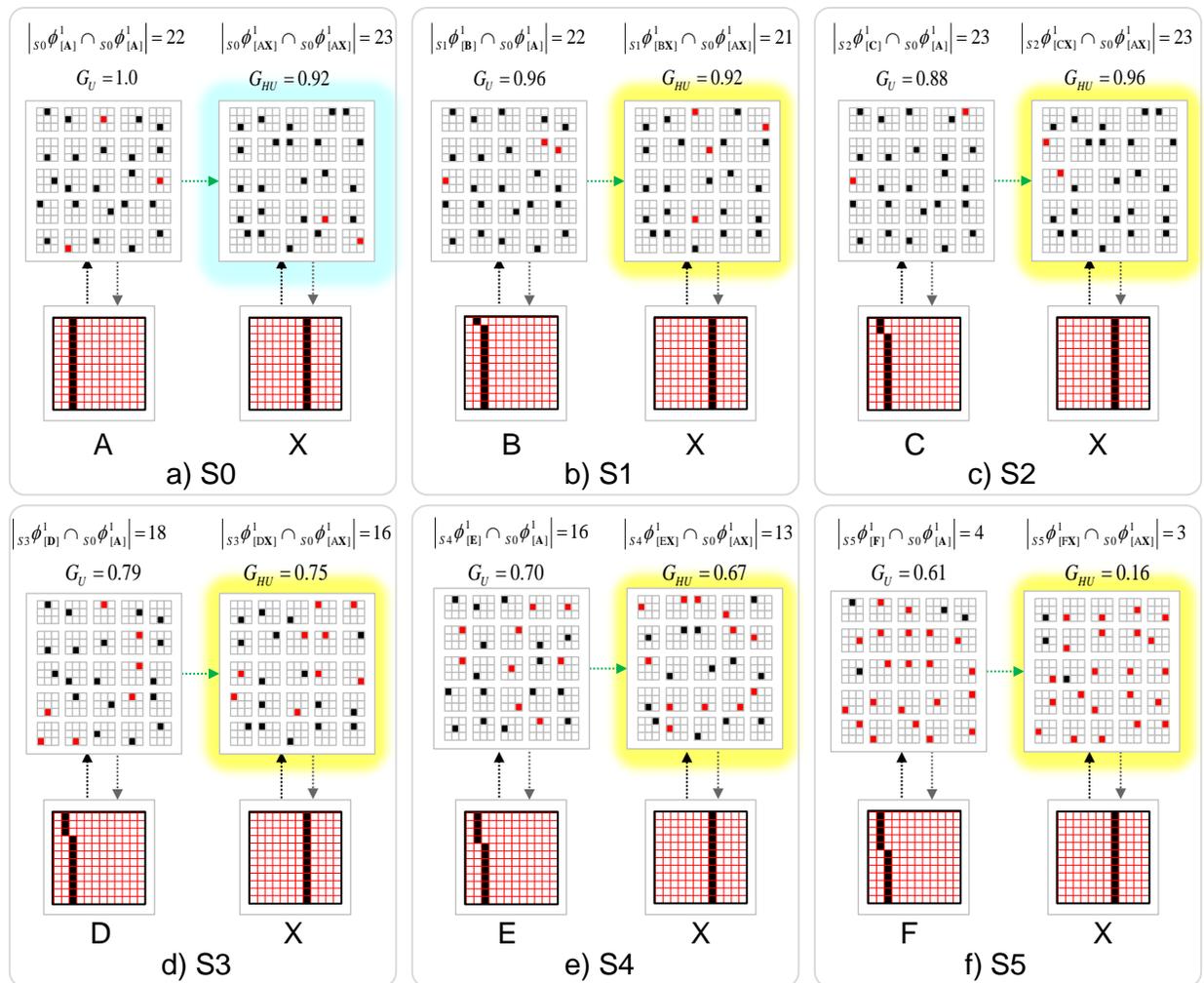

**Figure III-2:** Memory traces assigned to specific instances of the six sequences of Study 1. The basic SISC property can be seen in the decreasing intersection size of the L1 codes assigned to the second moment of each sequence (highlighted in yellow) to the L1 code assigned to the second moment of S0 (in Figure III-1c), [A**X**] (black units are those that do intersect, red are those that do not). The $G$ values are the model's estimates of spatiotemporal similarity of the current moment. Note that the same trend of intersection size decreasing with similarity can be seen in comparing the first moments of each sequence, S1-S5, with the first moment of S0. However, strictly that is a purely spatial similarity measure since no temporal context signals present on the first moment of a sequence.

The main result visible in Figure III-2 is that in comparing the L1 codes assigned to frame 2 of each sequence, S1 to S5, to the L1 code assigned to frame 2 of S0 (in Figure III-1c), we see progressively smaller intersection. These five L1 codes are highlighted in yellow. Black units are units which are the same as for frame 2 of sequence [AX] (Figure III-1c); red units are different.[3] Thus, on the second moment, [B**X**],

---

[3] If we viewed the presentations of S1 to S5 as recognition trials in which we were presenting progressively more perturbed variants of [AX], then these red units would be considered errors. However, in this case, we are viewing these as presentations of similar but not identical sequences to S0, in which case it is appropriate for the model to assign unique codes. In this case, the red units are not errors, but simply just different from the unit chosen in the corresponding CM in frame 2 of S0.

 http://dx.doi.org/10.3389/fncom.2014.00160

of sequence S1, the code assigned, $_{S1}\phi^1_{[BX]}$, has 21 out of the maximum possible 25 units in common with the code, $_{S0}\phi^1_{[AX]}$, assigned to the second moment, [A**X**], of S0, i.e., $\left|_{S1}\phi^1_{[BX]} \cap \, _{S0}\phi^1_{[AX]}\right| = 21$. Note that we have slightly generalized the code name convention: the lead subscript indicates the sequence in which the code occurs. As the spatiotemporal similarity of the second sequence moment with [A**X**] decreases further across panels c-f, the intersection of the assigned code with $_{S0}\phi^1_{[AX]}$ trends downward, despite the fact that in this particular instance, $\left|_{S2}\phi^1_{[CX]} \cap \, _{S0}\phi^1_{[AX]}\right| = 23$ even though [C**X**] must clearly be considered less similar to [A**X**] than [B**X**] is to [A*X*]. Despite this statistical blip, the codes assigned for the remaining progressively less spatiotemporally similar moments, [D**X**], [E**X**], and [F**X**], have monotonically decreasing intersection with $_{S0}\phi^1_{[AX]}$ as summarized in the right column of Table III-1. In fact, the same trend obtains with respect to the first sequence moment as well (left column). However, note that in the latter case, it is purely spatial similarity in the input space that is relevant (since no temporal context information is present on the first moment of a sequence).

**Table III-1:** Code similarity decreases with spatiotemporal similarity of moments.

| **Decreasing Similarity of 1st Moment** | **Decreasing Similarity of 2nd Moment** |
|---|---|
| $\left|_{S0}\phi^1_{[A]} \cap \, _{S0}\phi^1_{[A]}\right| = 22$ (88%) | $\left|_{S0}\phi^1_{[AX]} \cap \, _{S0}\phi^1_{[AX]}\right| = 23$ (92%) |
| $\left|_{S1}\phi^1_{[B]} \cap \, _{S0}\phi^1_{[A]}\right| = 22$ (88%) | $\left|_{S1}\phi^1_{[BX]} \cap \, _{S0}\phi^1_{[AX]}\right| = 21$ (84%) |
| $\left|_{S2}\phi^1_{[C]} \cap \, _{S0}\phi^1_{[A]}\right| = 23$ (92%) | $\left|_{S2}\phi^1_{[CX]} \cap \, _{S0}\phi^1_{[AX]}\right| = 23$ (92%) |
| $\left|_{S3}\phi^1_{[D]} \cap \, _{S0}\phi^1_{[A]}\right| = 18$ (72%) | $\left|_{S3}\phi^1_{[DX]} \cap \, _{S0}\phi^1_{[AX]}\right| = 16$ (64%) |
| $\left|_{S4}\phi^1_{[E]} \cap \, _{S0}\phi^1_{[A]}\right| = 16$ (64%) | $\left|_{S4}\phi^1_{[EX]} \cap \, _{S0}\phi^1_{[AX]}\right| = 13$ (52%) |
| $\left|_{S5}\phi^1_{[F]} \cap \, _{S0}\phi^1_{[A]}\right| = 4$ (16%) | $\left|_{S5}\phi^1_{[FX]} \cap \, _{S0}\phi^1_{[AX]}\right| = 3$ (12%) (~chance) |

We emphasize that each of the memory traces shown in Figure III-2 is a particular instance. The winner in a CM is chosen as a draw from a likelihood distribution over the CM's units, i.e., "softmax" (CSA Step 12), *not* by simply choosing the max likelihood unit, i.e., plain ("hard") max. Thus, we will generally see some variation in the chosen codes across instances of the same experiment and the amount of variation will increase as the similarity of the test sequence to the learned sequence, [AX], decreases. This statistical variation, for example, is why the memory trace in Figure III-2a is not perfect. Due to the statistical nature of Sparsey's CSA, demonstration of the SISC property requires running many instances of each of the experiments shown in Figure III-2 and reporting average results. Such a protocol was followed in Study 2.

## III.B. STUDY 2: SINGLE-TRIAL LEARNING OF SETS OF LONGER SEQUENCES

Study 2 demonstrates single-trial learning of longer and more complex sequences, derived from natural video, by a model with multiple internal levels. We presented eight 20-frame 24x24-pixel, natural-derived, snippets (movies), produced from the KTH Video data set (Schuldt, Laptev et al. 2004). All 160 frames of the eight snippets are shown in Figure III-3. These are taken from instances of people waving their arms. See example video. The snippets were presented once each.

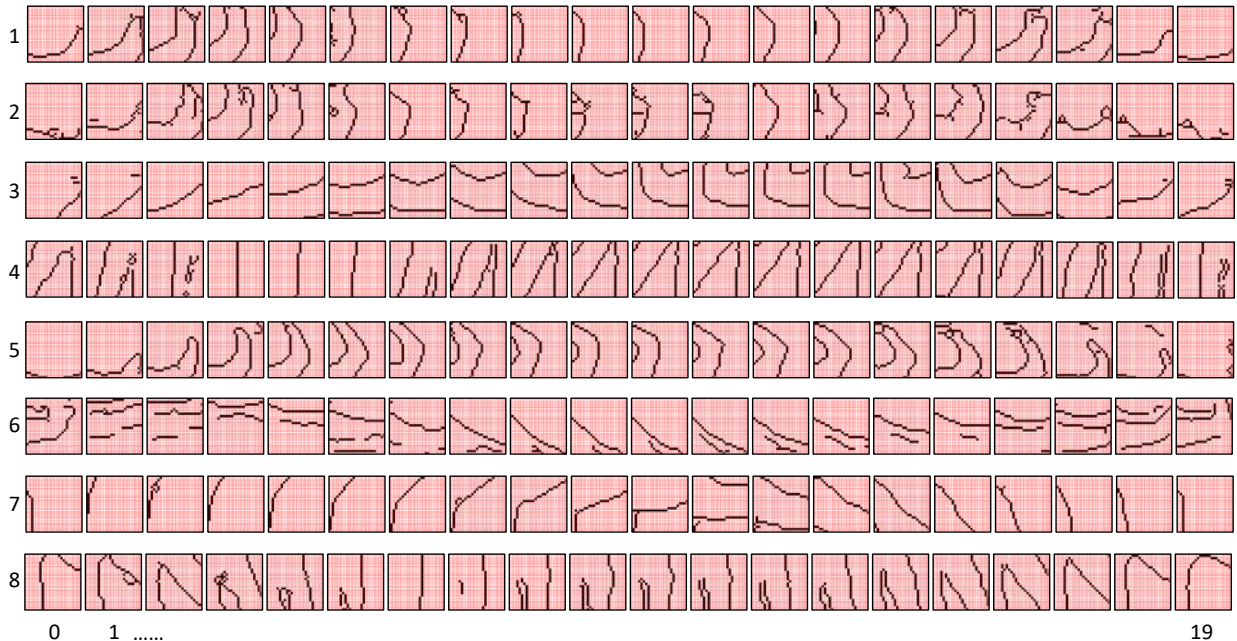

**Figure III-3:** The frames of the eight snippets used in Study 2.

The model in Study 2 had 4 levels, a total of 21 macs, 3,285 cells ("neurons"), and 1,880,568 synapses.[4] As shown in Figure III-4, the first internal level (L1) had sixteen macs, each consisting of $Q_1=9$ CMs, each having $K_1=16$ cells. L2 had 4 macs, each having of $Q_2=9$ CMs, each having of $K_2=9$ cells. The top level (L3) consisted of one mac, consisting of $Q_3=9$ CMs, each with $K_3=9$ cells. The semi-transparent blue prisms indicate the bottom-up (U) wiring scheme. Each 6x6-pixel *aperture* of the input level, L0, *U-connects* to all 9×16=144 cells in the corresponding L1 mac. Each L1 mac U-connects to all 9×9=81 cells in the overlying L2 mac. All four L1 macs forming one quadrant of level L1 U-connect to the same overlying L2 mac (i.e., convergence). All four L2 macs U-connect to all 9×9=81 L3 cells (more convergence). The figure is a snapshot of the model while processing frame 15 of Snippet 1. L1 mac activation criteria were set in this study so that an L1 mac would only become active if between 5 and 7 (of the 36) pixels in its aperture were active: apertures with too few or too many active pixels are grayed out. Criteria were set to allow an L2 mac to become active if between 1 and 4 of its four afferent L1 macs were active, and to allow the L3 mac to become active if between 1 and 4 of its four afferent L2 macs were active.

---

[4] The model included an additional 82,944 top-down (D) synapses from cells at the first internal level (L1) to cells at the input level (L0). However, these synapses are neither required for nor used during recognition and thus, are not counted in computation of information storage capacity in bits/synapse.

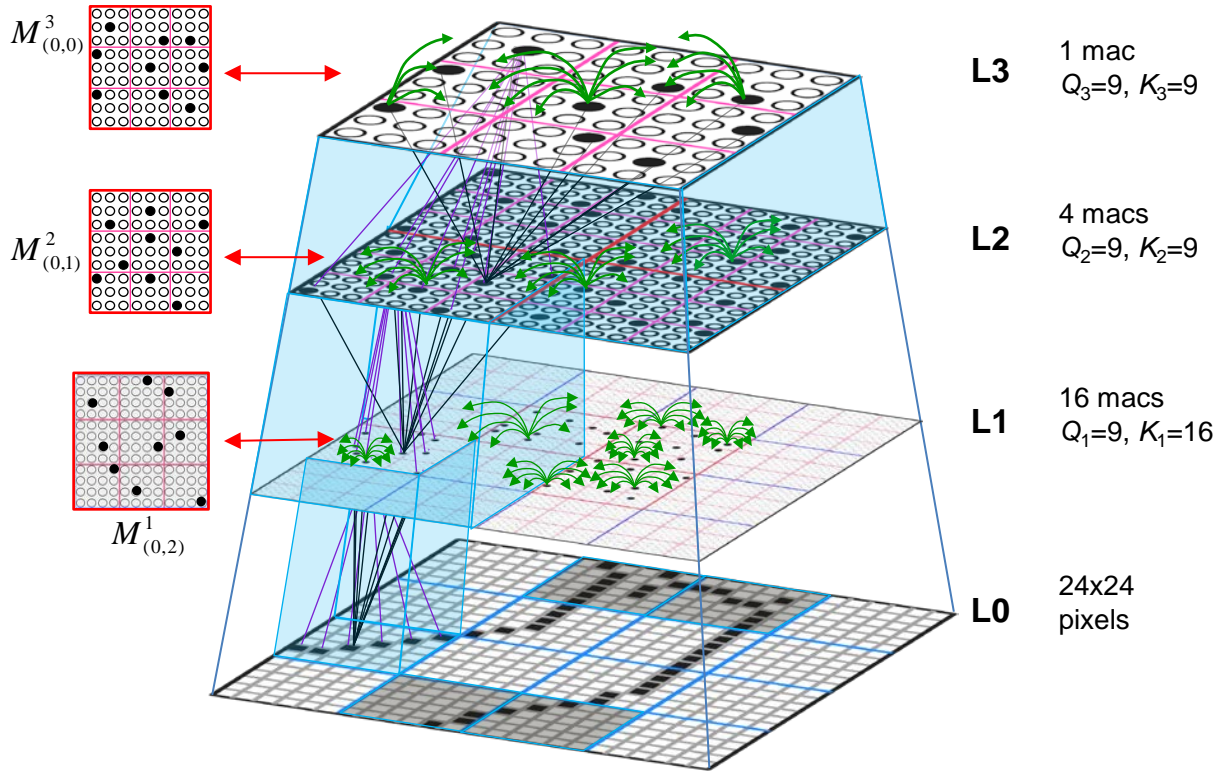

**Figure III-4:** The 4-level model used in Study 2. Blue semi-transparent prisms show the U-RFs of the macs at the prism tops, i.e., the L3 mac $M^3_{(0,0)}$'s U-RF contains all four L2 macs, each L2 mac's U-RF contains the four underlying L1 macs, and each L1 mac's U-RF is the underlying 6x6-pixel L0 aperture. At left, we show plan views of individual macs showing their active codes at the particular spatiotemporal moment depicted (T=15 of Snippet 1). The four gray L0 apertures have too few/many active pixels for their associated L1 macs to activate. Subsets of the U (black), H (green), and D (magenta) wts are shown.

Before discussing the features learned by several of the model's macs, we first report the recognition accuracy. The core accuracy measure, $\Gamma(x, x')$, is the similarity (normalized intersection) of the codes (SDCs) active in a given mac $M_i^j$ on a given frame $t$ during the learning and test presentations of a snippet $x$ as in Eq. 15, where we normalize by the fixed size $Q_j$ of codes in macs at level $j$. Note that the test presentation of $x$ is denoted as $x'$. We can then average over all macs at all levels to get the recognition accuracy for the whole network on frame $t$ of the test trial for $x'$, $R_t(x')$, as in Eq. 16. (Note that since these studies involve exact-match recognition, where the test and training snippets are identical, we can drop $x$ from the notation.) We can then average over all $T$ frames of the test trial to get the recognition accuracy of the entire hierarchical spatiotemporal memory trace for snippet $x'$, $R^*(x')$, as in Eq. 17. We also report the full network accuracy on just the last frame of the test snippet, $R^\Omega(x')$, which is just Eq. 16 with $t$ equal to the final frame of the snippet.

$$\Gamma_t^{j,i}(x, x') = \left| {}_x\phi_t^{j,i} \cap {}_{x'}\phi_t^{j,i} \right| / Q_j \qquad \text{(Eq. 15)}$$

$$R_t(x') = \sum_{j=1}^{J} \sum_{i=1}^{M_j} \Gamma_t^{j,i}(x, x') \qquad \text{(Eq. 16)}$$

$$R^*(x') = \sum_t R_t(x')\Big/T \qquad \text{(Eq. 17)}$$

Table III-2 reports $R^*(x')$ and $R^\Omega(x')$ for all snippets (and broken down by level as well) and averaged across all snippets (bottom row). It provides these results using the two CSA retrieval modes described in Section II, the *probabilistic* mode (columns 5 and 6), which is identical to the learning mode except that it uses the back-off protocol, and the *simple* mode (columns 3 and 4), which simply chooses the cell with the maximum *V* in each CM as winner (i.e., without using the mac-global information, *G*). The first point to make regarding Sparsey's performance on this set is that using the simple retrieval mode, it achieves an overall accuracy across all frames of all episodes of 85% and across all final frames of 91%. One can readily see that the simple retrieval mode does far better than the probabilistic mode. But again, the simple mode presumes that the model "knows" that it is operating purely in retrieval mode.

As noted above, these are *exact-match* recognition tests: the test sequences are identical to the training sequences. One might therefore be underwhelmed by anything less than 100% recognition. After all, in classification experiments, perfect classification of all training inputs is typically considered a basic sanity test. However, our *R* measures are not reporting the *class* of the test sequences: they are reporting the detailed comparison of the hierarchical, spatiotemporal patterns of activation that occur during the test and training trials. (Note: we refer to the activation pattern that transpires on the test trial as a *memory trace* and to to the one that transpires on the training trial as the *learning trace*.) In this study, these traces span four levels, 20 time steps, involve precisely ordered activation of 1-2 thousand neurons, and are formed with one trial. Figure III-5 gives some idea of this complexity: it shows the full 4-level learning trace for the first four frames of Sequence (Snippet) 1.

Thus, despite being less than perfect on this exact-match recognition experiment, we consider this performance (in the simple retrieval mode) to be good. Bear in mind that these experiments reflect very little in the way of parameter optimization: the model parameter space is very large and its exploration will be ongoing for quite some time. Moreover, we anticipate that there are many possible straightforward model modifications that would likely boost performance without increasing the model's time complexity for either learning or retrieval. For example, many of the static parameters in the CSA equations could be made dynamic, e.g., to depend on temporal offset from start of sequence, or on degrees of saturation of weight matrices, etc. There is a very large landscape to explore here. Furthermore, as noted, this study involved only unsupervised learning. As discussed in Section II.A.14, the addition of supervised learning to the model greatly increases its capabilities, i.e., to learning arbitrarily nonlinear (spatiotemporal) categories. However, we do not report supervised learning studies in this paper.

 

**Table III-2:** Recognition Accuracy for Study 2

| Snippet | Level | Larger Model | | | | Smaller Model | |
|---|---|---|---|---|---|---|---|
| | | Simple Mode | | Probabilistic Mode | | Simple Mode | |
| | | $R^*(x')$ | $R^\Omega(x')$ | $R^*(x')$ | $R^\Omega(x')$ | $R^*(x')$ | $R^\Omega(x')$ |
| 1 | L3 | 0.56 | 0.78 | 0.46 | 0.67 | 0.53 | 1.00 |
| | L2 | 0.90 | 1.00 | 0.73 | 0.50 | 0.77 | 1.00 |
| | L1 | 0.95 | 0.97 | 0.71 | 0.53 | 0.82 | 0.88 |
| 2 | L3 | 0.53 | 0.33 | 0.29 | 0 | 0.39 | 0.50 |
| | L2 | 0.84 | 1.00 | 0.39 | 0 | 0.55 | 0.50 |
| | L1 | 0.86 | 1.00 | 0.55 | 0.37 | 0.72 | 0.75 |
| 3 | L3 | 0.57 | 0.56 | 0.26 | 0.67 | 0.69 | 0.75 |
| | L2 | 0.92 | 1.00 | 0.65 | 1.00 | 0.92 | 1.00 |
| | L1 | 0.98 | 1.00 | 0.74 | 0.89 | 0.98 | 1.00 |
| 4 | L3 | 0.85 | 0.78 | 0.86 | 0.78 | 0.90 | 1.00 |
| | L2 | 0.94 | 1.00 | 0.86 | 1.00 | 0.91 | 1.00 |
| | L1 | 0.94 | 1.00 | 0.83 | 0.95 | 0.93 | 1.00 |
| 5 | L3 | 0.63 | 0.89 | 0.46 | 0.56 | 0.84 | 1.00 |
| | L2 | 0.91 | 1.00 | 0.86 | 1.00 | 0.90 | 1.00 |
| | L1 | 0.94 | 1.00 | 0.85 | 0.93 | 0.93 | 1.00 |
| 6 | L3 | 0.86 | 1.00 | 0.69 | 1.00 | 0.81 | 1.00 |
| | L2 | 0.97 | 1.00 | 0.82 | 0.83 | 0.92 | 1.00 |
| | L1 | 0.96 | 1.00 | 0.82 | 0.93 | 0.95 | 1.00 |
| 7 | L3 | 0.64 | 0.78 | 0.55 | 0.44 | 0.84 | 0.75 |
| | L2 | 0.95 | 1.00 | 0.91 | 0.33 | 0.99 | 0.88 |
| | L1 | 0.98 | 1.00 | 0.84 | 0.41 | 1.00 | 1.00 |
| 8 | L3 | 0.79 | 0.78 | 0.55 | 0.78 | 0.93 | 1.00 |
| | L2 | 0.95 | 1.00 | 0.85 | 0.89 | 0.94 | 1.00 |
| | L1 | 0.94 | 1.00 | 0.84 | 0.81 | 0.93 | 1.00 |
| All Snippets | All Levels | 85.0 | 91.0 | 0.68 | 0.68 | 0.84 | 0.92 |

**Key:** All accuracies expressed as decimal (between 0 and 1). $R^*(x')$ is averaged over *all* 20 frames of snippet, $x'$, where in this case (the exact-match test case), $x'$ is identical to the training snippet, $x$. $R^\Omega(x')$ is the accuracy only on the final frame of snippet $x'$.

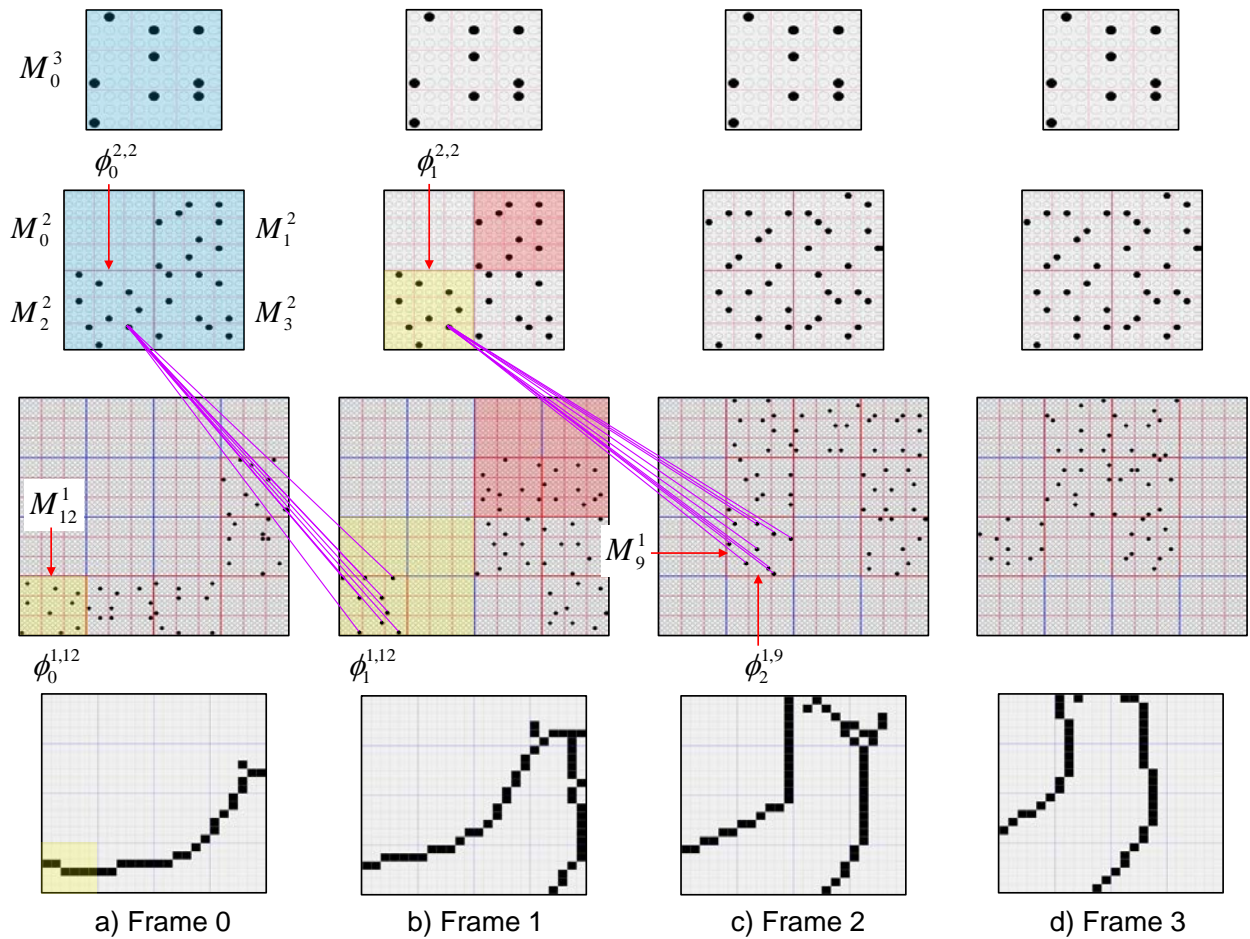

**Figure III-5:** The inputs (L0) and codes that become active at all three internal levels on the first four frames of Snippet 1. Note that the mac codes that become active at L2 persist for two time steps. Thus, the code active in mac, $M_2^2$, on frames 0 and 1 can be referenced by two names, $\phi_0^{2,2}$ and $\phi_1^{2,2}$. The magenta lines show the D-wts from one of the cells comprising $\phi_0^{2,2}$, which are increased onto $\phi_1^{1,12}$ on frame 1 and onto $\phi_2^{1,9}$ on frame 2. See text for further discussion.

While the simple mode performance is good, we note that the model in this case has almost 1.9 million weights. Thus, the information storage capacity here is rather low, approximately 0.018 bits/synapse. However, in the course of our investigations, we were routinely able to achieve the same or better performance on this data set with much smaller networks, e.g., 4-level networks with a total of 331,000 weights[5], yielding a storage capacity of ~0.1 bits/synapse, which is within an order of magnitude of the theoretical maximum for associative memory, ~0.69 bits/synapse (Willshaw, Buneman et al. 1969). Those results are given in the last two columns Table III-2.

---

[5] The smaller model had 4 levels, a total of 21 macrocolumns (macs), 1,692 cells ("neurons"), and 343,116 synapses. It had an additional 32,256 D synapses from L1 cells to L0 cells. However, these synapses are neither required for nor used during recognition and thus, are not counted in the computation of information storage capacity in bits/synapse. L1 consisted of 16 macs, each with of $Q_1=4$ CMs, and each CM consisting of $K_1=14$ cells. L2 had 4 macs, each having of $Q_2=4$ CMs, each having of $K_2=12$ cells. The top level (L3) consisted of one mac, consisting of $Q_3=4$ CMs, each with $K_3=7$ cells.

While this unsupervised learning study involves only the exact-match condition (the test inputs are identical to the training inputs), the more typical goal of an unsupervised learning study is to show that the model learns the *higher-order statistical structure* of the input space, or in terms we used earlier, that the model maps similar inputs to similar codes (SISC). Study 3 involves the non-exact-match condition (the test inputs differ from the training inputs) and directly demonstrates that the model retrieves the spatiotemporally best matching stored input given a novel input.

The effect of the lower and upper mac activation bounds on the number of active features needed for a mac to activate (see Section II.A.3) can also be seen in Figure III-5. For L1, $\pi_U^{1,-} = 5$ and $\pi_U^{1,+} = 7$ (we've added the level index to the superscript since these parameters can vary by level): thus only a few of the 16 L1 macs become active on each frame, e.g., five on Frame 0, six on Frame 1, etc. One such criterion-meeting L1 mac, $M_{12}^1$, and its L0 aperture (with six active pixels) are highlighted in yellow in Frame 0. As noted in Section II.A.3, for L2 and higher, the number of active features equals the number of active macs in a mac's U-RF. In this simulation, the bounds for L2 macs were $\pi_U^{2,-} = 1$ and $\pi_U^{2,+} = 4$ and the bounds for the L3 mac were $\pi_U^{3,-} = 1$ and $\pi_U^{3,+} = 3$. Thus, on Frame 1, we can see that L2 mac $M_2^2$ (yellow) is active because the number of active features in its U-RF, $\pi_U^{2,2}(1) = 1$, meets the criteria:

$$\pi_U^{2,-} = 1 \leq \pi_U^{2,2}(1) = 1 \leq \pi_U^{2,+} = 4$$

$M_1^2$ (rose) and $M_3^2$ (no color) also activate because they also meet the criteria:

$$\pi_U^{2,-} = 1 \leq \pi_U^{2,1}(1) = 2 \leq \pi_U^{2,+} = 4$$
$$\pi_U^{2,-} = 1 \leq \pi_U^{2,3}(1) = 3 \leq \pi_U^{2,+} = 4$$

The blue boxes indicate that L3 mac $M_0^3$'s U-RF is the entire L2 level; $M_0^3$ is active on all four frames because it meets its mac activation bound criteria on all for frames.

The *progressive persistence* property can also be seen in Figure III-5. The persistence at L2 is two frames, i.e., $\delta^2 = 2$. Thus, the L2 code (the set of 9 black cells) that becomes active in $M_2^2$ on Frame 0 remains active on Frame 1. That same L2 code, which (following earlier notation) we can denote, $\phi_0^{2,2}$, becomes D-associated with the L1 codes active in its U-RF on Frames 0 and 1, denoted $\phi_1^{1,12}$ and $\phi_2^{1,9}$, respectively. Magenta lines show the increased D-wts from one of the cells in $\phi_0^{2,2}$ to the L1 codes, $\phi_1^{1,12}$ and $\phi_2^{1,9}$, though the same increases would occur from the other eight cells comprising $\phi_0^{2,2}$ ($= \phi_1^{2,2}$) as well. Similarly, the code that becomes active in $M_2^2$ on Frame 2 remains active on Frame 3. L3 persistence is $\delta^3 = 4$, thus the code activated in $M_0^3$ on Frame 0 remains active until Frame 3.

The reader may note a discrepancy at L3 between the progressive persistence policy, which says that (during learning) once active, an L3 code will remain active for 4 frames, and the activation bounds, which in this simulation says that an L3 mac will only become active if it has between 1 and 3 active features in its U-RF, whereas on Frames 3 and 4, there are four active features in $M_0^3$ 's U-RF. The resolution is that persistence trumps the activation criteria: that is, the policy, during learning, is to allow a mac that has *already* become active to remain active for its full persistence regardless of how the number of active features in its U-RF changes throughout its persistence.

We also note that though not shown in Figure III-5, large numbers of (U, H, and D) synapses are increased within/between macs on each of these frames. This is especially true early in the system's life, when most input patterns that occur will be novel. In general, as more and more frames are experienced, fewer and fewer synapses are increased with each new frame. However, as described in Section II.A.13, the model has a "freezing" policy wherein, once a critical fraction of the weights of any of a mac's three afferent projections (U, H, or D) have been increased, *all* of that mac's afferent projections are frozen, preventing any further codes (i.e., features) from being stored in its basis. Freezing is necessary in order to avoid oversaturating the weight matrices, which would lead to information (memory) loss. Once a mac's learning is frozen, the set of features that has been stored in it, remains its permanent lexicon, or *basis*, for perceiving/recognizing all future inputs to it. Note that even if a mac's afferent matrices are frozen, its efferent matrices are not, meaning that previously stored codes in a frozen mac can still be efferently-associated with other codes following freezing.

Although none of the macs in the model in this study became frozen, the codes that were stored in the various macs across the 160 frames of the input set still constitute their learned feature bases. Figure III-6 shows the complete set of criteria-meeting inputs, i.e., having between $\pi_U^{1,-} = 5$ and $\pi_U^{1,+} = 7$ active pixels, which present to L0 Aperture 0 across all 160 frames. These 45 inputs constitute the learned feature basis of L1 mac $M_0^1$. Note the near-canonical nature of many of the patterns, e.g., perfect, or near-perfect vertical, horizontal, diagonal edges.

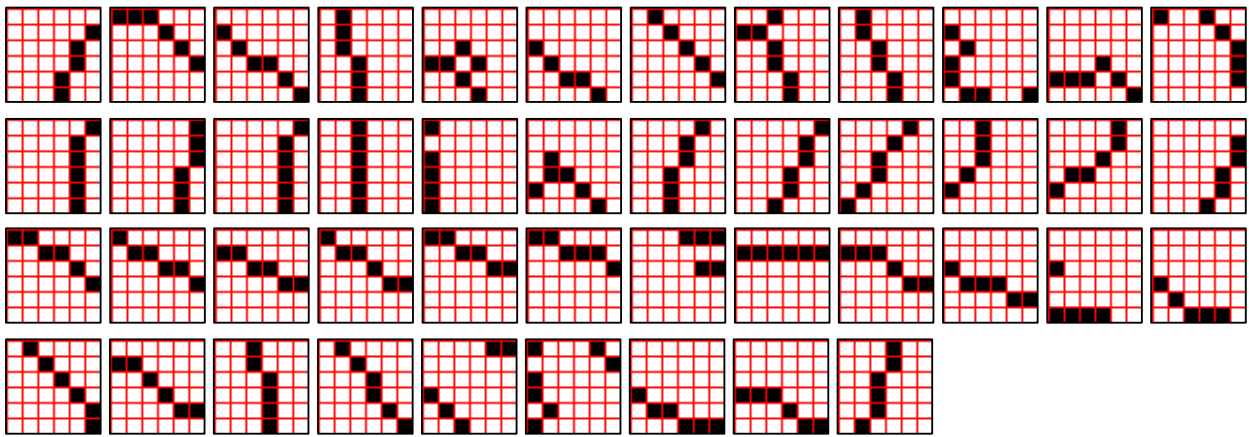

**Figure III-6:** The set of all unique patterns with between $\pi_U^{1,-} = 5$ and $\pi_U^{1,+} = 7$ active pixels that occurred in L0 Aperture 0 (and are stored in mac $M_0^1$) throughout the 160 frames of the training set.

As another example, Figure III-7 shows the complete set of unique, criteria-meeting patterns that occurred in Aperture 8, and were stored in $M_8^1$ over the course of the training set. Here, we manually ordered them so as to emphasize the "canonicalness" of the resulting features. In this case, seven of these features (blue underbars) occurred at least twice during the 160 frames. It is perhaps surprising that given such a small number of frames derived from natural video, the resulting basis can be so canonical. Moreover, several of these features are already beginning to recur in the input stream even within the first 160 frames of this model's experience. These phenomena are due to the conjunction of the preprocessing (1-pixel wide edges and binarization), the small aperture size, and the L1 mac activation criteria. Similar bases were learned in the other 14 L1 macs as well. These findings give us confidence that freezing L1 macs even very early in the "life" of the model, e.g., after a few hundred features have been stored, will allow the macs to parse/recognize all future inputs with quite sufficient fidelity. We feel these results provide an illuminating framework for understanding the various *critical period* phenomena observed in the visual and other modalities of biological brains (Wiesel and Hubel 1963, Barkat, Polley et al. 2011, Pandipati and Schoppa 2012).

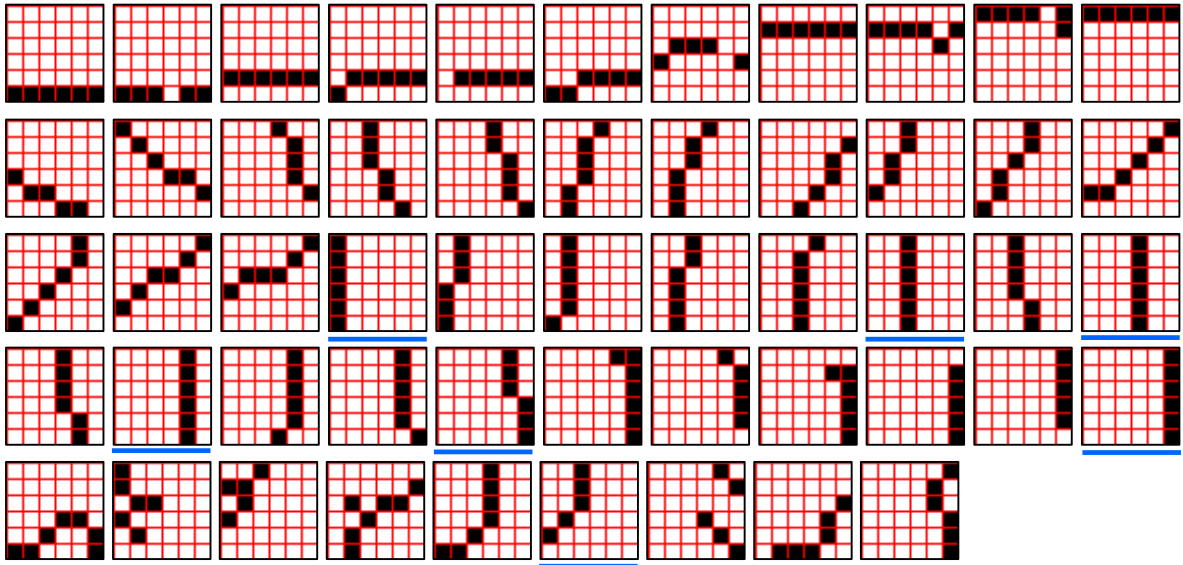

**Figure III-7:** The set of all unique patterns with between $\pi_U^{1,-} = 5$ and $\pi_U^{1,+} = 7$ active pixels that occur in L0 Aperture 8 (and are stored in mac $M_8^1$) throughout the 160 frames of the training.

We used the same protocol as above to catalog the input patterns learned by, and stored in, the L2 macs and in the L3 mac. Figure III-8 shows 78 of the 112 unique, criteria-meeting patterns that occurred in the 12x12-pixel region comprising the U-RF of L2 mac $M_0^2$, throughout the 160 frames of the training set (the thich-outlined green and red pairs are duplicates). This region is the union of the U-RFs of the four L1 macs, $M_0^1$, $M_1^1$, $M_4^1$, and $M_5^1$. The gray / yellow 6x6 quadrants are L0 apertures in which too many ($> \pi_U^{1,+} = 7$) / too few ($< \pi_U^{1,-} = 5$) pixels were active for the L1 mac to activate. Thus, when any of the 12x12 patterns in the figure occurs, the actual input passed up to $M_0^2$ will be from codes active only in the L1 macs whose 6x6 RFs are *not* gray or yellow.

As can be seen in Figure III-8, the spatial extent of the L2 RF has doubled in width and height compared to L1 RF. Thus, the space of possible inputs in such an RF is exponentially larger. Nevertheless, most of these larger features still have low intrinsic dimensionality, e.g., an essentially straight or low-curvature edge across the whole 12x12 RF. Even the more complex features such as the angle features in the bottom row, i.e., two straight/low-curvature segments with a single "elbow" point (thick pink outline), have rather low intrinsic dimensionality (i.e., we can give short verbal descriptions of them). Again, these are canonical-looking features, and they end up in the basis of $M_0^2$, but they were not hand-engineered. The number of active features (quadrants that are neither gray nor yellow) in each 12x12 pattern varies from 0 (in which case, $M_0^2$ will not become active on that frame, thick blue outline) to 4. Thus, $M_0^2$ learns input patterns having varying numbers of features (varying complexities). Thus, it is also the case that during retrievals, all these features, of varying complexities, formally compete with each other. In general, this argues for narrower mac activation ranges, [$\pi_U^-, \pi_U^+$], because narrower ranges make normalization easier. Exploration of the interaction of mac activation ranges across levels and with other parameters is another ongoing effort of our research.

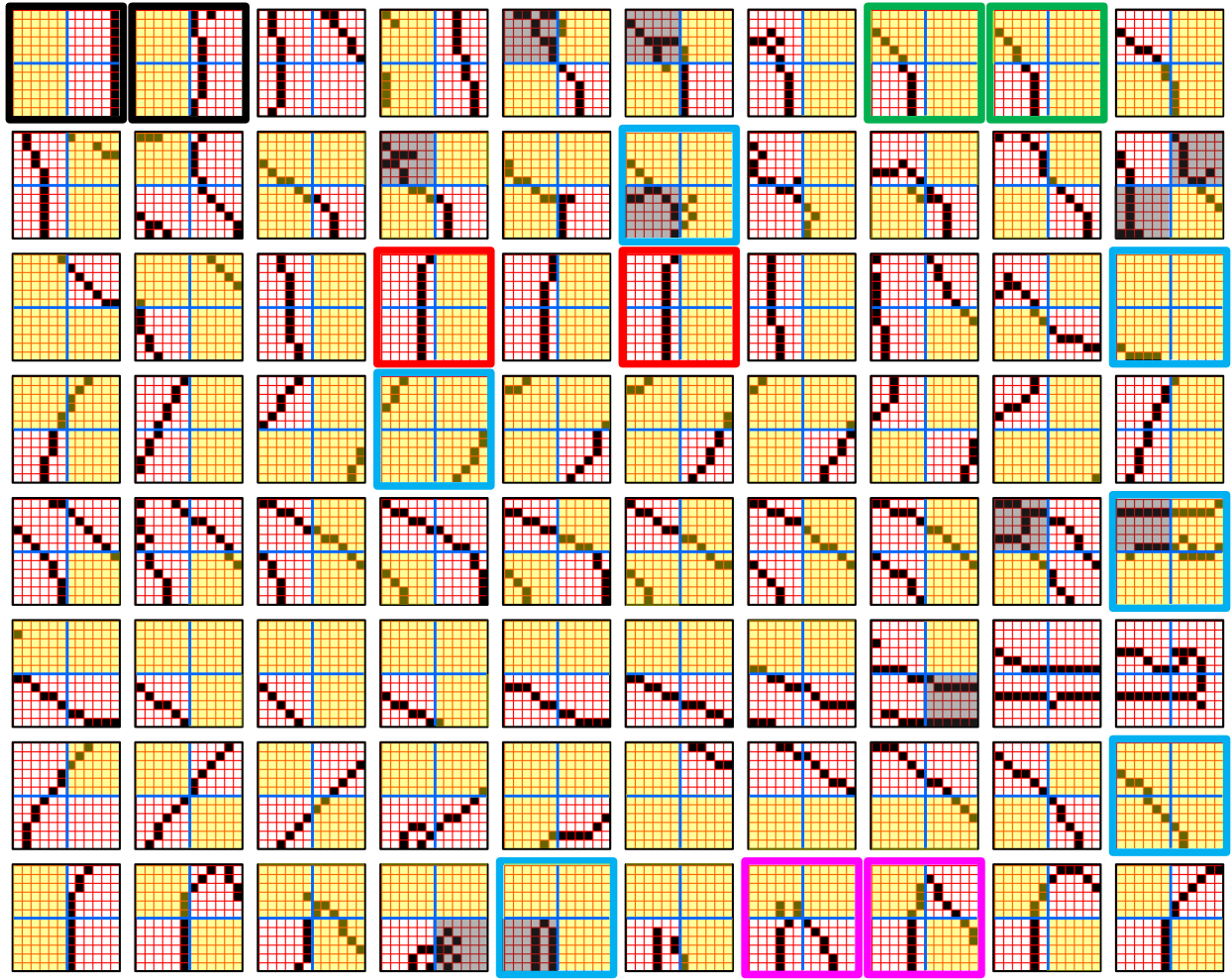

**Figure III-8:** 78 of the 112 unique patterns with between $\pi_U^{2,-}=1$ and $\pi_U^{2,+}=4$ active features that occurred in the U-RF of L2 mac $M_0^2$ throughout the 160 frames of the training set. Gray / yellow quadrants are ones in which too many / few pixels were active for the corresponding L1 mac to activate.

Note that since L2 codes persist for two frames, these input patterns, or to be more precise the SDCs in the corresponding L1 macs, will be associated to only roughly half as many codes in $M_0^2$. Thus, each consecutive pair of two 12x12 panels (in row-major order) would become associated with the same $M_0^2$ code. Figure III-9 illustrates this concept for the first pair of 12x12 panels of Figure III-8(thick black outline). Thus, the L2 codes formally represent spatiotemporal patterns. Given the discrete nature of our overall framework, i.e., discrete frames, binary pixels, constrained wiring schemes, these larger-scale (both spatially and temporally) spatiotemporal patterns, i.e., L2 features, can be viewed as spatiotemporal *compositions* of lower-level features. A detailed development and analysis of this spatiotemporal compositional aspect is one major focus of current and future studies.

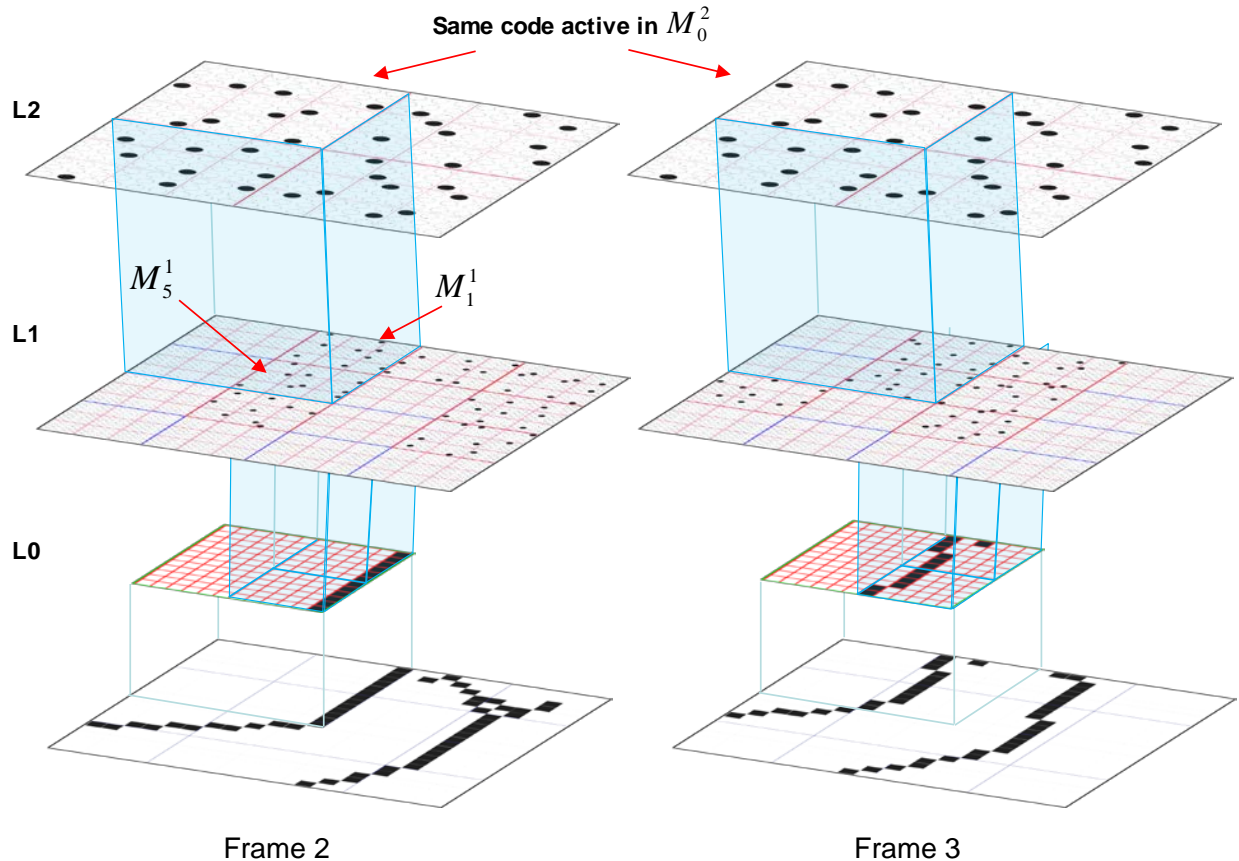

**Figure III-9:** All L1 codes that become active in $M_0^2$'s U-RF on the two frames depicted will associate with the same $M_0^2$ code. Notice that that same code is active for both frames. Thus, a total of four unique L1 codes—the two active in $M_1^1$ on Frames 2 and 3 and the two active in $M_5^1$ on Frames 2 and 3—will associate with the $M_0^2$ code shown. There will in addition be H-association from the L1 codes active on Frame 2 to those active on Frame 3 and also from the L2 code recurrently to itself (since it is on for two consecutive frames), and D-associations as well.

The concept of operation during learning and also during recognition, is one in which all of the macs across all levels operate, in parallel, on the particular spatiotemporal fragments of the input that they receive, dealing with variation on a fragment-by-fragment basis. Support for this view comes from recent experimental work (Bart and Hegdé 2012). In subsequent work, we will be quantitatively assessing the similarity of features that occur, over the long time frame of experience, following the initial period in which many of the lower-level macs become frozen, within apertures of the different scales corresponding to the model's different levels, to the (frozen) bases of those macs. The goal will be to assess how well the model is able to represent (and if novel, learn) future inputs using the fixed lexicon of features stored in its lower levels.

Finally, before leaving this section, we want to underscore the very different concept of feature basis present in Sparsey than that present in localist models such as (Olshausen and Field 1997). This difference is summarized in terms of four characteristics in Table III-3.

53                                                                 http://dx.doi.org/10.3389/fncom.2014.00160

**Table III-3: Comparison of the concept of "feature basis" present in Sparsey and localist models**

| 1. Origin of any single basis feature | |
|---|---|
| Sparsey | A single input pattern experienced, even with a single trial. |
| Localist | An average of many inputs experienced. |
| **2. Content of any single feature** | |
| Sparsey | Multiple spatial phases (i.e., multiple edge segments at different locations in the aperture), as apparent, e.g. in Figure III-8, multiple spatial frequencies, multiple orientations (i.e., each of the multiple possible edge segments in the aperture can have a different orientation), and multiple temporal frequencies. Because each Sparsey feature is derived from a single event (and not an average over multiple events), it's not really appropriate to speak of a Sparsey feature as having multiple modes on each of the encoded stimulus dimensions. |
| Localist | A single spatial phase, a single spatial frequency, a single orientation, and a single temporal frequency. In general, a localist feature such as these is unimodal (e.g., Gaussian, Gabor) on each of the encoded stimulus dimensions. |
| **3. Number of units in the code of any single feature** | |
| Sparsey | Many. $Q$, where, in real macrocolumns, $Q$ is order 100. |
| Localist | One |
| **4. Number of basis features participating in the representation of any single input (moment).** | |
| Sparsey | One. But again, that one active feature is represented by $Q$ active units. Thus, this type of representation is called "sparse" specifically because the number of physical units active in representing any one input is small compared to the total number of physical units. But, these representations can also be sparse in the senses typically used for localist models (below). As noted above, any single active SDC represents the presence of multiple (*but, for most natural inputs, a smallish number of*) spatial phases, spatial frequencies, orientations, and temporal frequencies. |
| Localist | Few, several. These representations are called "sparse" for two reasons.<br>• The number of features in a sufficient basis is small compared to the number of all possible features definable on the input space.<br>• The number of features active in the representation of any one input is small compared to the number of features in the basis. |

### III.C. STUDY 3: SPATIOTEMPORAL BEST-MATCH RETRIEVAL

In this study, we demonstrate spatiotemporal best-match retrieval as follows. In this case, we are again using a model with one internal level (L1) consisting of one mac with $Q=9$ CMs; $K$ varies across experiments. In each experimental run, we train the model on a set of random sequences. We then create a noisy version of each training sequence by randomly changing some fraction of the pixels in each of its frames. Figure III-10 (middle) shows a typical training sequence. Figure III-10 (top) shows the corresponding randomly produced noisy version of that sequence: one pixel was randomly changed in each frame, which actually yields two pixel-level differences between the original and the noisy frame. Each frame in the training set had between 9 and 12 active pixels, which yields noise levels from $2/9 = 22.2\%$ to $2/12 = 16.7\%$. Figure III-10 (bottom) shows a sequence produced from the middle one by randomly changing two pixels in each frame, which yields four pixel-level differences and thus noise levels, from $4/9 = 44.4\%$ to $4/12 = 33.3\%$. In this study, we ran one series of experiments testing with the 1-pixel-changed frames (columns 5-7 of Table III-3) and one series testing with the 2-pixels-changed frames (columns 8-10 of Table III-3).

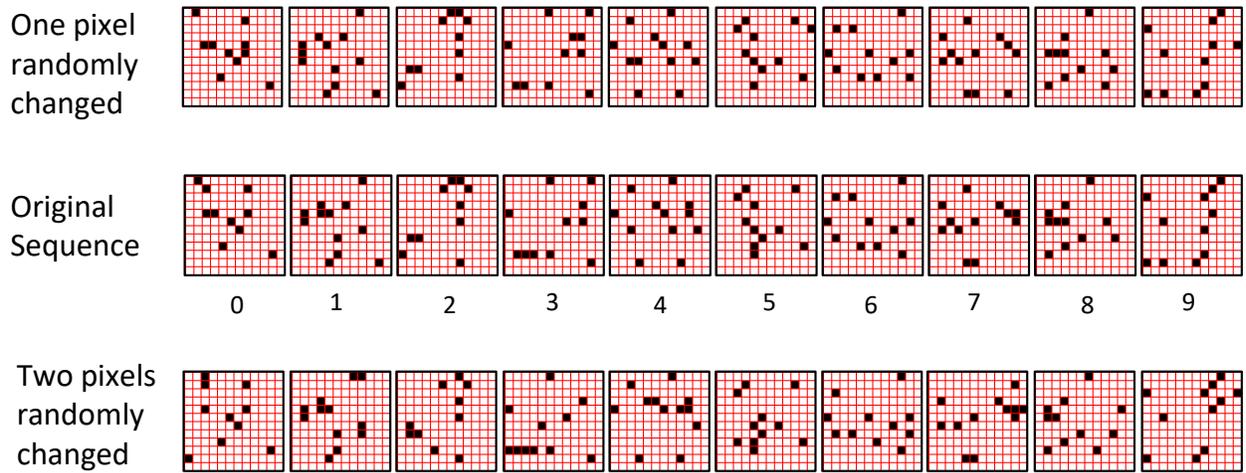

**Figure III-10:** (Middle row) An example 10-frame training sequence used in Study 3. (Top row) A noisy version of the training sequence in which one pixel was randomly changed in each frame. The resulting frame has two pixel-level differences from the original. (Bottom Row) A noisy version of the training sequence in which two pixels were randomly changed in each frame.

Given the random method of creating individual frames of the training set and the high input dimension involved (144), if the fraction of changed pixels is small enough, e.g., < 10-20%, then the probability that a changed frame, $x'$ will end up closer to (having higher intersection with) any other frame in the training set than to the frame, $x$, from which it was created, is extremely small. Moreover, remember that Sparsey actually "sees" each input frame *in the context of the sequence frames leading up to it*, i.e., it computes the spatiotemporal similarity of particular *moments* in time (by virtue of its combining of U and H signals on each time step), not simply the spatial similarity between isolated *snapshots*. Thus, the relevant point is that the probability that a changed moment, e.g., [x′,y′,**z**′], with its exponentially higher dimensionality ($144^3$), will end up closer to any other moment in the training set than to the moment, [x,y,**z**], from which it was created, is vanishingly small.[6] This condition is required to validate the testing protocol/criterion described above, which compares the L1 code on each test frame to the L1 code on the corresponding training frame. Thus, if we can show that the model activates the exact same sequence of L1 codes in response to the noisy sequence, then we will have shown that the model is doing spatiotemporal *best-match* retrieval.

---

[6] The notation [x,y,**z**], with z bolded, indicates the moment on which frame z is being presented as the third frame of a sequence after x and y have been presented as the first and second frames of the sequence.

Columns 5-7 of Table III-4 show that the model is able to recognize a set of training sequences despite significant noise on every frame. The absolute capacity increases with network size. For example, the network of Row 1 had 6,336 weights and showed good recognition of two 10-frame random sequences despite 16.7%-22.2% noise on each frame, while the larger network of Row 4 had 39,168 weights and showed very good recognition for 10, similarly noisy, 10-frame sequences, and so on. Columns 8-10 of Table III-4 show that the model still performs well even for much larger per-frame noise levels. In these tests, which involved randomly changing two pixels on every frame, the frame-wise noise levels varied from 33.3% (on frames which had 12 active pixels) to 44.4% (on frames with 9 active pixels). A key point to note in Table III-4 is that while the absolute capacities (the number of sequences that are can be stored) are lower for the 2-pixel-changed series compared to the 1-pixel-chaged series, capacity still remains large. The primary reason for lower storage capacity in the 2-pixel-changed case is that because the test input frames are less similar to the training input frames (than in the 1-pixel-changed case), the $\rho$ distributions from which winners in the CMs are chosen (Eqs. 11-12 of the CSA) will be flatter, yielding more single-unit errors, thus reducing $R^*(x')$.

**Table III-4: Best-Match Recognition Testing**

| Exp | $K$ | $Z$ | $W$ | 16.7% (1-pixel-changed) | | | 33.3% (2-pixels-changed) | | |
|---|---|---|---|---|---|---|---|---|---|
| | | | | $S$ | $R^*(x')$ | $R^\Omega(x')$ | $S$ | $R^*(x')$ | $R^\Omega(x')$ |
| 1 | 4 | 36 | 6,336 | 2 | 83.0 | 67.0 | 2 | 83.0 | 76.0 |
| 2 | 8 | 72 | 14,976 | 5 | 91.0 | 86.0 | 4 | 98.0 | 97.0 |
| 3 | 12 | 108 | 25,920 | 8 | 96.0 | 96.0 | 7 | 94.0 | 93.0 |
| 4 | 16 | 144 | 39,168 | 10 | 95.0 | 94.0 | 8 | 92.0 | 89.0 |
| 5 | 20 | 180 | 54,720 | 11 | 87.0 | 84.0 | 9 | 90.0 | 84.0 |
| 6 | 24 | 216 | 72,576 | 12 | 88.0 | 84.0 | 10 | 86.0 | 79.0 |
| 7 | 28 | 252 | 92,736 | 13 | 88.0 | 84.0 | 10 | 89.0 | 82.0 |
| 8 | 32 | 288 | 115,200 | 15 | 88.0 | 86.0 | 10 | 91.0 | 83.0 |

Key: The *R* measures (defined in the text) are in %. $Z=Q \times K$ is the total number of L1 units. W is the total number of U and H weights in the model. *S* is the number of sequences in the training set. All sequences were 10 frames long. $R^*(x')$ and $R^\Omega(x')$ are averages over the 10 runs of an experiment.

Table III-5 gives the detailed (frame-by-frame) accuracies for all sequences for individual runs of Experiments 1 and 4. The top two rows are for Experiment 1 in which the small network could store only two sequences while maintaining reasonably high recognition accuracy. The bottom ten rows are for a run of Experiment 4 in which the network had $Z=144$ L1 units and 39,168 weights. The rightmost column, $R^*(x')$, is the average over all 10 frames of a given sequence presentation. It is important to note how the model fails as it is stressed by having to store additional sequences. Specifically, even as accuracy averaged over all sequences falls, a subset of the stored sequences is still recognized perfectly. This can be seen even in the small network example: Seq. 1 is retrieved virtually perfectly. Only a single unit-level error is made on frame 6. Seq. 2 starts out being recalled perfectly for the first few frames but then begins picking up errors in frame 4 and hobbles along for the rest of the sequence. Nevertheless, note that even on the last frame of Seq. 2, the L1 code is still correct in 5 of the 9 CMs. In Experiment 4, we see that 9 of the 10

sequences are recalled virtually perfectly, while one (Seq. 9) begins perfectly but then picks up some errors on frame 5 and then degrades to 0% accuracy by the last frame. It is also important to realize that while the model occasionally makes mistakes, it generally recovers by the next frame. In other examples (not shown here), the model can often recover from more significant errors.

**Table III-5: Detailed Frame-by-Frame Accuracies**

| Seq | 0 | 1 | 2 | 3 | 4 | 5 | 6 | 7 | 8 | 9 | $R^*(x')$ |
|---|---|---|---|---|---|---|---|---|---|---|---|
| 1 | 100 | 100 | 100 | 100 | 100 | 100 | 88.9 | 100 | 100 | 100 | 99 |
| 2 | 100 | 100 | 100 | 100 | 55.6 | 66.7 | 66.7 | 66.7 | 55.6 | 55.6 | 77 |
|  |  |  |  |  |  |  |  |  |  |  |  |
| 1 | 88.9 | 100 | 100 | 100 | 100 | 100 | 100 | 100 | 88.9 | 100 | 98 |
| 2 | 100 | 100 | 100 | 100 | 100 | 100 | 100 | 100 | 100 | 88.9 | 99 |
| 3 | 100 | 100 | 100 | 100 | 100 | 100 | 100 | 100 | 100 | 100 | 100 |
| 4 | 88.9 | 100 | 88.9 | 88.9 | 100 | 100 | 100 | 100 | 100 | 100 | 97 |
| 5 | 100 | 100 | 100 | 100 | 100 | 100 | 100 | 100 | 100 | 100 | 100 |
| 6 | 100 | 88.9 | 100 | 100 | 100 | 100 | 100 | 88.9 | 100 | 100 | 98 |
| 7 | 100 | 100 | 100 | 100 | 100 | 100 | 88.9 | 100 | 100 | 100 | 99 |
| 8 | 88.9 | 100 | 100 | 100 | 100 | 100 | 100 | 100 | 100 | 100 | 99 |
| 9 | 100 | 100 | 100 | 100 | 100 | 77.8 | 66.6 | 22.2 | 11.1 | 0 | 68 |
| 10 | 100 | 100 | 100 | 100 | 100 | 100 | 100 | 100 | 100 | 100 | 100 |

Key: All table cells give accuracies as percent. Last column is average of columns indexed 0-9.

Figure III-11 shows the pair-wise L1 code intersections over the full set of frames experienced over all training and test frames (moments) of the experimental run described in the top two lines of Table III-5. Since there were two 10-frame sequences, this is a total of 40 frames. The upper yellow triangle shows the intersections between all codes assigned on the 10 frames of the training presentation of Seq. 1. Similarly for the other triangles down the main diagonal. The top value the green triangle (row 20, col. 1) shows that L1 code "20", i.e., the code activated on the first frame of the *test* presentation of Seq.1 intersects completely (in all $Q$=9 CMs) with L1 code 0, i.e., the code activated on the first frame of the *training* presentation of Seq. 1. Similarly, for codes, 21 and 1, 22 and 2, etc. Reading down the minor diagonal (between the red lines) tells how well the model does: perfect recognition of all noisy frames of all sequences would yield "9"s all the way down.

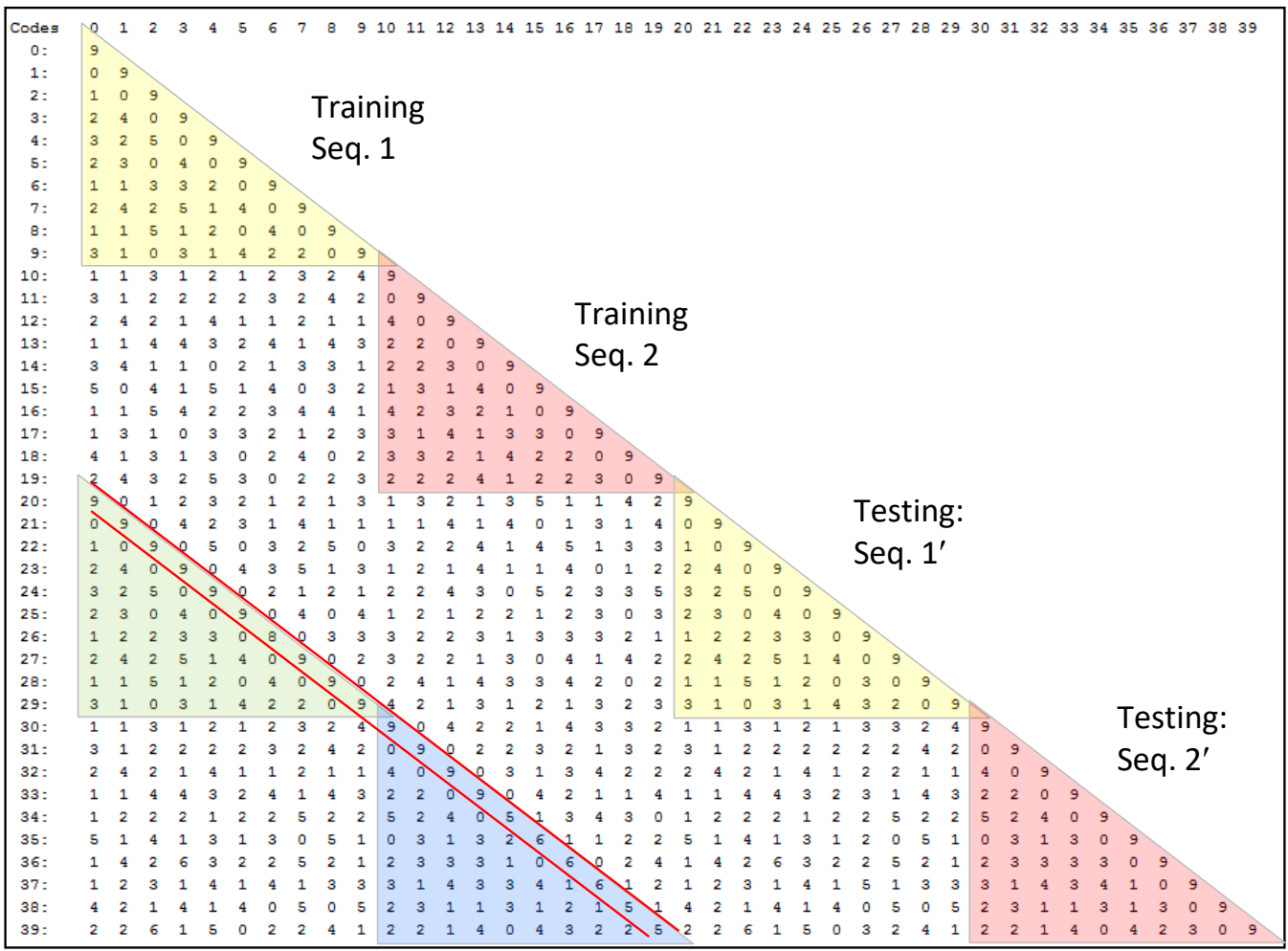

**Figure III-11:** Pair-wise intersections of all L1 codes assigned in one run of the 1-pixel-changed testing condition for the smallest model tested, which had $Q$=9, $K$=4, and 6,336 weights.

*Constant-Time Retrieval*

    When each frame is presented during a recognition test trial the likelihoods of all codes stored during the learning trial are formally evaluated. They are evaluated in parallel by the constant-time code selection algorithm (CSA). However, at no point does the model produce *explicit* representations of the likelihoods of the individual codes (hypotheses) stored. Such an explicit representation, e.g., a list, of likelihoods would constitute a localist representation of those likelihoods. What the model actually does is make $Q$ draws, one in each CM. However, the *net effect* of making these $Q$ draws (*soft-max* operations) is that a *hard-max* over *all* stored hypotheses is evaluated. This is true whether the model has stored a single 5-frame sequence, or a single 500-frame sequence, or 100 5-frame sequences. And crucially, because the numbers of CMs, and thus units, and weights, are fixed, the time it takes to make those $Q$ draws *remains constant* as additional codes (hypotheses) are stored.

    What the results in this report say is that that hard-max, i.e., the max-likelihood hypothesis, is returned with probability that can be very close to 1 if the amount of information (i.e., number of hypotheses) stored remains below a soft threshold, and which decreases as we move beyond that threshold. For example, looking at Table III-5, we see that for the second experiment (bottom 10 rows), the model chooses the correct, i.e., maximum likelihood, hypothesis on almost all of the 100 frames (moments) of test phase. These are 100 independent decisions, in each of which, all 100 stored hypotheses competed and had some

non-zero possibility of being activated. Yet, almost all 100 whole-code-level decisions were correct. And, at the finer scale of the individual CMs, where the actual decision process, albeit a soft decision process, takes place, almost all (861) of the 900 decisions were correct.

Table III-6 shows what happens when we move past or perhaps *through*, the aforementioned soft threshold. In these two experiments, we again used the network with 36,198 weights and the 1-pixel-changed test, but the training set contained 11 sequences (upper 11 rows) and 12 sequences (lower 12 rows), compared to only 10 in the experiment reported in Table III-5. For the 11-sequence case, the model still performs very well on six of the sequences, but adding another sequence degrades performance substantially more.

**Table III-6: Detailed Frame-by-Frame Accuracies. Overloaded Case.**

| Seq | 0 | 1 | 2 | 3 | 4 | 5 | 6 | 7 | 8 | 9 | $R^*(x')$ |
|---|---|---|---|---|---|---|---|---|---|---|---|
| 1 | 88.9 | 88.9 | 100 | 100 | 88.9 | 100 | 100 | 100 | 100 | 100 | 97 |
| 2 | 66.7 | 77.8 | 88.9 | 100 | 77.8 | 55.6 | 22.2 | 22.2 | 11.1 | 0 | 52 |
| 3 | 100 | 88.9 | 100 | 100 | 88.9 | 100 | 100 | 100 | 100 | 100 | 99 |
| 4 | 66.7 | 77.8 | 77.8 | 66.7 | 44.4 | 33.3 | 11.1 | 44.4 | 0 | 0 | 42 |
| 5 | 88.9 | 100 | 100 | 100 | 100 | 88.9 | 77.8 | 66.7 | 33.3 | 33.3 | 79 |
| 6 | 88.9 | 100 | 100 | 100 | 100 | 100 | 100 | 100 | 100 | 88.9 | 98 |
| 7 | 66.7 | 55.6 | 22.2 | 22.2 | 11.1 | 11.1 | 33.3 | 22.2 | 0 | 11.1 | 26 |
| 8 | 66.7 | 33.3 | 11.1 | 11.1 | 22.2 | 0 | 11.1 | 11.1 | 22.2 | 33.3 | 22 |
| 9 | 100 | 100 | 100 | 100 | 88.9 | 100 | 100 | 88.9 | 88.9 | 88.9 | 96 |
| 10 | 88.9 | 100 | 88.9 | 88.9 | 88.9 | 100 | 100 | 100 | 88.9 | 66.7 | 91 |
| 11 | 88.9 | 100 | 100 | 100 | 100 | 100 | 100 | 100 | 88.9 | 88.9 | 97 |
|  |  |  |  |  |  |  |  |  |  |  |  |
| 1 | 66.7 | 66.7 | 44.4 | 44.4 | 11.1 | 44.4 | 44.4 | 11.1 | 22.2 | 11.1 | 37 |
| 2 | 88.9 | 100 | 100 | 100 | 100 | 88.9 | 88.9 | 77.8 | 77.8 | 66.7 | 89 |
| 3 | 88.9 | 88.9 | 100 | 100 | 100 | 100 | 88.9 | 100 | 88.9 | 66.7 | 92 |
| 4 | 100 | 100 | 88.9 | 100 | 88.9 | 88.9 | 88.9 | 88.9 | 88.9 | 100 | 93 |
| 5 | 77.8 | 77.8 | 66.7 | 55.6 | 44.4 | 44.4 | 44.4 | 22.2 | 0 | 0 | 43 |
| 6 | 100 | 77.8 | 88.9 | 77.8 | 66.7 | 66.7 | 22.2 | 33.3 | 44.4 | 33.3 | 61 |
| 7 | 88.9 | 77.8 | 88.9 | 88.9 | 88.9 | 100 | 88.9 | 100 | 88.9 | 100 | 91 |
| 8 | 55.6 | 100 | 100 | 100 | 88.9 | 66.7 | 66.7 | 44.4 | 33.3 | 11.1 | 67 |
| 9 | 88.9 | 77.8 | 100 | 88.9 | 77.8 | 66.7 | 66.7 | 44.4 | 22.2 | 22.2 | 66 |
| 10 | 66.7 | 88.9 | 77.8 | 88.9 | 77.8 | 77.8 | 66.7 | 33.3 | 55.6 | 66.7 | 70 |
| 11 | 100 | 100 | 88.9 | 77.8 | 55.6 | 55.6 | 33.3 | 11.1 | 11.1 | 11.1 | 54 |
| 12 | 88.9 | 77.8 | 77.8 | 77.8 | 77.8 | 77.8 | 66.7 | 55.6 | 88.9 | 77.8 | 77 |

Key: All table cells give accuracies as percent. Last column is average of columns indexed 0-9.

## IV. SUMMARY AND CONCLUSION

In this paper, we described the hierarchical and spatiotemporal elaboration of the SDC-based macro/mini-column model of cortical computation described in Rinkus (2010), named Sparsey. The notion that *hierarchical representation* is essential to event recognition and intelligence more generally, has been present in models for decades (Fukushima 1984, Damasio 1989, Edelman and Poggio 1991, Riesenhuber and Poggio 1999, Lucke 2004, George and Hawkins 2005, Dean 2006, Jitsev 2010) including in the recent "Deep Learning" motif (LeCun and Bengio 1995, Hinton, Osindero et al. 2006, Hinton 2007a, Taylor, Fergus et al. 2010, Le, Zou et al. 2011). The representational and processing economy/efficiency of learning and recognition (inference) that is afforded by hierarchical decomposition of concepts/events has been understood (at least implicitly) for thousands of years, e.g., the game of "Twenty Questions", which works because of hierarchical way in which information is organized in our brains.

The hierarchical models noted above and many more all realize the benefit of compositional representation. However, most of those models use localist representations in which, in any given cortical patch, each feature/concept/event is represented by a single unit. In contrast, Sparsey uses sparse distributed codes (SDCs) in every cortical patch. As stated at the outset, the most important distinction between *localism* and SDC is that SDC allows the two essential operations of associative (content-addressable) memory, storing new inputs and retrieving the *best-matching* stored input, to be done *in fixed time for the life of the model*, which is essential for scalability to the huge problem sizes increasingly associated with label, "Big Data". The basis for this fixed-time capability was explained in Section I.A.

1. Because SDCs physically overlap, if one particular SDC, $\phi$ (and thus, the hypothesis that it represents), stored in a mac is *fully* active, i.e., if all $Q$ of $\phi$'s cells are active, *then all other codes (and thus, their associated hypotheses) stored in that mac are also simultaneously physically partially active in proportion to the size of their intersections with $\phi$.*[7]
2. Because the process/algorithm that assigns the codes to inputs (the code selection algorithm, CSA) enforces the *similar-inputs-to-similar-codes* (SISC) property, it follows that all stored inputs (hypotheses) are active with strength in descending order of similarity to (likelihood of) the hypothesis represented by $\phi$.

Crucially, since the $Q$ active (spiking) cells represent *all* stored hypotheses (with varying strengths), not just the single most likely hypothesis, $\phi$, it follows that *all of these hypotheses physically influence the next time step's decision processes*. Specifically, any stored hypothesis whose code has even one cell in common with $\phi$, will physically influence:

a) the $V$ distributions (and ultimately the $\rho$ distributions) in all CMs of all downstream macs on the next time step, and thus
b) the resulting likelihood distributions over all the stored hypotheses in each of the downstream macs on the next time step.

We emphasize that the representation of a hypothesis's likelihood (or probability) in our model—i.e., as the *fraction of the its code (of Q cells) that is active*—differs fundamentally from existing representations in which single neurons encode such probabilities in their (typically real-valued) scalar strengths of activation (e.g., firing rates) as described in the recent review of (Pouget, Beck et al. 2013).

---

[7] There is a nuance here. Although we say "all" stored hypotheses physically influence the next time step's decision processes, there may generally be a significant number of hypotheses stored in a mac, which have zero intersection with the current fully active code, $\phi$. One might therefore assert that these hypotheses do not physically influence the next time step's decision processes. While this is true, it still makes sense to say that *all* stored hypotheses are physically influencing subsequent decisions; it's just that the hypotheses having zero intersection with $\phi$ are so different from $\phi$ that they are appropriately viewed as having zero likelihood and thus as having no causal influence on subsequent decisions.

Another way of understanding the advantage of SDC over localism is that an *individual* machine operation on a single unit (cell), and moreover, on a single synapse—e.g., the addition of a synapse's weight into the input summation of a postsynaptic cell—transmits information about *multiple* items (hypotheses) represented in the synapse's presynaptic cell's mac. In stark contrast, in a localist model in which the presynaptic cell represents only one hypothesis, adding the synapse's weight into the input summation of a postsynaptic cell necessarily transmits information only about that *one* hypothesis. We believe this aspect of SDC—which qualifies as an instance of what has been termed *algorithmic*, or *representational*, *parallelism*—to be at the core of the biological brain's remarkable efficiency at processing information.

We also described several other important computational principles/mechanisms used in Sparsey:

1. How a single SDC code active in a mac can simultaneously represent two or more equally likely hypotheses and how information entering that mac on subsequent time steps can pare down the set of equally likely hypotheses (Section II.A.5).
2. How an important type of invariance, nonlinear time invariance, can be computed via a "back-off" policy that does not increase the time complexity of recognition (inference) (Section II.B). Essentially, on each frame, a mac computes a series of estimates of the match of the current temporal-context-dependent input (i.e., the current spatiotemporal *moment*) not just to the set of actual moments it experienced during learning (which constitute its explicit spatiotemporal basis), but to a much larger (encompassing) space of variants of the basis moments that were not actually experienced. This is similar in spirit to dynamic time warping (DTW) (Sakoe and Chiba 1978), but is far more efficient, again because of the underlying algorithmic parallelism.
3. How Sparsey can learn arbitrarily nonlinear and intertwined, i.e., "tangled", classes via supervised learning of associations between codes in different macs (Section II.A.14). That categories in the physical world are smooth in the neighborhood around any single exemplar but possibly very nonlinear and intertwined, i.e., "tangled", with other classes at the global scale has been pointed out by many, e.g., (Saul and Roweis 2002, Bengio 2007, Bengio, Courville et al. 2012). In particular, DiCarlo, Zoccolan et al. (2012) state as a next step the need to formally specify what is meant by "untangling local" subspace. We believe that Sparsey addresses this need. First, the CSA's two functions of storing (learning) and (best-match) retrieval of stored memories, can be viewed as a SISC-respecting content-addressable memory. Thus, individual macs handle the smooth category structure around individual exemplars: i.e., a novel input that is sufficiently similar to a known exemplar should activate an SDC with high intersection with the known exemplar's code and therefore exert similar downstream influence to that which would be produced by the familiar exemplar's code. The global highly nonlinear category structure is untangled by the hierarchy of macs, and specifically, by the ability (strongly subserved by progressive persistence) for multiple arbitrarily different codes in one cortical patch (e.g., one mac or set of macs) to be associated with a single code in another patch.
4. That, during learning, the CSA formally involves two rounds of competition amongst the mac's cells. In the first round, CSA Step 8, the $Q$ cells with the maximal V values in their respective CMs are determined and must activate (i.e., spike) so that their outputs can be summed and averaged to yield $G$. In the second round, CSA Step 12, a final winner is chosen in each CM according to the $\rho$ distribution in that CM, i.e., soft max. In general, the second round winners may differ (perhaps substantially, especially when $G \approx 0$) from the first round winners. This hypothesis that the canonical cortical computation involves two rounds of competition is a strong and falsifiable prediction.

5. And, that the concept of feature basis present in Sparsey differs markedly from that present in localist models such as (Olshausen and Field 1997), summarized in in Table III-3.

A great deal of work remains, particularly in understanding and mechanistically explaining the learning and usage (as in on-line rapid recognition/inference) of a hierarchy of spatiotemporal features. Even though Sparsey centers around a *single* canonical algorithm/circuit, the CSA [much of which was described (Rinkus 1996)], the ultimate algorithmic solution of cortex lives in what DiCarlo, Zoccolan et al. (2012) term a "very, very large space of details", which will take quite some time to explore, as suggested by Study II (Sections. III.B), which itself only begins to scratch the surface of the myriad parameter interactions that we would like to understand.